\documentclass[12pt]{article}
\usepackage{amsmath,amssymb,amsthm,amscd}
\usepackage{tikz}
\usepackage{jheppub}
\usepackage{microtype}
\pdfoutput=1

\def\tm{\tilde{m}}
\def\tX{\tilde{X}}
\def\tM{\tilde{M}}

\newcommand{\normord}[1]{\,\mathopen{:}\,#1\,\mathclose{:}\,}

\def\e{\textrm{e}}

\newcommand{\bea}{\begin{eqnarray}}
\newcommand{\eea}{\end{eqnarray}}
\newcommand{\be}{\begin{eqnarray}}
\newcommand{\ee}{\end{eqnarray}}

\newcommand{\im}{\textrm{Im}\,}

\newcommand{\qPoc}[1]{\left(#1;q\right)_\infty}

\title{$T[SU(N)]$ duality webs: mirror symmetry, spectral duality and
  gauge/CFT correspondences}

\author[1]{Anton Nedelin}
\author[1]{Sara Pasquetti}
\author[1,2]{Yegor Zenkevich}

\affiliation[1]{Dipartimento di Fisica, Universit\`a di Milano-Bicocca \& INFN, Sezione di Milano-Bicocca, \\
I-20126 Milano, Italy
}
\affiliation[2]{ITEP, Moscow 117218, Russia\\
}
\emailAdd{anton.nedelin@physics.uu.se}
\emailAdd{sara.pasquetti@gmail.com}
\emailAdd{yegor.zenkevich@gmail.com}

\abstract{We study various duality webs involving the $3d$ $FT[SU(N)]$
  theory, a close relative of the $T[SU(N)]$ quiver tail. We first map
  the partition functions of $FT[SU(N)]$ and its $3d$ spectral dual to
  a pair of spectral dual $q$-Toda conformal blocks. Then we show how
  to obtain the $FT[SU(N)]$ partition function by Higgsing a $5d$
  linear quiver gauge theory, or equivalently from the refined
  topological string partition function on a certain toric Calabi-Yau
  three-fold. $3d$ spectral duality in this context descends from $5d$
  spectral duality. Finally we discuss the $2d$ reduction of the $3d$
  spectral dual pair and study the corresponding limits on the
  $q$-Toda side. In particular we obtain a new direct map between the
  partition function of the 2d $FT[SU(N)]$ GLSM and an $(N+2)$-point
  Toda conformal block.}

\begin{document}
\begin{flushright}
ITEP/TH-43/17
\end{flushright}
\vspace{-1cm}

\maketitle

\section{Introduction}

Over the last decade, following Nekrasov's~\cite{Nekrasov:2002qd} and
Pestun's~\cite{Pestun:2007rz} seminal works, the application of the
localization technique to SUSY gauge theories on various manifolds,
has produced an unprecedented amount of exact results (for a
comprehensive review see \cite{Pestun:2016zxk} and references
therein).

Localized partition functions (or vevs of BPS observables) depend
on various parameters such as fugacities for the global
symmetries and  data specifying the background.
For  certain backgrounds partition functions do not depend on the gauge coupling 
and can be used to test Seiberg-like dualities and mirror symmetry in various
dimensions. 

The exact results obtained via localization have also led to the
discovery of AGT-like correspondences which provide dictionaries to
map objects in the gauge theories (partition functions, Wilson loops
vevs etc\ldots) to objects in different systems such as $2d$ CFTs or
TQFTs \cite{Alday:2009aq,Gadde:2009kb}.

It is interesting to study what happens when we take different limits
of the parameters appearing in the partition functions triggering some
sort of RG flows. For example, focusing on the global symmetry
parameters we can explicitly check how certain dualities can be
obtained by taking massive deformations of other dualities. We can
also consider limits involving the data specifying the background. For
manifolds of the form $M_{d-1} \times S^1$ we can explore what happens
when the circle shrinks and in particular gather hints on the fate of
dualities in $d$ dimensions: do they reduce to dualities in $d-1$
dimensions? In recent years these questions have been reconsidered
systematically in a series of papers
\cite{Aharony:2013dha,Aharony:2013kma,Aharony:2016jki,Aharony:2017adm}.

Another interesting procedure, the so-called Higgsing, involves
turning on the vev of some operator in a certain UV theory $T'$ which
triggers an RG flow to an IR theory $T$ that contains a codimension
two defect, the prototypical case being a surface operator in a $4d$
theory.  At the level of localized partition functions this procedure
can be implemented very efficiently and involves tuning the gauge and
flavor parameters of the mother theory $T$ to specific values.  At
these values $T$ typically develops some poles and picking up their
residues we obtain the partition function of the theory $T$ with a
codimension two defect
\cite{Gaiotto:2012xa,Mironov:2009qt,Kozcaz:2010af,Dimofte:2010tz,Dorey:2011pa,Nieri:2013vba,Gaiotto:2014ina}.

In this note we provide a concrete example where all these ideas and
techniques come together.  We discuss $3d$ mirror symmetry, spectral
duality and gauge/CFT correspondences and explore how they behave
under dimensional reduction and how they arise via Higgsing.

Our starting point is the $3d$ $T[SU(N)]$ quiver theory introduced in~\cite{Gaiotto:2008ak} as boundary conditions for the
$4d$ ${\cal N}=4$ supersymmetric Yang-Mills theories.  $T[G]$ has a
global symmetry group $G\times G^L$ acting respectively on the Higgs
and Coulomb branch.  The $T[G]$ has the remarkable property of being
invariant (or self-mirror) under $3d$ mirror symmetry which acts by
exchanging the Higgs and the Coulomb branches of the theory.

In this work we will consider a closely related quiver theory, the
$FT[SU(N)]$ theory, which contains an additional set of gauge singlet
fields. The $FT[SU(N)]$ theory is also self-dual under a duality which
we call {\it 3d spectral duality} since it descends from $5d$ spectral
duality.

In particular we discuss three webs of dualities:
\begin{itemize}
\item In Duality web I, we relate the $3d$ spectral pair $FT[SU(N)]\leftrightarrow \hat{FT}[SU(N)]$   to a  pair of  spectral dual $q$-CFT blocks via gauge/CFT correspondence.
\item In Duality web II, we view the $3d$ spectral dual pair as the
  result of Higgsing a pair of $5d$ spectral dual theories and the CY
  three-folds which geometrically engineer them.
\item In Duality web III,  we  reduce  the $3d$ spectral dual pair to $2d$ and study the corresponding limit of the  $q$-CFT blocks.
\end{itemize}

\subsection*{Duality web I}
\begin{figure}[h]
  \centering
\includegraphics{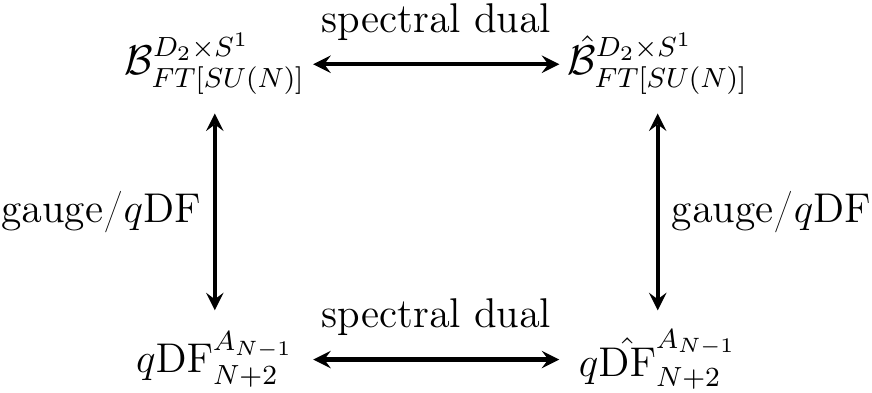}  
\caption{Duality web I represents the relation between $3d$ $FT[SU(N)]$
  quiver gauge theories and DF representations of the $N+2$-point
  $A_{N-1}$ $q$-Toda conformal blocks with $N$ degenerate
  primaries. $3d$ mirror symmetry of the gauge theories upstairs
  corresponds to the spectral duality of the CFTs downstairs.}
\label{fig1}
\end{figure}

Duality web I is shown in Fig.~\ref{fig1}.
In the top left corner we have~$\mathcal{B}^{D_2\times
  S^1}_{FT[SU(N)]}$, the $D_2\times S^1$ partition function, or
holomorphic block,\footnote{The background is actually twisted with
  twisting parameter $q$, i.e.\ $D_2$ is fibered over $S^1$ so that it
  gets rotated by $\ln q$ every time one turns around $S^1$. The
  notation $D_2 \times_q S^1$ would be more proper, however we omit
  the subscript $q$ for the sake of brevity.  The name holomorphic
  block is due to the fact that $D_2 \times_q S^1$ partition functions
  can be used to build partition functions on compact spaces, such as
  $S^3$ or $S^2\times S^1$
  \cite{Beem:2012mb,Pasquetti:2016dyl}\label{fn:1}.} of the
$FT[SU(N)]$ theory.  For this theory we turn real mass deformations
for all the flavors and topological symmetries, so that this theory
has $N!$ isolated vacua. As we will see the $FT[SU(N)]$ theory is
self-dual under the action of the spectral duality and correspondingly
in the top right corner we find the partition function $\hat{\cal
  B}^{D_2\times S^1}_{FT[SU(N)]}$ of the dual theory. This edge of the
web is a genuine duality between two theories flowing to the same IR
SCFT. However, here we are only looking at the map of the \emph{mass
  deformed} $D_2\times S^1$ partition functions which can be regarded
as a refinement of the map between the effective twisted
super-potential evaluated on the Bethe
vacua~\cite{Nekrasov:2009uh,Nekrasov:2009ui} of the two
theories~\cite{Gaiotto:2013bwa}. A thorough discussion of this duality
will be provided in~\cite{APZ}.

In Section \ref{secweb1} we discuss in detail the nontrivial map of
the $T[SU(N)]$ and $FT[SU(N)]$  holomorphic blocks under mirror symmetry and spectral duality using various
approaches including direct residue computations, the relation of the
holomorphic blocks to the integrable Ruijsenaars-Schneider (RS) system as in \cite{Bullimore:2014awa} 
and, in  Sec.~\ref{sec:higgsing-5d-gauge}, using the relation between holomorphic blocks, $5d$ gauge theories and refined topological strings.

The vertical edges of the web in Fig. \ref{fig1} represent
correspondences between gauge theories and conformal blocks akin to
the AGT
correspondence~\cite{Alday:2009aq,Wyllard:2009hg,Mironov:2009by}. One
can observe that the holomorphic block integrals
$\mathcal{B}^{D_2\times S^1}$ of $3d$ quiver theories can be directly
identified with the Dotsenko-Fateev (DF) integral representation of
the conformal blocks in $q$-deformed Toda theory. This correspondence
is part of the so called triality proposed
in~\cite{Aganagic:2013tta,Aganagic:2014kja} and generalized
in~\cite{Nieri:2015dts,Iqbal:2015fvd,Nedelin:2016gwu,Lodin:2017lrc}.

In the particular case of the $FT[SU(N)]$ theory we find that the
holomorphic block ${\cal B}^{D_2\times S^1}_{FT[SU(N)]}$ can be mapped
to the conformal block ${q{\rm DF}}^{A_{N-1}}_{N+2}$ involving $N$
fully-degenerate and two generic primaries, and a particular choice of
screening charges in the $q$-deformed $A_{N-1}$ Toda theory.  The dual
holomorphic block $\hat{\cal B}^{D_2\times S^1}_{FT[SU(N)]}$ is also
mapped to a ${q{\rm DF}}$ integral block $\hat{q{\rm
    DF}}^{A_{N-1}}_{N+2}$ which is related to ${q{\rm
    DF}}^{A_{N-1}}_{N+2}$ by a ``degenerate'' version of spectral
duality.  An exact meaning of this statement should become clear at
the end of the discussion of the Duality web II.

The details of the correspondence between holomorphic blocks of the
$FT[SU(N)]$ theory and $q$-Toda integral blocks as well as spectral
duality are presented in the Sec.~\ref{secweb1}.

\subsection*{Duality web II}

\begin{figure}[h]
\centering
\includegraphics{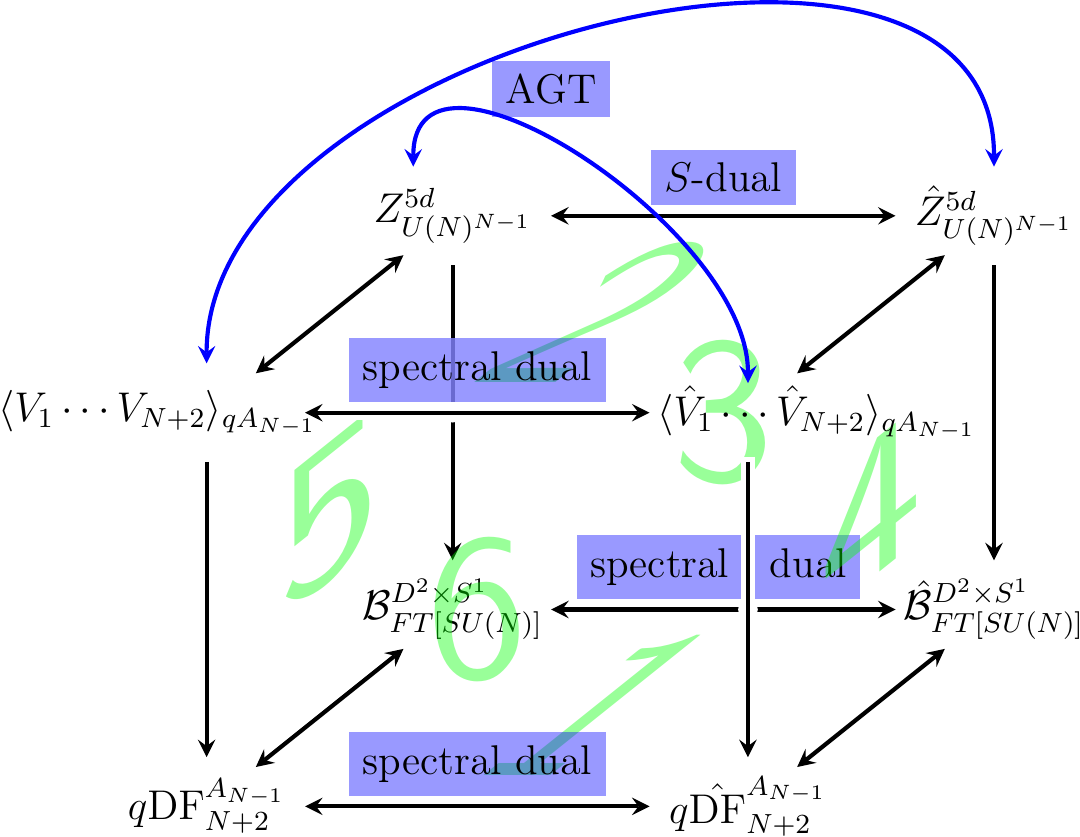}
\caption{Duality web II incorporates duality web I (face~{\color{green!60!black}\textsf{1}} of the
  cube) in a more general context of $3d$-$5d$-CFT triality.}
\label{fig2}
\end{figure}

Duality web I in fig. \ref{fig1} can actually be understood as a
consequence of another web of dualities involving $5d$ $\mathcal{N}=1$
quiver theories and correlators of generic (non-degenerate)
$q$-deformed Toda vertex operators. More precisely, we consider the
duality web II shown in Fig.~\ref{fig2} where duality web I
corresponds to the bottom face
(face~{\color{green!60!black}\textsf{1}}) of the cube.
In the top left corner we have the $5d$ $\mathcal{N}=1$ linear quiver
gauge theory with $(N-1)$ $U(N)$ gauge nodes and $N$
(anti-)fundamental matter hypermultiplets on each end of the
quiver. This theory is self-dual under $5d$ spectral duality which
relates $U(N)^{M-1}$ to $U(M)^{N-1}$ linear quiver theories
compactified on a circle.

This is a duality between two low energy descriptions of the same
strongly interacting UV SCFT which can be conveniently understood
using brane setup~\cite{Aharony:1997bh}. The details of the maps of
the parameters of the two theories are nontrivial and have been
recently discussed in~\cite{Bao:2011rc} and~\cite{Bergman:2013aca}.
This duality has been studied also in the context of integrability in
\cite{Mironov:2013xva,Mironov:2012uh,Mitev:2014isa,Isachenkov:2014eya,Zenkevich:2014lca}. The
term \emph{spectral} for this duality comes from this interpretation.

We will be focusing on the $\mathbb{R}^4\times S^1$ instanton
partition function which can be realised using geometric engineering
as the refined topological string partition function
$Z_{\mathrm{top}}$ associated to the square toric diagram depicted in
Fig.~\ref{fig:4}~a). Then one can immediately understand invariance of
the square quiver theory under spectral duality as the fiber-base
duality corresponding to the reflection along the diagonal of the
diagram.

The instanton or topological string partition functions are actually
based on $U(N)$ quivers, so if we are interested in the $SU(N)$ case,
we should strip off the $U(1)$ contribution. This procedure is
discussed for example in~\cite{Bergman:2013aca}. However, for the
purpose of this paper, where we discuss instanton partition functions,
we can keep the $U(1)$ parts and work with the duality relating
$U(N)^{M-1}$ to $U(M)^{N-1}$ theories.

In the other two vertices of face~{\color{green!60!black}\textsf{2}}
we have an $(N+2)$-point correlator in the the $q$-deformed $A_{N-1}$
Toda theory and its spectral dual\footnote{In the conformal block
  $\langle V_1\cdots V_{N+2}\rangle_{qA_{N-1}}$ the primaries $V_1$
  and $V_{N+2}$ have generic momenta while all the others have momenta
  proportional to the same fundamental weight and correspond to simple
  punctures in the AGT language.}.

The $q$-Toda correlators also enjoy the spectral duality relating
$(K+2)$-point correlators in $A_{N-1}$ $q$-Toda to $(N+2)$-point
correlators in $A_{K-1}$ $q$-Toda theory~\cite{Zenkevich:2014lca,
  Morozov:2015xya} which is the avatar of the $5d$ spectral duality
relating $U(N)^{M-1}$ to $ U(M)^{N-1}$ $5d$ quivers. The
identification between $5d$ instanton partition functions and $q$-Toda
correlators is the $5d$ uplift of the AGT correspondence
\cite{Awata:2010yy,Awata:2009ur}.  More precisely, the AGT map
corresponds to the diagonal edges (shown in blue in Fig.~\ref{fig2}),
while the map along the edges of
face~{\color{green!60!black}\textsf{2}} are from the triality
approach~\cite{Aganagic:2013tta,Aganagic:2014kja}.

The vertical arrows going down from the $5d$ web (face
{\color{green!60!black}\textsf{2}}) to the $3d$ web
(face~{\color{green!60!black}\textsf{1}}) indicate a \emph{tuning
  procedure} where the parameters are fixed to specific discrete
values. On the gauge theory side
(face~{\color{green!60!black}\textsf{3}}) this tuning corresponds to
the so called Higgsing procedure~\cite{Gaiotto:2012xa,
  Mironov:2009qt,Kozcaz:2010af,Dimofte:2010tz,Dorey:2011pa,Nieri:2013vba,Gaiotto:2014ina}. By
tuning the $5d$ Coulomb branch parameters one can degenerate the $5d$
partition function into the partition function of a coupled $5d$--$3d$
system describing co-dimension two defect coupled to the remaining
$5d$ bulk theory. We consider particular tuning of the parameters so
that the square $5d$ quiver is Higgsed \emph{completely,} i.e.\ it
reduces to the $3d$ $FT[SU(N)]$ theory coupled to some free $5d$
hypers\footnote{The $T[SU(N)]$ vortex partition function has also been
  related to a ramified surface defect in the $5d$ $\mathcal{N}=2^*$
  theory in~\cite{Bullimore:2014awa}.}. We demonstrate this in
Sec.~\ref{sec:higgsing-5d-gauge}.  Repeating the Higgsing procedure on
the spectral dual side we land on the $3d$ spectral dual $FT[SU(N)]$
theory.  We then see that $3d$ (self)-duality for $FT[SU(N)]$ follows
via Higgsing from the $5d$ spectral duality for the square quiver.

On the $q$-Toda side (face~{\color{green!60!black}\textsf{6}}) the
tuning procedure corresponds to the tuning of the momenta of the
vertex operators to special values (corresponding to fully degenerate
vertex operators) and to a given assignment of screening charges
(corresponding to conditions on the internal momenta, or Coulomb
branch parameters). In this way the $q$-Toda $A_{N-1}$ correlator with
$N$ semi-degenerate and two full primary operators reduces to the
$q$-DF representation of the conformal block.

This explains our previous statement that the integral blocks ${q{\rm
    DF}}^{A_{N-1}}_{N+2}$ and $\hat{q{\rm DF}}^{A_{N-1}}_{N+2}$ are
related by a degenerate version of spectral duality.

\subsection*{Duality web III}

\begin{figure}[h]
\centering
 \includegraphics{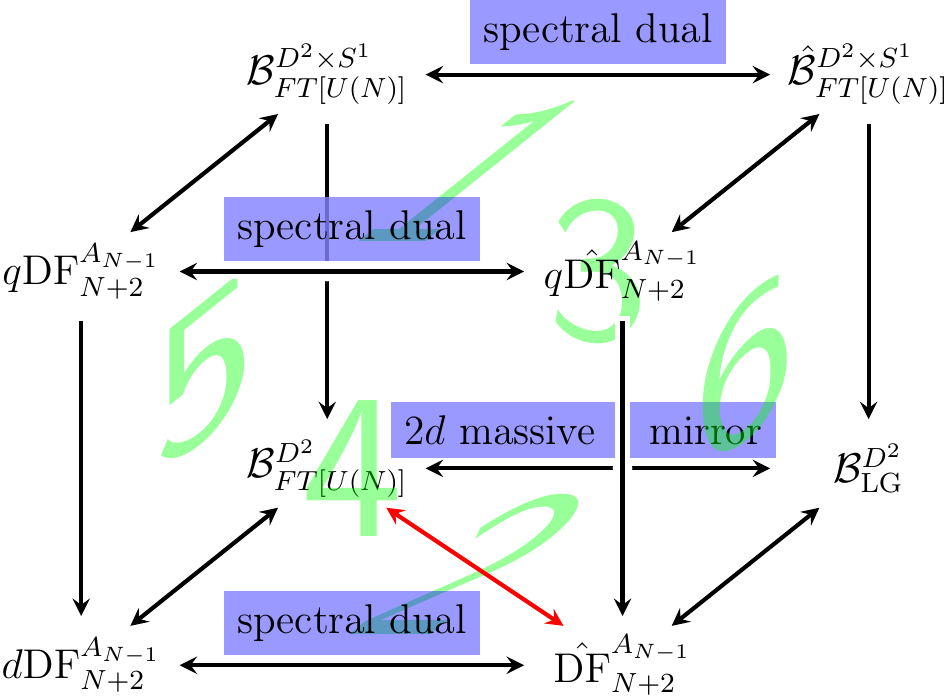}
 \caption{Duality web III. Fig.~\ref{fig1} is the top face (face
   ~{\color{green!60!black}\textsf{1}}) of the cube. The arrows going
   downstairs correspond to $q\to 1$ limits. Notice that the two
   theories related by the spectral duality tend to different theories
   under $q\to 1$. This asymmetry appears because one needs to choose
   the scaling of the parameters with $q$ and the spectral duality map
   relates two \emph{different} choices.}
\label{fig3}
\end{figure}

Finally, starting from duality web I in Fig.~\ref{fig1} we can obtain
another interesting duality web by taking a suitable limit $q\to 1$ as
shown in Fig.~\ref{fig3}, where the duality web I corresponds to face
{\color{green!60!black}\textsf{1}} of the cube.
Let's consider face~{\color{green!60!black}\textsf{3}} in
Fig.~\ref{fig3}. Here we are performing the reduction of a $3d$
spectral pair of theories on $D_2\times S^1$ from $3d$ to $2d$ by
considering the $q\to 1$ limit, which corresponds to shrinking the
$S^1$ radius.  Taking this limit is subtle, as recently discussed in~\cite{Aharony:2017adm} (and before in~\cite{Aganagic:2001uw}), since
there exist in fact \emph{several} meaningful limits. Concretely, one
can consider the situation when some of the $3d$ real mass parameters
are scaled to infinity when going from $3d$ to $2d$ so that $m_{3d}
R=m_{2d}$ remains finite as $R\to 0$.

Starting from $\mathcal{B}^{D_2\times S^1}_{FT[SU(N)]}$ we take the so
called Higgs limits which reduces it to the $\mathcal{N}=(2,2)$ gauged
linear sigma-model (GLSM) $\mathcal{B}^{D_2}_{FT[SU(N)]}$. In the
\emph{Higgs} limit the real mass scaled to infinity is the FI
parameter, while the matter remains light, hence the name. This limit
generally reduces a $3d$ gauge theory to a $2d$ gauge theory. However,
here we want to lift also the Higgs branch and we turn on all the mass
deformations so that the $2d$ gauge theory is massive and has $N!$
isolated vacua.

Since spectral duality, similarly to mirror symmetry, swaps Higgs and
Coulomb branch parameters, on the dual side the limit has a very
different effect. The dual block $\hat{\cal B}^{D_2\times
  S^1}_{FT[SU(N)]}$ in the $q\to 1$ limit (which is now a
\emph{Coulomb} limit) reduces to the partition function of a theory of
twisted chiral multiplets with twisted Landau-Ginzburg superpotential
on $D_2$. The horizontal link in
face~{\color{green!60!black}\textsf{2}} of the cube in Fig.~\ref{fig3}
is, therefore, a duality of Hori-Vafa type~\cite{Hori:2000kt} for mass
deformed theories.

In general claiming that a duality for mass deformed theories implies
a duality for massless theories is dangerous. In particular, in this
context the subtleties of inferring a genuine IR $2d$ duality from a
duality for $2d$ mass deformed theories obtained from the reduction of
pairs of dual theories have been discussed in
\cite{Aharony:2016jki,Aharony:2017adm}. Here we are not interested in
removing the mass deformations since, as we are about to see, the
holomorphic blocks for the mass deformed theories can be directly
mapped to CFT conformal blocks.

Indeed if we look at face~{\color{green!60!black}\textsf{4}} of the
duality web III in Fig.~\ref{fig3}, we see that we are taking
\emph{several} different $q\to 1$ limits of the $q$-Toda conformal
blocks in DF representation. Similarly to the gauge theory side there
are \emph{several} possible ways to take the limit. The limit when we
scale the momenta of the vertex operators and keep the insertion
points fixed is natural from the CFT point of view and reduces
$q$-Toda conformal blocks to conformal blocks of the undeformed Toda
CFT.  This is exactly the limit we take when we reduce the spectral
dual block $q\hat{{\rm DF}}^{A_{N-1}}_{N+2}$ down to the undeformed
conformal block $\hat{{\rm DF}}^{A_{N-1}}_{N+2}$ in $2d$ Toda
theory. Therefore, we have just discovered that the $2d$ $FT[SU(N)]$
GLSM holomorphic block $\mathcal{B}^{D_2}_{FT[SU(N)]}$ is mapped to a
$2d$ CFT conformal block $\hat{{\rm DF}}^{A_{N-1}}_{N+2}$ (red
diagonal on the face~{\color{green!60!black}\textsf{4}}). In other
terms we have derived the familiar gauge/CFT correspondence between
$S^2$ partition functions and degenerate CFT correlators discussed in
\cite{Gomis:2014eya,Gomis:2016ljm,Doroud:2012xw} as a limit of our
$3d$ spectral duality web.

Finally to complete the picture we study what is the effect of the
$q\to 1$ limit on the $q{\rm DF}^{A_{N-1}}_{N+2}$ conformal block. This is
a less familiar limit which reduces the $q{\rm DF}^{A_{N-1}}_{N+2}$ to
a block in the channel with the vertex operators of certain bosonized
algebra, which we denote by $d$-$W_N$, where $d$ stands for
\emph{difference} in the same way as $q$ in $q$-$W_N$ is for
\emph{quantum}. The algebra\footnote{We thank A.~Torrielli for
  pointing out a paper~\cite{Hou:1996fx} in which a similar algebra
  has appeared earlier in a very different context.} $d$-$W_N$ is a
particular limit of the $q$-$W_N$ algebra when $q\to 1$.  We briefly
describe the algebra, correlators and screening charges, leaving a
more detailed investigation for the future~\cite{vertex:future}.

The Duality web III is discussed in Sec.~\ref{secweb4}.

\section{Duality web I: $3d$ $FT[SU(N)]$  and $q$-Toda  blocks}\label{secweb1}


In this section we study Duality web I shown in Fig. \ref{fig1}. We
first introduce the $3d$ holomorphic block ${\cal B}^{D_2\times
  S^1}_{T[SU(N)]}(\vec{\mu}, \vec{\tau}, q, t)$, then we
show the effect of adding the flipping fields and discuss the mirror
and spectral duals of the theory. Finally we introduce the DF
representation for the $q$-Toda blocks and determine the gauge/$q$-DF
dual to ${\cal B}^{D_2\times S^1}_{FT[SU(N)]}(\vec{\mu},
\vec{\tau}, q, t)$.

\subsection{$3d$ blocks for $T[SU(N)]$, flipping fields, mirror and
  spectral duals}\label{3dblocks}

\subsubsection{$3d$ holomorphic blocks}
\label{sec:3d-block}
\begin{figure}[h]
  \centering
  \includegraphics[width=8cm]{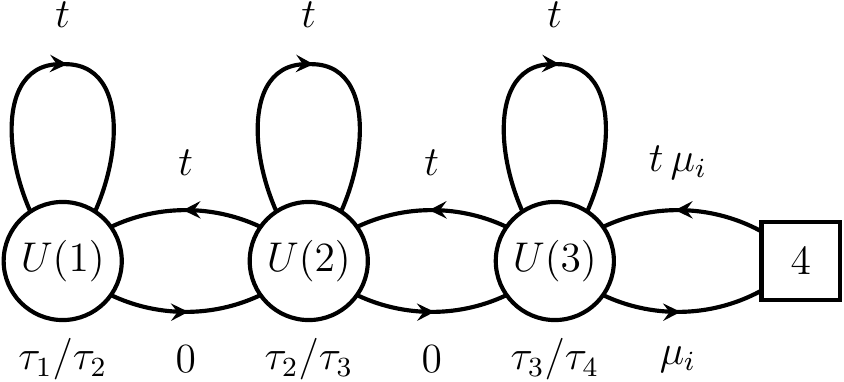}
  \caption{$3d$ $T[SU(4)]$ gauge theory.  $\tau_a$ are the FI
    parameters. The masses of the chirals are indicated over or under
    the corresponding arrows.}
  \label{fig:2}
\end{figure}

We begin by introducing our main character ${\cal B}^{D_2\times
  S^1}_{T[SU(N)]}(\vec{\mu}, \vec{\tau}, q, t)$, the $D_2\times S^1$
partition function, or $3d$ holomorphic block integral for the
$T[SU(N)]$ theory. The $\mathcal{N}=4$ $T[SU(N)]$ theory is a quiver
theory with gauge group $U(1) \times
U(2) \times \cdots \times U(N-1)$, with bifundamental hypers
connecting the $U(N_a)$ and $U(N_{a+1})$ nodes for $a=1,\cdots, N-2$
and $N$ hypermultiplets at the final node. As an example we present
the quiver diagram of the $T[SU(4)]$ theory on Fig.~\ref{fig:2}.
We turn on real masses $M^{3d}_a$ in the Cartan of the $SU(N)_H$
symmetry rotating the Higgs branch and $T^{3d}_a$ in the Cartan of the
$SU(N)_C$ symmetry rotating the Coulomb branch. We also turn on an
extra real axial mass deformation $m^{3d}$ for $U(1)_A$, the
anti-diagonal combination of $U(1)_C\times U(1)_H\in SU(2)_C\times
SU(2)_H$, which breaks the super-symmetry down to ${\cal N}=2^*$.  We
define the dimensionless mass parameters $M'_a=R M^{3d}_a$,
$T'_a=RT^{3d}_a$ and $m'=R m^{3d}$ and the parameter
$q=e^{\hbar}=e^{R\epsilon}$, where $R$ is the $S^1$ circle radius and
$\epsilon$ is the equivariant parameter rotating the cigar $D_2$ (see
footnote $^{\ref{fn:1}}$).

The holomorphic block integral for this theory can be constructed as
explained in~\cite{Beem:2012mb} and reads:
\begin{multline}
  {\cal B}^{D_2\times S^1}_{T[SU(N)]}(\vec{\mu}, \vec{\tau},
  q, t)= F(q,t,\vec{\tau}) \int_{\Gamma}
  \prod\limits_{a=1}^{N-1}\prod\limits_{i=1}^{a} \left(
    \frac{dx_i^{(a)}}{x_i^{(a)}} e^{ X_i^{(a)}
      \left(T_{a}-T_{a+1}\right)/\hbar}\, t^{-X_i^{(a)}/\hbar}\right) \times\\
  \prod\limits_{a=1}^{N-1} \frac{ \prod\limits_{i\neq j}^{a} \Big(
    \frac{x^{(a)}_j}{x^{(a)}_i} ; q\Big)_\infty }{
    \prod\limits_{i,j=1}^{a} \Big( t \frac{x^{(a)}_j}{x^{(a)}_i};
    q\Big)_\infty }
  \prod\limits_{a=1}^{N-2}\prod\limits_{i=1}^{a}\prod\limits_{j=1}^{a+1}
  \frac{ \Big(t \frac{x^{(a+1)}_j}{x^{(a)}_i} ; q\Big)_\infty }{ \Big(
    \frac{x^{(a+1)}_j}{x^{(a)}_i}; q\Big)_\infty } \,
  \prod\limits_{p=1}^{N}\prod\limits_{i=1}^{N-1} \frac{ \Big(t
    \frac{\mu_p}{x^{(N-1)}_i} ; q\Big)_\infty }{ \Big(
    \frac{\mu_p}{x^{(N-1)}_i}; q\Big)_\infty } \,,
\label{TSUN:partition}
\end{multline}
where the prefactor $F(q,t,\vec{\tau})$ is given by
\begin{equation}
  F(q,t,\vec{\tau})=e^{-\frac{2}{3} N(N-1)(2N-1)\hbar  \beta(1-\beta)} 
  e^{-\frac{m'^2 N}{4\hbar}} e^{(1-\beta)  \sum_{a=1}^{N-1} \frac{a^2}{2} (T_{a+1} - T_a)}\,.
\end{equation}
The integral is performed over the Cartan of the gauge group. For each
gauge node we have the contribution of vector and adjoint chiral
multiplets (first factor in the second line) given by a ratio of
$q$-Pochhammer symbols defined as
\begin{equation}
(x;\,q)_\infty=\prod\limits_{k=0}^\infty \left( 1- x q^k \right)\,.
\label{qPoc:symbol}
\end{equation}
The other factors in the second line are the contributions of the
bifundamental chirals and of the fundamentals attached to the last
node.

More precisely $(qx^{-1};q)_\infty$ is the contribution to the block
integral of a chiral multiplet of zero $r$-charge and charge $+1$
under a flavor symmetry with associated real mass $x$, plus a
$-\frac{1}{2}$ Chern-Simons unit. This corresponds to a chiral
multiplet with Dirichlet boundary conditions along $\partial (D_2
\times S^1) = T^2$ in \cite{Yoshida:2014ssa}. A chiral multiplet of
$r$-charge $+2$, charge $-1$ and $\frac{1}{2}$ Chern-Simons unit
contributes as $(x;q)^{-1}_\infty$ and corresponds to Neumann boundary
conditions\footnote{There is a relation between these two setups:
\begin{equation}
  \label{eq:2}
  (qx^{-1};q)_\infty = \frac{\theta_q(x)}{(q;q)_{\infty}} (x;q)^{-1}_\infty,
\end{equation}
which can be explained by viewing the $3d$ theory of a single chiral
multiplet $\phi$ as a linear sigma model with target
$\mathbb{C}$. Dirichlet boundary conditions on $\partial (D_2 \times
S^1) = T^2$ correspond to a D-brane at a point $\phi=0$ in the
target. However, one can view the D-brane in a different way, as (an
equivalence class of) a complex of sheaves
\begin{equation}
  \label{eq:4}
  0 \to \mathcal{O} (\mathbb{C}) \stackrel{s}{\to}
  \Omega^{1,0}(\mathbb{C}) \to 0
\end{equation}
supported on the whole $\mathbb{C}$. Here $\mathcal{O}(\mathbb{C})$ is
the sheaf of functions on $\mathbb{C}$, and $\Omega^{1,0}(\mathbb{C})$
is that of $(1,0)$ differential forms, e.g.\ $g(\phi) \psi$, where
$\psi$ is an anticommuting coordinate on the fiber; the differential
$s = \phi \psi$ is nilpotent because $\psi^2 = 0$. The relation
between the brane at fixed $\phi=0$ and the complex is as
follows. Mnemonically, one can think that the two terms of the
complex~\eqref{eq:4} ``cancel'' everywhere outside the point
$\phi=0$. More concretely, the space of functions on a point $\{
\phi=0 \} \subset \mathbb{C}$ can be equivalently described by the
cohomology of the complex~\eqref{eq:4}:
\begin{align}
  \label{eq:7}
  H^0_s(\mathbb{C}) &= \ker s = 0 \\
  H^1_s(\mathbb{C}) &= \Omega^{1,0}(\mathbb{C})/\im s = \frac{\{ \psi g(\phi)
    \}}{ g(\phi) \sim g(\phi) + \phi f(\phi)} = \{ \mathrm{const} \}
  = \mathbb{C},
\end{align}
In the field theory language $\mathcal{O}(\mathbb{C})$ corresponds to
a $3d$ free chiral with \emph{Neumann} boundary conditions, while to
get the whole complex $\Omega^{\bullet,0}(\mathbb{C})$
from~\eqref{eq:4} one needs to add a $2d$ free chiral fermion $\psi$
living on $T^2$ whose partition function is precisely given by
$\frac{\theta_q(x)}{(q;q)_{\infty}}$. The identity~\eqref{eq:2} is
therefore just the equivalence between two views on the D-brane.}.

If we assemble the matter contribution to the block integrals taking
some  chirals with Dirichlet and some with Neumann boundary
conditions we induce \emph{mixed} Chern-Simons couplings (because of
the attached $\frac{1}{2}$ units) which we might need to compensate by
adding extra Chern-Simons terms to the action.
With our symmetric choice of boundary conditions the induced dynamical
Chern-Simons couplings vanishes automatically, the induced mixed
gauge-flavor couplings (the $t^{\frac{-X^{(a)}_i}{\hbar}}$ factor in
the integrand~\eqref{TSUN:partition}) renormalize the FI parameters,
while the background mixed couplings contribute as the prefactor
$F(q,t,\vec{\tau})$.

To present the block in a form more convenient for the following we
have shifted the integration variables and identified a new set of
exponentiated mass parameters\footnote{The shifted mass parameters
  satisfy:
  \begin{equation}
  \label{eq:57}
  \sum_{i=1}^N M_i = \hbar (1-\beta) \frac{N^2}{2},\qquad \sum_{i=1}^N T_i = \hbar \beta \frac{N^2}{2}.
\end{equation}
The parameter $t$ introduced in this section will be identified with
the parameter of the $5d$ $\Omega$-background
$\mathbb{R}^4_{q,t}\times S^1$ and with the $(q,t)$-Toda parameter in
the next sections. }
\begin{gather}
  x^{(a)}_i=e^{X^{(a)}_i}, \quad
  \mu_p=e^{M_p}=e^{M'_p}\left( \frac{q}{t}\right)^{N/2},\notag \\
  \quad \tau_p=e^{T_p}=e^{T'_p}t^{N/2}, \quad t= q^{\beta}=-q^{1/2}
  e^{-m'}.\label{eq:58}
\end{gather}
An alternative procedure to write down the block integrand $\Upsilon$
is to view it as ``square root'' of the integrand of the partition
function on a compact manifold. Details of this procedure are
presented in the Appendix~\ref{appendix:block}. This construction also
indicates that the contribution of (mixed) Chern-Simons coupling to
the partition function should actually be expressed in terms of ratios
of theta functions rather than exponents
\begin{equation}
\label{eq:56}
e^{\frac{AB}{\hbar}} \leadsto
\frac{\theta_q(e^A)\theta_q(e^B)}{\theta_q(e^{A+B}) (q;q)_{\infty}}\,
\end{equation}
as we discuss in Appendix~\ref{appendix:block}. However, for the
purpose of this paper we can avoid introducing theta functions and
work with the exponents provided that on the integration contours on
which we are going to evaluate the blocks, the theta functions and the
exponents have no poles and contribute with the same residue. One can
check that for $T[SU(N)]$ blocks~\eqref{TSUN:partition} this will
indeed be the case. As will be shown below, the residues of the
integrals~\eqref{TSUN:partition} come in geometric progressions, i.e.\
a pole $x_{*}$ is accompanied by a string of poles at $q^k x_{*}$ with
$k\in \mathbb{Z}_{\geq 0}$. Notice then that both sides of
Eq.~\eqref{eq:56} transform in the same way under the $\hbar$-shifts
of $A$ and $B$ variables, i.e.\ under $q$-shifts of $e^A$ and
$e^B$. Thus, their contributions to the residues in the string differ
only by an \emph{overall} constant factor, independent of $k$. This
overall constant can be factored out of the integral and included in
the normalization factors.

Finally we need to discuss the integration contour on which we
evaluate the block integral. The integration in
Eq.~(\ref{TSUN:partition}) is performed over a basis of integration
contours $\Gamma=\Gamma_\alpha$, with $\alpha=1,\ldots, N!$ which are
in one to one correspondence with the SUSY vacua, the critical points
of the one-loop twisted superpotential
$\mathcal{W}^{\mathbb{R}^2\times S^1}$. The label of the integration
contour $\alpha$ is essentially an element of the permutation group
$\mathfrak{S}_N$. One can understand the origin of the contours
$\Gamma_{\alpha}$ as follows.

The integrations in Eq.~\eqref{TSUN:partition} can be done
step by step starting from $x_i^{(N-1)}$ and proceeding to
$x_1^{(1)}$. There are $(N-1)$ integration variables at the first
step. The poles of the integrand in $x_i^{(N-1)}$ correspond to zeroes
of $\prod_{p=1}^N\prod_{i=1}^{N-1} ( \mu_p/x^{(N-1)}_i;
q)_\infty$. Moreover, upon closer examination one can see that each
$x_i^{(N-1)}$ should be of the form $q^{k_i^{(N-1)}} \mu_p$ with
integer $k_i^{(N-1)}$ and \emph{distinct} values of $p$, i.e.\ each of
$(N-1)$ variables $x_i^{(N-1)}$ settles at a pole close to its own
mass $\mu_p$ and no two of the variables can sit near the same
mass. Therefore, there are $N$ possible configurations with $(N-1)$
variables filling $N$ places (the integrand is symmetric in
$x_i^{(N-1)}$). Evaluating the residues in $x_i^{(N-1)}$ we can
proceed to the next step of integration. Here the situation is
repeated: there are $(N-2)$ integration variables $x_i^{(N-2)}$ and
$(N-1)$ variables $x_i^{(N-1)}$ from the previous step play the role
of $\mu_p$ for them. The poles in $x_i^{(N-2)}$ are located at
$q^{k_i^{(N-2)} - k_i^{(N-1)}} x_j^{(N-1)}$ with integer $k_i^{(N-2)}
\geq k_i^{(N-1)}$ and again no two variables $x_i^{(N-2)}$ can sit
near the same $x_j^{(N-1)}$. There are $(N-1)$ possibilities at this
step. Proceeding further, one notices the general pattern: the poles
at each step sit near the poles of the previous step with one free
place. Equivalently, there are $(N-1)$ strings of poles with lengths
$1$, $2$,~\ldots,~$(N-1)$, in each of which the poles are close
together, e.g.\ for a string of length $a$ we get
\begin{equation}
  \label{eq:6}
  x^{(a)}_a = q^{k_a^{(a)}-k_a^{(a+1)}} x^{(a+1)}_a = q^{k_a^{(a)}-k_a^{(a+2)}} x^{(a+2)}_a =\ldots =q^{k_a^{(a)}-k_a^{(N-1)}}
  x^{(N-1)}_a =  q^{k_a^{(a)}} \mu_p
\end{equation}
where $k^{(a)}_i$ are all integers and we have used the symmetry of
the integrand in $x_1^{(a)}$, \ldots, $x_a^{(a)}$ to set all the lower
indices in the string to $a$. Each string terminates at the the free
place, not filled by the pole on the next step. Choosing the
integration contour is equivalent to specifying which string (of
length $a$) sits near which mass $\mu_p$. Evidently, any choice can be
obtained from a given one by the unique permutaiton of masses
$\mu_p$. There are therefore $N!$ choices in total, each one
corresponding to an element of the symmetric group
$\mathfrak{S}_N$.

We will do the calculations for a certain convenient reference choice
of contour $\alpha = \alpha_0$, i.e.\ in the reference vacuum in which
\begin{equation}
  \label{eq:18}
  x_i^{(a)} = q^{k_i^{(a)}} \mu_i \qquad \text{(in vacuum $\alpha_0$).}
\end{equation}
In this vacuum one can expand the vortex partition function as a
double series in $\frac{\mu_i}{\mu_{i+1}}$ and
$\frac{\tau_i}{\tau_{i+1}}$, i.e.\ it is implicitly assumed that the
theory sits in the chamber of the moduli space where
$\frac{\tau_i}{\tau_{i+1}} \ll 1$. Blocks for other vacua can be
obtained from the block in the reference vacuum by analytic
continuation in $\frac{\tau_i}{\tau_{i+1}}$, taking into account the
intricate (theta-function) connection coefficients. Let us also notice
that since the block is self-dual under mirror symmetry, analytic
continuation in $\mu_i$ and $\tau_i$ will give the same results.

The integration over $\Gamma_{\alpha_0}$ yields
\begin{equation}
{\cal B}^{D_2\times S^1,\, (\alpha_0)}_{T[SU(N)]}=Z^{3d,\, (\alpha_0)}_{\textrm{cl}} 
Z^{3d,\, (\alpha_0)}_{1\mathrm{-loop}}  Z^{3d,\, (\alpha_0)}_{\mathrm{vort}}
\end{equation}
where $Z^{3d,\, (\alpha_0)}_{\textrm{cl}}$, $Z^{3d,\,
  (\alpha_0)}_{1\mathrm{-loop}}$ and $Z^{3d,\,
  (\alpha_0)}_{\mathrm{vort}}$ denote the classical, perturbative
one-loop and nonperturbative vortex contributions respectively. We
have\footnote{We can trade the theta-functions for exponents using the
  equivalence~\eqref{eq:56}, but we retain the exact answer for the
  integration for the sake of completeness.}
\begin{equation}
  \label{eq:27}
  Z^{3d,\, (\alpha_0)}_{\textrm{cl}}(\vec{\mu}, \vec{\tau}, q, t) = F(q,t,\vec{\tau})\,
  (q;q)_{\infty} ^{-\frac{N(N-1)}{2}} \prod_{i=1}^N e^{\frac{(T_i - T_N)
      M_i}{\hbar}} t^{-\frac{(N-i)M_i}{\hbar}} \prod_{i<j}^N \frac{\theta_q
    \left( t \frac{\mu_j}{\mu_i} \right)}{\theta_q
    \left(\frac{\mu_j}{\mu_i} \right)}. 
\end{equation}
The one-loop factor is given by:
  \begin{equation}
    Z_{1-\mathrm{loop}}^{3d,\,(\alpha_0)}(\vec{\mu}, \vec{\tau}, q, t)
    = \prod_{i<j} \frac{\left( q \frac{\mu_i}{\mu_j}; q
      \right)_{\infty}}{\left( \frac{q}{t} \frac{\mu_i}{\mu_j}; q
      \right)_{\infty}}.
    \label{eq:59}
  \end{equation}
  Notice that there are cancellations between the theta-functions in
  classical part and the $q$-Pochhammer functions in the one-loop
  part. The vortex part reads\footnote{There are several ways to write
    the instanton contributions in this sum connected to each other by
    identities involving products of $q$-Pochhammer symbols. For
    example, the middle factor can be rewritten as:
  \[
  \prod_{i\neq j}^a \frac{\left(t
        \frac{\mu_i}{\mu_j};q\right)_{k^{(a)}_i - k^{(a)}_j}}{\left( 
        \frac{\mu_i}{\mu_j};q\right)_{k^{(a)}_i - k^{(a)}_j}}
= \prod_{i\neq j}^a \frac{\left(q
        \frac{\mu_i}{\mu_j};q\right)_{k^{(a)}_i - k^{(a)}_j}}{\left( \frac{q}{t}
        \frac{\mu_i}{\mu_j};q\right)_{k^{(a)}_i - k^{(a)}_j}}    
\]}
\begin{multline}
  \label{eq:15}
  Z_{\mathrm{vort}}^{3d,\,(\alpha_0)} (\vec{\mu}, \vec{\tau}, q, t)
  =\\
  = \sum_{k^{(a)}_i \in (\ref{eq:411s})} \prod_{a=1}^{N-1} \left[\left( t
      \frac{\tau_a}{\tau_{a+1}} \right)^{\sum_{i=1}^a k^{(a)}_i}
    \prod_{i\neq j}^a \frac{\left(t
        \frac{\mu_i}{\mu_j};q\right)_{k^{(a)}_i - k^{(a)}_j}}{\left(
        \frac{\mu_i}{\mu_j};q\right)_{k^{(a)}_i - k^{(a)}_j}}
    \prod_{i=1}^a \prod_{j=1}^{a+1} \frac{\left( \frac{q}{t}
        \frac{\mu_i}{\mu_j}; q \right)_{k_i^{(a)}-k_j^{(a+1)}}}{\left(
        q \frac{\mu_i}{\mu_j}; q \right)_{k_i^{(a)}-k_j^{(a+1)}}}
  \right]
\end{multline}
where we assume $k^{(N)}_i = 0$ and the sum is over sets of integers
$k_i^{(a)}$ satisfying the inequalities
\begin{equation}
  \label{eq:411s}
  \begin{array}{ccccc}
    k^{(1)}_1 \geq & k^{(2)}_1 \geq & k^{(3)}_1 \geq & \cdots \geq &
    k^{(N-1)}_1 \geq 0\\
    & k^{(2)}_2 \geq & k^{(3)}_2 \geq &\cdots \geq & k^{(N-1)}_2 \geq
 0\\
    &  & k^{(3)}_3 \geq &\cdots \geq & k^{(N-1)}_3 \geq
 0\\
 & & &\ddots  & \vdots\\
 &  &  & & k^{(N-1)}_{N-1} \geq 0\\   
  \end{array}
\end{equation}
The block can actually be expressed through higher $q$-hypergeometric
functions. This representation also allows one to deduce the monodromy
properties of the block under the permutation of parameters $\tau_i
\to \tau_{\sigma(i)}$. However, these issues will not be considered in
the present work. In the semiclassical limit ${\cal B}^{D_2\times
  S^1,\, (\alpha)}_{T[SU(N)]} \sim
e^{\mathcal{W}^{\mathbb{R}^2\times S^1}_{(\alpha)}/\hbar }$ where
$\mathcal{W}^{\mathbb{R}^2\times S^1}_{(\alpha)}$ is the one-loop
twisted superpotential evaluated on the $\alpha$-th vacuum.

\subsubsection{Mirror duality}
\label{sec:mirror-duality}
We will consider two similar but subtly different dualities of the
$T[SU(N)]$ theory: the mirror duality and the spectral duality,
explaining the relationship between them and their differences. 

The mirror duality~\cite{Intriligator:1996ex} (which in this case is a
self-duality~\cite{Gaiotto:2008ak}) swaps the Higgs and Coulomb
branches and consequently the vector masses and FI parameters
$M_i\leftrightarrow T_i$ and sends $m\leftrightarrow -m$ or in terms
of the exponentiated parameters:
\begin{equation}
  \label{mirror:map}
  \mu_i\leftrightarrow \tau_i\,, \quad  t\to \frac{q}{t} \,.
\end{equation}
The mirror block is given by:
\begin{multline}
  \check{\cal B}^{D_2\times S^1,\,(\alpha)}_{T[SU(N)]}
  (\vec{\mu}, \vec{\tau}, q, t) = \mathcal{B}^{D_2\times
    S^1,\,(\alpha)}_{T[SU(N)]}
  \left( \vec{\tau}, \vec{\mu}, q, \frac{q}{t}\right) =\\
  =\check{F}(q,t,\vec{\tau}) \int_{\Gamma_{\alpha}}
  \prod\limits_{a=1}^{N-1}\prod\limits_{i=1}^{a}\frac{dx_i^{(a)}}{x_i^{(a)}}
  \prod\limits_{a=1}^{N-1} \prod\limits_{i=1}^{a}e^{ X_i^{(a)}
    \left(M_{a}-M_{a+1}\right)/\hbar} \left(
    \frac{t}{q}\right)^{X_j^{(a)}/\hbar}
  \,\times\\
  \times \prod\limits_{a=1}^{N-1}\frac{ \prod\limits_{i\neq j}^{a}
    \Big( \frac{x^{(a)}_j}{x^{(a)}_i} ; q\Big)_\infty }{
    \prod\limits_{i,j=1}^{a} \Big( \tfrac{q}{t}
    \frac{x^{(a)}_j}{x^{(a)}_i}; q\Big)_\infty }
  \prod\limits_{a=1}^{N-2}\prod\limits_{i=1}^{a}\prod\limits_{j=1}^{a+1}
  \frac{ \Big( \tfrac{q}{t} \frac{x^{(a+1)}_j}{x^{(a)}_i} ;
    q\Big)_\infty }{ \Big( \frac{x^{(a+1)}_j}{x^{(a)}_i};
    q\Big)_\infty } \, \prod\limits_{p=1}^{N}\prod\limits_{i=1}^{N-1}
  \frac{ \Big(\tfrac{q}{t} \frac{\tau_p}{x^{(N-1)}_i} ; q\Big)_\infty
  }{ \Big( \frac{\tau_p}{x^{(N-1)}_i}; q\Big)_\infty } \,.
\label{dTSUN:partition}
\end{multline}
Showing that the $T[SU(N)]$ holomorphic block is self-dual, i.e.\ that
\begin{equation}
  \label{eq:28}
  {\cal B}^{D_2\times
    S^1,\,(\alpha)}_{T[SU(N)]} (\vec{\mu}, \vec{\tau}, q, t)=\check{\cal B}^{D_2\times
    S^1,\,(\alpha')}_{T[SU(N)]}(\vec{\mu}, \vec{\tau}, q, t)
\end{equation}
or equivalently
\begin{equation}
  \label{eq:29}
  {\cal B}^{D_2\times
    S^1,\,(\alpha)}_{T[SU(N)]} (\vec{\mu}, \vec{\tau}, q,
  t) = \mathcal{B}^{D_2\times
    S^1,\,(\alpha')}_{T[SU(N)]}\left( \vec{\tau},\vec{\mu}, q, \frac{q}{t}\right)
\end{equation}
is fairly complicated. As discussed in \cite{Beem:2012mb} if we want
to describe how the bases of contours $\alpha$ and $\alpha'$ are
related we need to take into account Stokes phenomena. One approach is
to use mirror-invariant combinations of blocks, e.g.\ squashed sphere
partition functions.

Here we take a different approach showing that the \emph{space of
  blocks} is invariant under the mirror
map. Following~\cite{Bullimore:2014awa} we can view the space of
blocks for the $T[SU(N)]$ theory as the space of solutions to a system
of linear difference equations:
\begin{equation}
  \label{eq:30}
  H_r(\mu_i, q^{\mu_i \partial_{\mu_i}}, q, t) {\cal B}^{D_2\times
    S^1,\,(\alpha)}_{T[SU(N)]}(\vec{\mu}, \vec{\tau}, q, t)
  = e_r(\vec{\tau}){\cal B}^{D_2\times
    S^1,\,(\alpha)}_{T[SU(N)]} (\vec{\mu}, \vec{\tau}, q, t)
\end{equation}
where the difference operators are quantum Ruijsenaars-Schneider
Hamiltonians \cite{Ruijsenaars:1986vq,Ruijsenaars:1986pp}:
\begin{equation}
  \label{eq:31}
  H_r = t^{\frac{r(r-1)}{2}} \sum_{
    \begin{smallmatrix}
      I \subset \{ 1, \ldots, N\}\\
      |I|=r
    \end{smallmatrix}
} \prod_{
  \begin{smallmatrix}
    i \in I\\
    j \notin I
  \end{smallmatrix}
} \frac{t \mu_i - \mu_j}{\mu_i - \mu_j}  q^{\sum_{i \in I} \mu_i \partial_{\mu_i}} 
\end{equation}
and the eigenvalues $e_r(\vec{\tau})$ are elementary symmetric
polynomials
\begin{gather}
  \label{eq:33}
  e_r(\vec{\tau}) = \sum_{i_1<\cdots <i_r}^N \tau_{i_1} \cdots
  \tau_{i_r}.\qquad
\end{gather}
We give a short proof of Eq.~\eqref{eq:30} for $r=1$ in
Appendix~\ref{sec:hamilt-vecmu-vari}. From the theory of integrable
systems it is known that Ruijsenaars-Schneider system has a peculiar
duality symmetry called $p$-$q$ duality. It implies that for certain
choice of normalization the eigenfunctions of the
Ruijsenaars-Schneider Hamiltonian are actually also eigenfunctions of
the \emph{dual} Ruijsenaars-Schneider Hamiltonian. The dual operator
is obtained by the mirror map: $\vec{\tau}$ and $\vec{\mu}$ are
exchanged as are $t$ and $\frac{q}{t}$. We therefore have:
\begin{equation}
  \label{eq:55}
  H_r\left(\tau_i, q^{\tau_i \partial_{\tau_i}}, q, \frac{q}{t}\right) {\cal B}^{D_2\times
    S^1,\,(\alpha)}_{T[SU(N)]}(\vec{\mu}, \vec{\tau}, q, t)
  = e_r(\vec{\mu}){\cal B}^{D_2\times
    S^1,\,(\alpha)}_{T[SU(N)]} (\vec{\mu}, \vec{\tau}, q, t)
\end{equation}
We prove the simplest case of Eq.~\eqref{eq:55} for $r=1$, $N=2$ in
Appendix~\ref{sec:hamilt-vect-vari}. The self-mirror property of the
blocks (\ref{eq:29}) follows from Eqs. (\ref{eq:30}) and
(\ref{eq:55}).

Alternatively we can check mirror symmetry by ``brute force''
computation of the partition function. Using explicit
expressions~\eqref{eq:27},~\eqref{eq:59} and~\eqref{eq:15} for the
one-loop and vortex parts of the partition function we can see  that 
\begin{equation}
  \label{eq:711}
  Z_{1\mathrm{-loop}}^{3d,\,(\alpha_0)}(\vec{\mu}, \vec{\tau}, q, t ) Z_{\mathrm{vort}}^{3d,\,(\alpha_0)}(\vec{\mu}, \vec{\tau}, q, t ) =
  Z_{1\mathrm{-loop}}^{3d,\,(\alpha_0)} \left( \vec{\tau}, \vec{\mu}, q, \frac{q}{t}
  \right) Z_{\mathrm{vort}}^{3d\,(\alpha_0)}\left( \vec{\tau}, \vec{\mu}, q, \frac{q}{t} \right)
\end{equation}
and using the conditions~\eqref{eq:57} for the sum of masses and FI
  parameters (up to the  equivalence~\eqref{eq:56}):
\begin{equation}
  \label{eq:60}
  Z_{\mathrm{cl}}^{3d,\,(\alpha_0)}(\vec{\mu}, \vec{\tau}, q, t ) =
  Z_{\mathrm{cl}}^{3d,\,(\alpha_0)} \left( \vec{\tau}, \vec{\mu}, q, \frac{q}{t}
  \right)\,.
\end{equation}

As a very simple test of the mirror symmetry~\eqref{eq:711} consider
two degenerate limits of the $T[SU(N)]$ block:
\begin{enumerate}
\item $t=q$. In this case all the terms of the vortex
  series, except the first one vanish:
\begin{equation}
  \label{eq:81}
  Z_{\mathrm{vort}}^{3d,(\alpha_0)} (\vec{\mu}, \vec{\tau}, q, q) = 1.
\end{equation}
The one-loop factor also simplifies and reads
\begin{equation}
  \label{eq:61}
  Z_{1\mathrm{-loop}}^{3d,(\alpha_0)} (\vec{\mu}, \vec{\tau}, q, q) =
  \prod_{i<j} \frac{1}{1 - \frac{\mu_i}{\mu_j}}.
\end{equation}

\item $t=1$. In this case the vortex sum factorizes into a
  product of geometric progressions:
\begin{equation}
  \label{eq:9}
  Z_{\mathrm{vort}}^{3d,(\alpha_0)} (\vec{\mu}, \vec{\tau}, q, 1) = \prod_{i<j} \frac{1}{1 - \frac{\tau_i}{\tau_j}}.
\end{equation}
The one-loop part becomes trivial:
\begin{equation}
  \label{eq:62}
  Z_{1\mathrm{-loop}}^{3d,(\alpha_0)} (\vec{\mu}, \vec{\tau}, q, 1) = 1.
\end{equation}

\end{enumerate}
Two degenerate cases are mirror dual to each other and one immediately
sees that Eq.~\eqref{eq:711} indeed holds in this limit.

\subsubsection{Flipping fields and spectral duality}
\label{sec:flipping-fields}
We now introduce the modification of the $T[SU(N)]$ model, in which
we add $N^2$ singlets fields, the {\it flipping flieds}, transforming  in the adjoint of the $SU(N)$ flavor symmetry group.
These fields modify the  $T[SU(N)]$ superpotential by the  extra term
\begin{equation}
  \label{eq:16}
  \mathcal{W} = F_{ij} Q_{ij},
\end{equation}
where $Q_{ij}$ is the meson matrix, built from the bifundamental
chirals $q_i^a$ at the rightmost node of the quiver from
Fig.~\ref{fig:2}:
\begin{equation}
  \label{eq:17}
  Q_{ij} = \sum_{a=1}^{N-1} q_i^a \bar{q}_j^a,
\end{equation}
so that if the bifundamental has $R$-charge $r$, then $R[Q_{ij}] = 2r$
and $R[F_{ij}] = 2-2r$ . We
call the resulting theory $FT[SU(N)]$, where $F$ indicates the
flipping of the Higgs branch operators (the meson).

Since flipping fields are gauge singlets, they simply modify the $D_2
\times S^1$ partition function of $T[SU(N)]$ by multiplicative factors in front of
the integral:
\begin{equation}
  \label{eq:19}
  {\cal B}^{D_2\times S^1,\,(\alpha)}_{FT[SU(N)]}
  (\vec{\mu}, \vec{\tau}, q, t) = f(\vec{\mu},q,t)^{-1}   {\cal B}^{D_2\times S^1,\,(\alpha)}_{T[SU(N)]}(\vec{\mu}, \vec{\tau}, q, t),
\end{equation}
where\footnote{We could equivalently use a combination of theta-functions
  instead of powers for the contact terms multiplying the $q$-factorials to make $f$ a $2\pi i$-periodic function of
  $M_i$. Notice also that $f(\vec{\mu},q,\frac{q}{t}) = f(\vec{\mu},q,t)^{-1}$. }
\begin{equation}
  \label{eq:66}
  f(\vec{\mu},q,t) = e^{(1-2\beta)\sum_{i=1}^N (i-1) M_i}   \prod_{k < l}^N
  \frac{\left( t \frac{\mu_k}{\mu_l} ; q \right)_{\infty}}{\left( \frac{q}{t} \frac{\mu_k}{\mu_l} ; q \right)_{\infty}}.
\end{equation}
The factor $f(\vec{\mu},q,t)$ crucially modifies the action of the RS
Hamiltonians~\eqref{eq:31} on the block. One proves by direct
computation that
\begin{equation}
  \label{eq:65}
H_r\left( \mu_i, q^{\mu_i \partial_{\mu_i}}, q, \frac{q}{t} \right) = f(\vec{\mu},q,t)^{-1}
  H_r(\mu_i, q^{\mu_i \partial_{\mu_i}}, q, t) f(\vec{\mu},q,t).
\end{equation}
Since the block $ {\cal B}^{D_2\times S^1,\,(\alpha)}_{T[SU(N)]}$ is
the eigenfunction of $H_r(\mu_i, q^{\mu_i \partial_{\mu_i}}, q, t)
f(\vec{\mu},q,t)$, the holomorphic blocks of the \emph{flipped} theory
$FT[SU(N)]$ are eigenfunctions of $H_r\left( \mu_i,
  q^{\mu_i \partial_{\mu_i}}, q, \frac{q}{t} \right)$ with the same
eigenvalues.  Moreover, since $f(\vec{\mu},q,t)$ does not depend on
$\tau$, the flipped block $ f(\vec{\mu},q,t)^{-1} {\cal B}^{D_2\times
  S^1}_{T[SU(N)]}(\vec{\mu}, \vec{\tau}, q, t)$ is still an
eigenfunction of the \emph{dual} RS Hamiltonians $H_r(\tau_i,
q^{\tau_i \partial_{\tau_i}}, q, \frac{q}{t})$. Hence we conclude that
because ${\cal B}^{D_2\times S^1}_{T[SU(N)]}(\vec{\mu}, \vec{\tau}, q,
t)$ is invariant under the \emph{mirror} duality~\eqref{mirror:map},
the \emph{flipped} block ${\cal B}^{D_2\times S^1}_{FT[SU(N)]}(\vec{\mu}, \vec{\tau}, q, t)$ is invariant under the
\emph{spectral} duality:
\begin{equation}
  \label{eq:20}
  \mu_i\leftrightarrow \tau_i\,, \quad  t\to t \,.
\end{equation}
We denote the spectral dual block by $\hat{\mathcal{B}}^{D_2\times
  S^1,(\alpha)}_{FT[SU(N)]}(\vec{\mu}, \vec{\tau}, q, t)$ (notice
the hat instead of the tick, which we have used for the mirror
block). For our special contour $\alpha_0$ we have
\begin{equation}
  \label{eq:21}
  \hat{\mathcal{B}}^{D_2\times
    S^1,(\alpha_0)}_{FT[SU(N)]}(\vec{\mu}, \vec{\tau}, q, t) =   \mathcal{B}^{D_2\times
    S^1,(\alpha_0)}_{FT[SU(N)]}(\vec{\tau}, \vec{\mu}, q, t).
\end{equation}

We will discuss the origin of the spectral duality when we come to the
brane description of the flipped theory in
sec.~\ref{sec:higgsing-5d-gauge}.  In \cite{APZ} we will examine in the detail this duality together with another duality obtained from $T[SU(N)]$ by flipping simultaneously  the Higgs and Coulomb branch operators.

\subsection{$q$-Toda blocks}\label{qtodaintro}

In a series of
works~\cite{Aganagic:2014kja,Aganagic:2013tta,Nedelin:2016gwu}
partition functions of $3d$ theories have been shown to match
conformal blocks in $q$-deformed Toda theories\footnote{An alternative
  map between $q$-CFT correlators and $3d$ partition functions have
  been discussed in~\cite{Nieri:2013yra}. This approach is similar to
  the map between $S^2$ partition functions and CFT correlators
  discussed in~\cite{Gomis:2014eya,Gomis:2016ljm,Doroud:2012xw}.}.  In
this section we will demonstrate the details of the correspondence
between holomorphic blocks of the $FT[SU(N)]$ theory and conformal
blocks of the $q$-Toda CFT. For this we will first review basic
aspects of $A_n$ Toda CFT and derive Dotsenko-Fateev (DF) integrals
describing conformal blocks in certain channel. In this part we will
closely
follow~\cite{Fateev:2007ab,Aganagic:2014kja,Aganagic:2013tta}. Then we
will briefly describe quantum deformation of the Toda theory and
corresponding $q\mathrm{DF}$ integrals. Finally we will describe the
map between parameters of $FT[SU(N)]$ theory and $q$-Toda CFT that
will allow us to manifestly match holomorphic blocks and
$q\mathrm{DF}$ integrals on two sides of the correspondence.

\subsubsection{Warm-up: conformal block of ordinary Toda}
We begin by quickly introducing the integral representation of the
Toda conformal blocks, for more detailed review see
\cite{Fateev:2007ab}. The action of the theory is given by
\begin{equation}
S_{\mathrm{Toda}}=\int dz\, d\bar{z}\, \sqrt{g}\left[ g^{z\bar{z}}( \partial_z\vec{\phi},\partial_{\bar{z}}\vec{\phi} )+{\cal Q}_\beta(\vec{\rho},\vec{\phi})\,R+
\sum\limits_{a=1}^{n} e^{\sqrt{\beta}(\vec{\phi},\,\vec{e}_{(a)}  )}\right]\,,
\label{Toda:action}
\end{equation}
where $\vec{\phi}$ is the $(n+1)-$component vector whose components
$\phi^{(a)}$ are bososnic fields in $2d$ Toda CFT. $\vec{\rho}$ and
$\vec{e}_{(a)}$ are the Weyl vector and the simple roots of the $A_n$
Lie algebra respectively:
\begin{gather}
	\left( \vec{\rho},\,\vec{\phi} \right)=\frac{1}{2}\sum\limits_{a=1}^{n+1}\left( n-2a+2 \right)\phi^{(a)},\notag\\
	\left( \vec{\phi},\,\vec{e}_{(a)} \right)=\phi^{(a)}-\phi^{(a+1)}.
\label{eq:5}
\end{gather}
The first term in the action (\ref{Toda:action}) is just the canonical
kinetic term with (inverse) background metric $g^{z\bar{z}}$, while
$(.\,,.)$ denote the standard scalar product on ${\mathbb
  R}^{n+1}$. Second term in the action is responsible for the
nonminimal coupling of $\vec{\phi}$ to the background curvature
$R$. Coefficient of the coupling ${\cal Q}_\beta$ is
\begin{equation}
{\cal Q}_\beta = \sqrt{\beta}-\frac{1}{\sqrt{\beta}}\,,
\end{equation}
where $\beta$ is a convenient parameter, which will be used throughout
this paper. Finally the last term in (\ref{Toda:action}) is the Toda
potential. The theory described above possess $W_{n+1}$ symmetry,
which has Virasoro subalgebra with the central charge parametrized by
$\beta$ in the following way:
\begin{equation}
c = n - n(n+1)(n+2)\mathcal{Q}_{\beta}^2\,.
\label{central:charge}
\end{equation}

Basic ingredients we will need for finding correlators in Toda CFT are
screening currents
\begin{multline}
S_{(a)}(x)= \normord{\exp\left[\sqrt{\beta}\sum\limits_{k\neq 0}\sum\limits_{b=1}^{n+1} \frac{1}{k}c_k^{(b)}\,e_b^{(a)}\,x^{-k}  \right]}
\exp\left[\sqrt{\beta}\sum\limits_{b=1}^{n+1} Q^{(b)}e_b^{(a)}  \right]\,x^{\sqrt{\beta}\sum\limits_{b=1}^{n+1}e_b^{(a)} P^{(b)} }
=\\
=\normord{\exp\left[ \sqrt{\beta}\sum\limits_{k\neq 0}\left( c_{k}^{(a)}-c_k^{(a+1)} \right)\frac{x^{-k}}{k}\right]}
e^{\sqrt{\beta}\left( Q^{(a)}-Q^{(a+1)} \right)}\,x^{\sqrt{\beta}\left(P^{(a)}-P^{(a+1)}  \right)},
\label{screening:current:toda}
\end{multline}
where the index $a$, which we call the \emph{sector} number, runs from
$1$ to $n$ for $A_n$ theory. We will also need vertex operators
defined as follows:
\begin{equation}
V_{\vec{\alpha}}(z)= \normord{\exp\left[\frac{1}{\sqrt{\beta}}\sum\limits_{k\neq 0}\sum\limits_{a=1}^{n+1} 
c_k^{(a)}\,\alpha_a\,\frac{z^{-k}}{k}  \right]}
e^{\frac{1}{\sqrt{\beta}}\sum\limits_{a=1}^{n+1} Q^{(a)}\alpha_a}  
\,z^{\frac{1}{\sqrt{\beta}}\sum\limits_{a=1}^{n+1}\alpha_a P^{(a)} },
\label{vertex:operator:toda}
\end{equation}
where $\vec{\alpha}$ is the $(n+1)-$component weight of the operator.
Bosonic operators $c_k^{(a)}$ satisfy the Heisenberg algebra
\begin{equation}
[c_k^{(a)},c_{m}^{(b)}]=k\,\delta_{k+m,0}\,\delta_{a,b}\,,
\label{heisenbeg:algebra}
\end{equation}
and $P^{(a)},\,Q^{(a)}$ are zero-modes satisfying usual commutation relations:
\begin{equation}
\left[P^{(a)},Q^{(b)}\right]=\delta_{a,b}\,.
\label{zero:modes:algebra}
\end{equation}

Now assume that we would like to calculate the following chiral half of the  correlator of $(l+2)$ primary 
vertex operators in Toda theory
\be
\langle V_{\vec{\alpha}^{(\infty)}}(\infty) \,&& \hspace{-3mm}V_{\vec{\alpha}^{(1)}}\left( z_{1} \right)\cdots 
V_{\vec{\alpha}^{(l)}}\left( z_l \right)\,V_{\vec{\alpha^{(0)}}}(0)\, \rangle_{\mathrm{Toda}}\,=\nonumber\\  
&&  \int D\vec{\phi}~V_{\vec{\alpha}^{(\infty)}}\left( \infty \right)\, V_{\vec{\alpha}^{(1)}}\left( z_{1} \right)\cdots 
  V_{\vec{\alpha}^{(l)}}\left( z_l \right)\,V_{\vec{\alpha}^{(0)}}\left( 0 \right)\, e^{-S_{\mathrm{Toda}}}\,,
\ee
where $z_k$ and  $\vec{\alpha}^{(k)}$ are positions and weights of corresponding vertex operator insertions. 
In general, due to the complicated interaction potential, evaluation of this correlator is extremely hard. 
However, one can treat Toda potential perturbatively. In this case the full answer for the correlator can be written 
as the sum of the following correlators in the theory of $(n+1)$ free bosons:
\begin{equation}
  \mathrm{DF}_{l+2}^{A_n}(z_1,\ldots, z_l, \vec{\alpha}^{(0)},
  \vec{\alpha}^{(1)},\ldots,
  \vec{\alpha}^{(l)}, \vec{N} , \beta ) \stackrel{\mathrm{def}}{=} \langle\vec{\alpha}^{(\infty)}|\,V_{\vec{\alpha}^{(1)}}\left( z_{1} \right)\dots 
  V_{\vec{\alpha}^{(l)}}\left( z_l \right)\,
  \prod_{a=1}^{n} Q_{(a)}^{N_a}|\vec{\alpha}^{(0)}\rangle_{\mathrm{free}}\,,
\label{correlator:toda}
\end{equation}
which play the role of the  conformal blocks in Toda theory and are usually
referred to as Dotsenko-Fateev  $\left( \mathrm{DF} \right)$ 
integrals \cite{Dotsenko:1984nm}. Here $Q_{(a)}$ are screening charges defined as the integrals of the
corresponding screening currents:
\begin{equation}
Q_{(a)} \stackrel{\mathrm{def}}{=} \oint dx\, S_{(a)}\left( x \right),
\label{screening:charge}
\end{equation}
and the states $|\vec{\alpha}^{(0)}\rangle$ and
$|\vec{\alpha}^{(\infty)}\rangle$ are defined as follows:
\begin{equation}
|\vec{\alpha}\rangle=\frac{1}{\sqrt{\beta}}e^{\sum\limits_{a=1}^{n+1}\alpha_a\,Q^{(a)}}|0\rangle
\label{state}
\end{equation}
so that it is the eigenstate of the momentum operators $P^{(a)}$ and
is annihilated by the positive modes:
\begin{gather}
  P^{(a)}|\vec{\alpha}\rangle=\frac{1}{\sqrt{\beta}}\alpha_{a}|\vec{\alpha}\rangle,\\
  c^{(a)}_n|\alpha\rangle=0,\qquad \forall n>0.
\end{gather}
Due to the operator-state correspondence the ket state
$|\vec{\alpha}\rangle$ can be created by the insertion of the vertex
operator~(\ref{vertex:operator:toda}) of weight $\vec{\alpha}$ at
point $z=0$. Bra state $\langle \vec{\alpha} |$ is created by
inserting the corresponding operator at $z=\infty$. We understand the
weight $\vec{\alpha}^{(0)}$ of the vertex operator at zero to be a
free parameter of the correlator. Then the weight
$\vec{\alpha}^{(\infty)}$ is determined uniquely by the momentum
conservation relation, which needs to be satisfied in order for the
correlator~(\ref{correlator:toda}) to be nonzero:
\begin{equation}
  2\sqrt{\beta}{\cal Q}_{\beta}\vec{\rho}=\vec{\alpha}^{(0)}+\vec{\alpha}^{(\infty)}+\sum\limits_{j=1}^l\vec{\alpha}^{(j)}+
  \beta\sum\limits_{a=1}^{n}N_a\vec{e}_{(a)}\,,
\label{charge:conservation}
\end{equation}
where $\vec{\rho}$ and $\vec{e}_{(k)}$ are given by
Eqs.~\eqref{eq:5}. The calculation of the free field
correlator~\eqref{correlator:toda} is presented in
Appendix~\ref{sec:toda-theory} and results in 
\begin{multline}
  {\rm DF}^{A_n}_{l+2}(z_1,\ldots, z_l, \vec{\alpha}^{(0)},
  \vec{\alpha}^{(1)},\ldots,
  \vec{\alpha}^{(l)}, \vec{N} , \beta ) \sim \\
  \sim \prod\limits_{p<k}^l\left( z_p-z_k\right)^{\frac{1}{\beta}\left(\vec{\alpha}^{(p)},\,\vec{\alpha}^{(k)}\right)}
  \oint \prod\limits_{a=1}^n\prod\limits_{i=1}^{N_a} d x^{(a)}_i
  \prod\limits_{a=1}^n\prod\limits_{i=1}^{N_a}\left( x_i^{(a)}
  \right)^{\beta(N_a-N_{a+1}-1)+
    (\alpha_a^{(0)}-\alpha_{a+1}^{(0)})}\times\\
  \times \prod\limits_{a=1}^n\prod\limits_{i\neq j}^{N_a}\left(
    1-\frac{x_j^{(a)}}{x_i^{(a)}}\right)^\beta
  \prod_{a=1}^{n-1}\prod\limits_{i=1}^{N_a}\prod\limits_{j=1}^{N_{a+1}}\left(
    1-\frac{x_j^{(a+1)}}{x_i^{(a)}}\right)^{-\beta}
  \prod\limits_{p=1}^l\prod\limits_{a=1}^n\prod\limits_{i=1}^{N_a}
  \left(
    1-\frac{x_i^{(a)}}{z_p}\right)^{\alpha_a^{(p)}-\alpha_{a+1}^{(p)}}.
\label{correlator1:toda}
\end{multline}
In the Virasoro $(A_1)$ case, the free field integrals are of Selberg type and can be calculated. In the higher rank case
the situation is much more complicated and it is known how to evaluate the integrals only for special values of the momenta of the vertex operators.
As we will see in this paper we are indeed interested in special value of the momenta for which we can calculate the integrals. 

\subsubsection{$q$-Toda conformal blocks}

The  $A_n$ Toda theory admits a $q$-deformation which is described in detail
in~\cite{Feigin:1995sf,Shiraishi:1995rp,Awata:1996dx}. Below we will
use free boson representation of this deformed algebra in order to
derive the corresponding conformal blocks of the $A_n$ $q$-Toda CFT.
For our calculations we use screening currents and vertex operators
from~\cite{Aganagic:2013tta,Aganagic:2014kja}, which are given by
\begin{multline}
S^q_{(a)}(x)=\, \mathopen{:}\,\exp\left( -\sum\limits_{k>0}\frac{1-t^k}{1-q^k}c_k^{(a)}\frac{x^{-k}}{k}+\sum\limits_{k>0}c_{-k}^{(a)}
\frac{x^k}{k} \right)\times\\
\times\exp\left(\sum\limits_{k>0}\frac{1-t^k}{1-q^k}v^k\,c_k^{(a+1)}\frac{x^{-k}}{k}-
\sum\limits_{k>0}v^k\,c_{-k}^{(a+1)}\frac{x^k}{k}\right)\,\mathclose{:}\,\times\\
\times e^{\sqrt{\beta}Q^{(a)}}\,x^{\sqrt{\beta}P^{(a)}}\,e^{-\sqrt{\beta}Q^{(a+1)}}\,x^{-\sqrt{\beta}P^{(a+1)}}\,
\label{screening:current}
\end{multline}
where $t=q^{\beta}$ and we have introduced $v= \sqrt{\frac{q}{t}}$.
Similarly to the undeformed case the sector index $a$ runs between $1$
and $n$.  Bosonic operators $c_k^{(a)},\, Q^{(a)},\, P^{(a)}$ satisfy
the Heisenberg algebra~(\ref{heisenbeg:algebra})

$q$-deformed primary vertex operator is chosen to have  the form
\be
  && V^q_{\vec{\alpha}}\left( z \right) =
  \normord{\exp\left(\sum\limits_{k>0}\sum\limits_{a=1}^{n+1}\frac{q^{k\alpha_a}-1}{1-q^k}c_k^{(a)}
      v^{-ka}\frac{z^{-k}}{k}+
     \right.\nonumber\\&&
     \hspace{6mm} \left.\sum\limits_{k>0}\sum\limits_{a=1}^{n+1}\frac{\left(q^{-k\alpha_a}-v^{2k\left( N-a-1 \right)} \right)}{1-t^k}c_{-k}^{(a)}
     v^{ka} \frac{z^k}{k}\right)} \times
     \e^{\frac{1}{\sqrt{\beta}}\sum\limits_{a=1}^{n+1}\alpha_a\,Q^{(a)}}\,
  z^{\frac{1}{\sqrt{\beta}}\sum\limits_{a=1}^{n+1}\alpha_a\,P^{(a)}
  }\,,
\label{vertex:operator}
\ee where $\vec{\alpha}$ is the weight vector just as in
Eq.~(\ref{vertex:operator:toda}). Essentially this is the vertex
operator of the same form\footnote{For precise matching of
  $\vec{\alpha}$-dependent part one also needs to perform shift of
  weights $\alpha_a\to \alpha_a+\frac{1}{2}a(1-\beta)$ for the vertex
  operators used in ~\cite{Aganagic:2013tta,Aganagic:2014kja} } as the
one that can be found
in~\cite{Aganagic:2013tta,Aganagic:2014kja}. However in the latter
case authors have omitted central part of the operator,
i.e. $\vec{\alpha}$-independent part that commutes with the screening
current $S(x)$ given in (\ref{screening:current}). As we will see this part
appears to be essential for us so we keep it.

As in the non-deformed case we are interested in the following free field correlator
\begin{equation}
  q\mathrm{DF}_{l+2}^{A_n}(z_1,\ldots, z_l, \vec{\alpha}^{(0)},
  \vec{\alpha}^{(1)},\ldots,
  \vec{\alpha}^{(l)}, \vec{N} , q,t)  \stackrel{\mathrm{def}}{=} \langle \vec{\alpha}^{(\infty)}|\,V^q_{\vec{\alpha}^{(1)}}\left( z_1 \right)\dots
  V^q_{\vec{\alpha}^{(l)}}\left( z_l \right)\,
  \prod_{a=1}^{n} Q_{(a)}^{N_a}|\vec{\alpha}^{(0)}\rangle_{\mathrm{free}}\,,
\label{correlator}
\end{equation}
where $Q_{(a)}$ are screening charges related to the screening
currents~(\ref{screening:current}) in the same way as in non-deformed
case~(\ref{screening:charge}). Initial and final states
$|\vec{\alpha}^{(0)}\rangle$, $|\vec{\alpha}^{(\infty)}\rangle$ are
defined in Eq.~(\ref{state}). Conservation
relation~(\ref{charge:conservation}) that constraints weights of the
vertex operators also holds in the $q$-deformed case.

The free field calculation in the $q$-Toda conformal block is similar
to the undeformed case and is presented in
Appendix~\ref{sec:q-toad-theory}. The final result is given by the following
matrix integral:
\begin{multline}
  q{\rm DF}^{A_n}_{l+2}(z_1,\ldots, z_l, \vec{\alpha}^{(0)},
  \vec{\alpha}^{(1)},\ldots,
  \vec{\alpha}^{(l)}, \vec{N} , q,t)\sim
  C_{{\rm vert}}^{q}\left( \vec{\alpha},z \right)\prod\limits_p^lz_p^{\frac{1}{\beta}\left( \vec{\alpha}^{(p)},\vec{\alpha}^{(0)} \right)+
\sum\limits_{a=1}^NN_a\left( \alpha_a^{(p)}-\alpha_{a+1}^{(p)} \right)}\times\\
 \oint \prod\limits_{a=1}^n\prod\limits_{i=1}^{N_a} d x^{(a)}_i
  \prod\limits_{a=1}^n\prod\limits_{i=1}^{N_a}\left( x_i^{(a)}
  \right)^{\beta(N_a-N_{a+1}-1)+(\alpha_a^{(0)}-\alpha_{a+1}^{(0)})
    +\sum\limits_{p=1}^l\left( \alpha_a^{(p)}-\alpha_{a+1}^{(p)}\right)}\times\\
  \times\prod\limits_{a=1}^n\prod\limits_{i\neq j}^{N_a}\frac{\left(
      \frac{x_j^{(a)}}{x_i^{(a)}};q \right)_\infty}
  {\left(t\frac{x_j^{(a)}}{x_i^{(a)}};q \right)_\infty}
  \prod_{a=1}^{n-1}\prod\limits_{i=1}^{N_a}\prod\limits_{j=1}^{N_{a+1}}\frac{\left(
      u \frac{x_j^{(a+1)}}{x_i^{(a)}}; q \right)_\infty} {\left( v
      \frac{x_j^{(a+1)}}{x_i^{(a)}};q\right)_\infty} \prod_{p=1}^l
  \prod_{a=1}^n\prod_{i=1}^{N_a} \frac{\left(
      q^{1-\alpha^{(p)}_{a}}v^a\frac{z_p}{x_i^{(a)}}; q
    \right)_\infty} {\left(
      q^{1-\alpha^{(p)}_{a+1}}v^{a}\frac{z_p}{x_i^{(a)}};q
  \right)_\infty}\,,
\label{correlator1}
\end{multline}
where $u=\sqrt{qt}$ and $C_{ {\rm vert}}$ is the prefactor coming from
ordering different vertex operators. Precise form of this prefactor is
given in~(\ref{q:toda:order:vert}). The expression appears to be very
complicated. However, as we will see further, in cases relevant for us,
in particular when some of the vertices are (semi-)degenerate, this
expression simplifies drastically.

\subsection{Map between $FT[SU(N)]$ and $q$-Toda blocks}\label{3dqdfmap}

The $q$-Toda blocks in DF representation have been shown to map to the
holomorphic blocks of the handsaw quiver theory
\cite{Aganagic:2013tta,Aganagic:2014kja}.  Here we are interested in
the simpler case of the $FT[SU(N)]$ holomorphic block which can be
mapped to a $A_{N-1}$ $q$-Toda block with full primary initial and
final states and $N$ fully degenerate primary vertex operators between
them (we again omit the prefactors in front of both integrals in
holomorphic and conformal blocks):
\begin{equation} {\cal B}^{D_2\times S^1}_{FT[SU(N)]}\sim
  q\mathrm{DF}^{A_{N-1}}_{N+2}\,.
\label{TSUN:map:1}
\end{equation}
with the identification of parameters which we give momentarily.

We begin by considering an $(N+2)-$point conformal block with the weights of
the vertex operators satisfying the following relation:
\begin{equation}
  \alpha_{a+1}^{(p)}=\alpha_a^{(p)},\qquad
  a=1,\dots, N-2, \quad p=1,\dots,N\,.
\label{weights:degeneration}
\end{equation}
The initial state has generic weight $\vec{\alpha}^{(0)}$ and the
weight $\vec{\alpha}^{(\infty)}$ is fixed by the charge conservation
condition~ (\ref{charge:conservation}). We also specify the number of
screening charges to be $N_a = a$ for $a = 1,\ldots, N-1 $.  With this
choice of momenta~(\ref{weights:degeneration}) and screening charges
$q$-Toda conformal block~(\ref{correlator1}) reduces to the following
expression

\begin{multline}
  \langle \vec\alpha^{(\infty)}| V^{q}_{\vec\alpha^{(1)}}(z_1)\cdots
  V^q_{\vec\alpha^{(N)}}(z_N)
  \prod\limits_{a=1}^{N-1}\left(Q_{(a)}^q\right)^a |\vec
  \alpha^{(0)}\rangle\sim
  \oint \prod\limits_{a=1}^{N-1}\prod\limits_{i=1}^{a} d y^{(a)}_i\times\\
  \times\prod\limits_{i=1}^{N-1}\left( y_i^{(N-1)}
  \right)^{\beta\left( N-2 \right)+\alpha_{N-1}^{(0)}-\alpha_N^{(0)}+
    \sum\limits_{p=1}^N\left( \alpha_{N-1}^{(p)}-\alpha_N^{(p)}
    \right)}
  \prod\limits_{a=1}^{N-2}\prod\limits_{i=1}^{a}\left( y_i^{(a)} \right)^{-2\beta+\alpha_a^{(0)}-\alpha_{a+1}^{(0)}}\times\\
  \times \prod\limits_{a=1}^{N-1}\prod\limits_{i\neq
    j}^{a}\frac{\left(\frac{y_j^{(a)}}{y_i^{(a)}};q \right)_\infty}
  {\left(t\frac{y_j^{(a)}}{y_i^{(a)}};q \right)_\infty}
  \prod_{a=1}^{N-2}\prod\limits_{i=1}^{a}\prod\limits_{j=1}^{a+1}\frac{\left(
      u \frac{y_j^{(a+1)}}{y_i^{(a)}}; q \right)_\infty}
  {\left( v \frac{y_j^{(a+1)}}{y_i^{(a)}}; q\right)_\infty} \prod\limits_{p=1}^N\prod\limits_{i=1}^{N-1}\frac{\left(
      q^{1-\alpha^{(p)}_{N-1}}v^{N-1} \frac{z_p}{y_i^{(N-1)}};q
    \right)_\infty} {\left(
      q^{1-\alpha^{(p)}_{N}} v^{N-1} \frac{z_p}{y_i^{(N-1)}}; q
  \right)_\infty}\,,
\label{fdeg}
\end{multline}
where we have omitted prefactors coming from the ordering of the vertices to concentrate only on
the integral for the moment. Expression on the r.h.s of~(\ref{fdeg}) is almost of the same form as the integral in ${\cal
  B}^{D_2\times S^1}_{FT[SU(N)]}$ block ~(\ref{TSUN:partition}).
To complete the map we need to impose a further restriction on the
$q$-Toda vertex operator parameters. First of all looking on the
one-loop contribution of the vector and adjoint multiplets in the
block integral~(\ref{TSUN:partition}) we can see that the gauge theory
parameter $t$ related to the $3d$ axial mass is identified with the
$t$-parameter of Toda CFT deformation. Then in order to match the
contribution of the bifundamental hypers with the corresponding term
in the correlator~(\ref{fdeg}) we need to  make the following
identification between the integration variables $y$ in the $q$-DF
integral~(\ref{fdeg}) and $x$ in the holomorphic block
integral~(\ref{TSUN:partition}):
\begin{equation}
y_i^{(a)}=x_i^{(a)}v^{-a}.
\end{equation}
To identify the last product in the third line of Eq.~(\ref{fdeg})
with the contribution of the fundamental chiral multiplets we need
\begin{gather}
  \mu_p=q^{1-\alpha_N^{(p)}}\,v^{2N-2}z_p\,,\notag\\
  t\mu_p=q^{1-\alpha_{N-1}^{(p)}}v^{2N-2}z_p\,,
\label{mass:map}
\end{gather}
which amounts to requiring
\begin{equation}
\alpha_N^{(p)}-\alpha_{N-1}^{(p)}=\beta\,.
\label{eq:10}
\end{equation}
Eq.~\eqref{eq:10} together with the
condition~(\ref{weights:degeneration}) completely fixes all the
components of the vertex weight vectors in terms of the last
components so that all weights have the form
\begin{equation}
        \vec{\alpha}^{(p)}=(g_p-\beta)\vec{1}+\beta\vec{\omega}_{N-1}\,,
\end{equation}
where $g_p$ is arbitrary constant and $\vec{\omega}_{N-1}$ is the
highest weight vector of $A_{N-1}$. The map~(\ref{mass:map}) give us
freedom to choose $g_p$ freely. For example we can absorbe it into the
definition of the insertion points $z'_p=q^{\beta-g_p}z_p$ and
consequently have $\mu_p=v^{2N}z'_p$.  Alternatively we can
simultaneously shift of all the components of the vertex operator
weight. This operation does not affect the $q$-DF
integral~(\ref{correlator1:toda}) as it only contribute an overall
factor in front of the integral which we omit anyway.  So we choose to
set $g_p=\beta$ corresponding to vertices with fully degenerate
momenta (corresponding to simple \emph{degenerate} punctures in the
AGT setup):
\begin{equation}
  \vec{\alpha}^{(p)}=\beta\vec{\omega}_{N-1}\,.
\label{full:degenration}
\end{equation}
Finally we need to identify the FI parameters of the $FT[SU(N)]$
theory with the components of the initial and final momenta of
$q$-Toda CFT $\alpha_a^{(0)}$. This can be done by looking at the
powers of $y_i^{(a)}$ in the $q$-DF integral~(\ref{fdeg}) and
$e^{X_i^{(a)}}$ powers in the block
integral~(\ref{TSUN:partition}). We arrive at the following relation:
\begin{equation}
  \alpha_a^{(0)}-\alpha_{a+1}^{(0)}+1-2\beta=
  \frac{T_a-T_{a+1}}{\hbar} - \beta,
\end{equation}
and thus
\begin{equation}
T_a=\hbar\left( \alpha_a^{(0)}+\left( \beta-1 \right)a \right).
\end{equation}
  
Summarizing, the dictionary between the ${\cal B}^{D_2\times
  S^1}_{FT[SU(N)]}$ block~(\ref{TSUN:partition}) and the
$q$-Toda block $q{\rm DF}^{A_{N-1}}_{N+2}\left( z_1,\ldots, z_N,
    \vec{\alpha}^{(0)}, \beta \vec{\omega}_N,\ldots, \beta
    \vec{\omega}_N, [1,2,\ldots,N-1], q, q^{\beta} \right)$ is given
in Table~\ref{table:map}.
\begin{table}[h!]
\centering
\begin{tabular}{|c|c|c|}
  \hline
  ${\cal B}^{D_2\times S^1}_{FT[SU(N)]}$ & Identification & $q\mathrm{DF}^{A_{N-1}}_{N+2}$ \\
  \hline
  Integration parameters $x_i^{(a)}$ &$y_i^{(a)}=x_i^{(a)}v^{-a}$ & Screening current positions $y_i^{(a)}$\\
  \hline
  Axial mass $t$ &  $t=q^\beta$ & Central charge parameter $\beta$ \\
  \hline
  Vector  masses $\mu_p$ &$\mu_p=v^{2N}z_p$ & Positions of the vertex
  operators $z_p$ \\
  \hline
  FI parameters $T_a$ &$T_a=\hbar \left( \alpha_a^{(0)}+(\beta-1) a
  \right)$ & Initial state momentum vector $\vec{\alpha}^{(0)}$\\
  \hline
\end{tabular}
\caption{Map between the parameter of the $FT[SU(N)]$ holomorphic block~(\ref{TSUN:partition})
and the conformal block~(\ref{correlator1}) of the $q$-Toda theory.}
\label{table:map}
\end{table}
It is important to notice here that with the choice (\ref{full:degenration}) of the vertex operator weights and the map of parameters specified in Table~\ref{table:map} 
  the prefactor $C_{ {\rm vert}}^q$ in the $q\mathrm{DF}$ integral (\ref{correlator1}) simplifies drastically and reduces to:
  \be
  C_{ {\rm vert}}^q\rightarrow \prod\limits_{p<r} \mu_p^\beta \frac{\qPoc{\frac{q}{t}\frac{\mu_r}{\mu_p}}}{\qPoc{t\frac{\mu_r}{\mu_p}}}\,.
  \ee
  As we can see ratio of $q$-Pochhammers in this expression exactly reproduces contribution (\ref{eq:66}) of the flipping singlets into holomorphic block
  of $FT\left[ SU(N) \right]$ theory. However we still omit mixed Chern-Simons terms since their matching would require more delicate
calculation of conformal blocks.

To complete the discussion of Duality web I in
  Fig. \ref{fig1} we need to comment on the counterpart of the $3d$
  spectral duality relating the $3d$ holomorphic blocks ${\cal
    B}^{D_2\times S^1}_{FT[SU(N)]}$ and $\hat{\cal B}^{D_2\times
    S^1}_{FT[SU(N)]}$ by the map~\eqref{eq:20}. In this context the
  spectral duality reads:
\begin{multline}
\hat{C}~
q\hat{{\rm DF}}^{A_{N-1}}_{N+2}\left(\hat{z}_1,\ldots,
    \hat{z}_N, \hat{\vec{\alpha}}^{(0)},
    \hat{\beta}\vec{\omega}_N,\ldots,
    \hat{\beta}\vec{\omega}_N, [1,2,\ldots,N-1], q, q^{\hat{\beta}}\right) =\\
  =C~
q{\rm DF}^{A_{N-1}}_{N+2}\left( z_1,\ldots, z_N,
    \vec{\alpha}^{(0)}, \beta \vec{\omega}_N,\ldots, \beta
    \vec{\omega}_N, [1,2,\ldots,N-1], q, q^{\beta} \right)\,.
\label{DF:duality}
\end{multline}
The parameters of the dual DF integrals are related
by the spectral duality map:
\begin{gather}
  \hat{\beta}=\beta,\notag\\
  \hat{\alpha}^{(0)}_p = \frac{1}{\hbar}\log z_p+(1-\beta)(N+ p),\label{DF:duality:map}\\
  \hat{z}_p=q^{\alpha_p^{(0)} + (N + p)(\beta-1)},\notag
\end{gather}
which swaps the \emph{coordinates} of the vertex operators with the
\emph{momenta}.  The prefactor
$C=C\left(z_1,\ldots,z_N,\vec{\alpha}^{(0)},\beta\right)$ (and
similarly $\hat{C}$) is given by the product of the omitted factor in
front of integral in the $q$-Toda conformal block~(\ref{correlator1})
and $F^{-1}(q,\,t,\,\vec{\tau})$ from the holomorphic
block~(\ref{TSUN:partition}).

\section{Duality web II: $FT[SU(N)]$ and its spectral dual via
  Higgsing}
\label{sec:higgsing-5d-gauge}

In this section we will describe how to obtain the partition function
of the $3d$ $\mathcal{N}=2$ $FT[SU(N)]$ gauge theory on
$\mathbb{R}^2_q \times S^1$ by Higgsing the $5d$ $\mathcal{N}=1$
square linear quiver theory on the $\Omega$-background
$\mathbb{R}^4_{q,t}\times S^1$.

We consider the $3d$-$5d$ version of the setup
of~\cite{Hanany:2003hp,Hanany:2004ea, Dorey:2011pa, Aganagic:2013tta}.
Physically the $3d$ theory lives on the worldvolume of the vortices
appearing in the Higgs phase of the $5d$ theory.  Using a Type IIB
brane setup with NS5 and D5 branes to engineer the $5d$ linear quiver
theory we can realise the $3d$ vortex theory as the low energy theory
on the D3 branes stretched between NS5 and D5 branes.  The spectral
self-duality of the $FT[SU(N)]$ theory descends from IIB S-duality
which swaps NS5 and D5 branes.

\subsection{$5d$ instanton partition function,  Higgs branch and the vortex theory}
\label{sec:5d-theory}
\begin{figure}[h]
  \centering
  \includegraphics{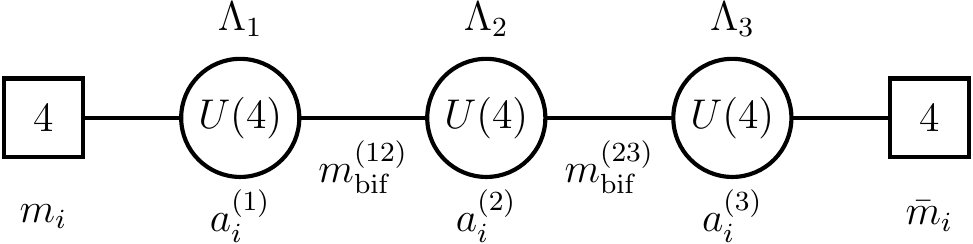}
  \caption{$5d$ linear ``square'' quiver gauge theory with gauge group
    $U(4)^3$. $a^{(a)}_i$ are the exponentiated vevs of the adjoint
    scalars.     $m_{\mathrm{bif}}^{(a,a+1)}$, $m_i$, $\bar{m}_i$ and $\Lambda^{(a)}$ are the exponentiated masses and gauge
    couplings.}
  \label{fig:1}
\end{figure}

Consider the $5d$ $\mathcal{N}=1$ square quiver gauge theory with
gauge group $U(N)^{N-1}$ in $\Omega$-background. An example of such
theory for $N=4$ is depicted in Fig.~\ref{fig:1}. The parameters of
the theory are:
\begin{enumerate}

\item Vacuum expectation values (vevs) of the adjoint scalar
  fields. We denote the exponentiated\footnote{The exponentiated vev
    (resp.\ mass, coupling) is related to the physical vev
    $A$ (resp.\ mass $M$, complexified coupling $T$) by the
    formula $a = e^A$ (resp.\ $m = e^M$, $\Lambda = e^T$). The masses
    $M$ and vevs $A$ in these formulas are made dimensionless, by
    measuring them in units of inverse radius $R^{-1}$ of the
    compactification circle $S^1$.} vev of the $i$-th diagonal component of the adjoint
  scalar of the $a$-the gauge group by $a_i^{(a)}$, $i=1,\ldots , N$,
  $a=1,\ldots,N-1$. 

\item Couplings $\Lambda_a$, $a=1,\ldots,(N-1)$ of the gauge groups.
  
\item Masses $m_i$ (resp.\ $\bar{m}_i$) of the fundamental (resp.\
  antifundamental) hypermultiplets coupled to the first (resp.\ the
  last) gauge groups in the linear quiver.

\item Bifundamental masses $m_{\mathrm{bif}}^{(a,a+1)}$.
 Since we consider  the $U(N)$ case these
  parameters  could be eliminated by
  shifting the ratio of trace parts $\prod_{i=1}^N
  \frac{a^{(a)}_i}{a^{(a+1)}_i} \to \prod_{i=1}^N
  \frac{a^{(a)}_i}{m_{\mathrm{bif}}^{(a,a+1)}a^{(a+1)}_i}$. However,
  we will keep them as separate parameters to make
  the formulas more symmetric.

\item Parameters $q$ and $t$ of the $\Omega$-deformation.

\end{enumerate}

The $\Omega$-background partition function of the theory is given by (the $5d$ version of) the
instanton partition function~\cite{Nekrasov:2002qd}. It is the product of three
factors: the classical piece $Z_{\mathrm{cl}}^{5d}$, the one-loop
determinant $Z_{1-\mathrm{loop}}^{5d}$ and the instanton part
$Z_{\mathrm{inst}}^{5d}$, of which we write down explicitly only the
last one:
\begin{multline}
  \label{eq:1}
  Z_{\mathrm{inst}}^{5d} (a^{(a)}_i, m_i, \bar{m}_i,
  m_{\mathrm{bif}}^{(a,a+1)}, \Lambda_a) =\sum_{\vec{Y}^{(1)}, \ldots ,
    \vec{Y}^{(N-1)}} \prod_{a=1}^{N-1}
  \Lambda_a^{|\vec{Y}^{(a)}|} \times \\
  \times z_{\mathrm{fund}}(\vec{m},\vec{a}^{(1)}, \vec{Y}^{(1)})
  \frac{\prod_{a=1}^{N-2} z_{\mathrm{bif}}(m_{\mathrm{bif}}^{(a,
      a+1)}, \vec{a}^{(a+1)}, \vec{a}^{(a)},
    \vec{Y}^{(a+1)}, \vec{Y}^{(a)}) }{\prod_{a=1}^{N-1}
    z_{\mathrm{vect}}(\vec{a}^{(a)}, \vec{Y}^{(a)})}\,
  z_{\overline{\mathrm{fund}}}(\vec{\bar{m}},\vec{a}^{(N-1)},
  \vec{Y}^{(N-1)})
\end{multline}
where $\vec{Y}^{(a)}$, $a=1,\ldots,(N-1)$ each denote the $N$-tuple of
Young diagrams, $\vec{Y}^{(a)} = \{ Y^{(a)}_1, \ldots, Y^{(a)}_N \}$
and
\begin{gather}
  z_{\mathrm{fund}}(\vec{m},\vec{a},\vec{Y}) = \prod_{r=1}^N
  \prod_{s=1}^N \prod_{(i,j)\in Y_r}\left( 1 - \frac{a_r}{m_s} q^{j-1}
    t^{1-i} \right),\\
  z_{\overline{\mathrm{fund}}}(\vec{m},\vec{a},\vec{Y}) =
  \prod_{r=1}^N \prod_{s=1}^N \prod_{(i,j)\in Y_r}\left( 1 -
    \frac{m_s}{a_r} q^{1-j}
    t^{i-1} \right),\\
  z_{\mathrm{bif}}(m_{\mathrm{bif}},\vec{a},\vec{b},\vec{Y}, \vec{W})
  = \prod_{r=1}^N \prod_{s=1}^N \prod_{(i,j)\in Y_r} \left( 1 -
    \frac{a_r}{m_{\mathrm{bif}} b_s} q^{Y_{r,i} - j}
    t^{W^{\mathrm{T}}_{s,j} -
      i + 1} \right) \times\\
  \times \prod_{(k,l)\in W_s} \left( 1 - \frac{a_r}{m_{\mathrm{bif}}
      b_s} q^{-W_{s,i} + j-1} t^{-Y^{\mathrm{T}}_{r,j} +
      i} \right),\\
  z_{\mathrm{vect}}(\vec{a}, \vec{Y}) =
  z_{\mathrm{bif}}(1,\vec{a},\vec{a},\vec{Y}, \vec{Y}).
\end{gather}
We are now going to show how the instanton partition functions can be
reduced via Higgsing to the $3d$ vortex partition function for the
$FT[SU(N)]$ theory.  As mentioned in the Introduction, Higgsing a
$5d$ partition function typically produces the partition function of a
coupled $5d$-$3d$ system describing a codimension-two surface operator
coupled to the bulk theory.  Here we are interested in the case where
the $5d$ bulk theory is trivial, consisting only of a bunch of
decoupled hypermultiplets and so rather than reducing to a coupled
system, the Higgsing directly yields the $3d$ vortex theory.

\subsubsection*{$FT[SU(2)]$ case}
Let us start with the simplest example of the square $U(2)$ theory,
i.e.\ the $U(2)$ gauge theory with two fundamental and two
antifundamental multiplets. The Higgs branch touches the Coulomb
branch at the point $a_i = m_i$, $i=1, 2$. The theory on the Higgs
branch contains nonabelian vortex strings with worldvolumes
spanning\footnote{Of course, since the $\Omega$-background localizes
  all the vortices at the origin of $\mathbb{R}^2_q$, one can
  equivalently view them as vortices spanning $\mathbb{R}_t^2 \times
  S^1 \subset \mathbb{R}^4_{q,t}\times S^1$. The equivalence between
  the two viewpoints gives rise to the famous Langlands
  correspondence~\cite{Kapustin:2006pk}.}  $\mathbb{R}_q^2 \times S^1
\subset \mathbb{R}^4_{q,t}\times S^1$. The $3d$ theory on charge $M$
vortices is the $\mathcal{N}=2$ $U(M)$ theory with \begin{itemize}
\item adjoint multiplet with mass $t$,
\item two fundamental multiplets with masses $\mu_i = m_i$,
\item two antifundamental multiplets with masses
  $\bar{\mu}_i\frac{q}{t} = \bar{m}_i$.
\end{itemize}
When the 3d FI parameter is turned on the $3d$ theory is on the Higgs
branch and the gauge group is broken to $U(M_1) \times U(M_2)$ with
$M_1 + M_2 =M$.
 
The vortex theory  is actually dual to  
the theory at certain discrete points on the Coulomb branch without any vortices~\cite{Dorey:2011pa}. 
In particular the $5d$ theory on the Higgs branch
with $U(M_1) \times U(M_2)$ vortex is equivalent to the theory on the
Coulomb branch with
\begin{equation}
  \label{eq:18}
  a_i = m_i\, t^{M_i},\qquad i=1,2.
\end{equation}
Indeed for $a_i = m_i\, t^{M_i}$ the sum over Young tableaux in the  instanton partition function  $Z_{\mathrm{inst}}^{5d}$~\eqref{eq:1}
 truncates so that only the diagrams $Y_i$ of length $l(Y_i) \leq M_i$
contribute. The surviving diagrams correspond to the values of the
$3d$ adjoint scalar fields fixed under the localization, which in the
IR are diagonal $M_1 \times M_1$ and $M_2 \times M_2$ matrices. The
instanton contributions for these diagrams indeed match those of the
vortex expansion $Z_{\mathrm{vort}}^{3d}$ of the $3d$ theory~\cite{Aganagic:2013tta}
with the following dictionary:
\begin{center}
  \begin{tabular}{cc|cc}
    \multicolumn{2}{c|}{$5d$ square $U(2)$} & \multicolumn{2}{|c}{$3d$
      vortex theory} \\
    \hline
    coupling &$\Lambda$ & $\tau 
    $ & FI parameter\\
    fundamental masses &$m_i$ & $\mu_i$ & fundamental masses\\
    antifundamental masses & $\bar{m}_i$ & $\bar{\mu}_i = \frac{q}{t} \bar{m}_i$ &antifundamental masses\\
    Coulomb parameters &$a_i = m_i t^{M_i}$ & $U(M_1) \times U(M_2)$ & rank of the gauge groups\\
    $\Omega$-parameter &$t$ & $t$ & adjoint mass \\
    $\Omega$-parameter &$q$ & $q$ & $\Omega$-parameter
\end{tabular}
\end{center}
To obtain the $FT[SU(2)]$ vortex partition functions one needs to
further tune the parameters of the $3d/5d$ setup choosing $M_1 = 1$
and $M_2 = 0$ and setting the antifundamental masses to $\bar{m}_i =
\frac{t^2}{q} m_i$ (consistent with the cubic superpotential of the
$\mathcal{N}=2^*$ theory).

\subsubsection*{$FT[SU(N)]$ case}

The Higgsing procedure described above can be easily extended to the
$FT[SU(N)]$ case which can be obtained from the square $U(N)^{N-1}$
$5d$ by tuning the masses and the Coulomb parameters to the following
values:
\begin{equation}
  \label{eq:222}
  \begin{array}{c}
    a^{(1)}_1 = m_1 t,\\
    a^{(1)}_2 = m_2,\\
    a^{(1)}_3 = m_3,\\
    \vdots\\
    a^{(1)}_{N-1} = m_{N-1},\\
    a^{(1)}_N = m_N,
  \end{array} \qquad \begin{array}{c}
    a^{(2)}_1 = m_1 t,\\
    a^{(2)}_2 = m_2 t,\\
    a^{(2)}_3 = m_3,\\
    \vdots\\
    a^{(2)}_{N-1} = m_{N-1},\\
    a^{(2)}_N = m_N,
  \end{array} \cdots \qquad \begin{array}{c}
    a^{(N-1)}_1 = m_1 t,\\
    a^{(N-1)}_2 = m_2 t,\\
    a^{(N-1)}_3 = m_3 t,\\
    \vdots\\
    a^{(N-1)}_{N-1} = m_{N-1} t,\\
    a^{(N-1)}_N = m_N,
  \end{array} \qquad \begin{array}{c}
    \bar{m}_1 = m_1 \frac{t^2}{q},\\
    \bar{m}_2 = m_2 \frac{t^2}{q},\\
    \bar{m}_3 = m_3 \frac{t^2}{q},\\
    \vdots\\
    \bar{m}_{N-1} = m_{N-1} \frac{t^2}{q},\\
    \bar{m}_N = m_N \frac{t^2}{q},
  \end{array}
\end{equation}
and $m_{\mathrm{bif}}^{(r,r+1)} = 1$ for all $r$. After this
specialization the instanton sum in~\eqref{eq:1} over $N$-tuples of
Young diagrams $\vec{Y}^{(a)}$ truncates: the only surviving terms are
those with $\vec{Y}^{(a)}$ of the following form:
\begin{equation}
  \label{eq:3}
    \begin{array}{c}
    Y^{(1)}_1 = [k^{(1)}_1],\\
    Y^{(1)}_2 = \varnothing,\\
    Y^{(1)}_3 = \varnothing,\\
    \vdots\\
    Y^{(1)}_{N-1} = \varnothing,\\
    Y^{(1)}_N = \varnothing,
  \end{array} \qquad \begin{array}{c}
    Y^{(2)}_1 = [k^{(2)}_1],\\
    Y^{(2)}_2 = [k^{(2)}_2],\\
    Y^{(2)}_3 = \varnothing,\\
    \vdots\\
    Y^{(2)}_{N-1} = \varnothing,\\
    Y^{(2)}_N = \varnothing,
  \end{array} \qquad \cdots \qquad \begin{array}{c}
    Y^{(N-1)}_1 = [k^{(N-1)}_1],\\
    Y^{(N-1)}_2 = [k^{(N-1)}_2],\\
    Y^{(N-1)}_3 = [k^{(N-1)}_3],\\
    \vdots\\
    Y^{(N-1)}_{N-1} = [k^{(N-1)}_{N-1}],\\
    Y^{(N-1)}_N = \varnothing,
  \end{array}
\end{equation}
where the integers $k^{(a)}_i$ satisfy the
constraints~(\ref{eq:411s}). We then see that  the instanton partition function reduces exactly to the 
 vortex series~\eqref{eq:15}.

%

\subsection{$(p,q)$-webs, topological strings and the geometric transition}
\label{sec:p-q-brane}
\begin{figure}[h]
  \centering
  \begin{tabular}{cc}
    \includegraphics
    {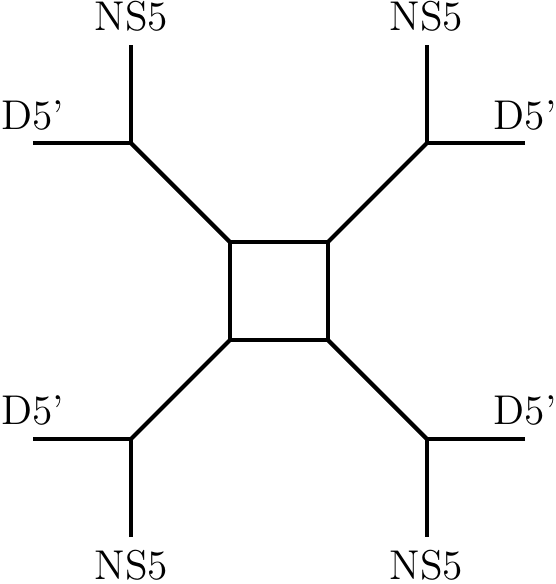}\hspace{2cm} & \includegraphics
    {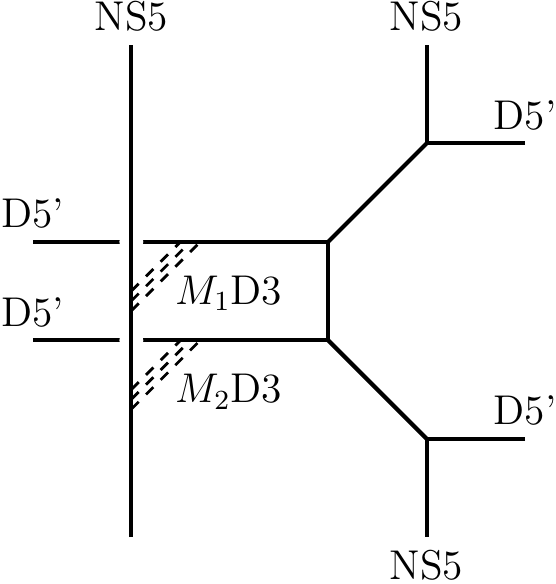}\\
    a) \hspace{2cm} & b)
  \end{tabular}
  \vspace{0.5cm}
  
    \begin{tabular}{ccc}
  \includegraphics
  {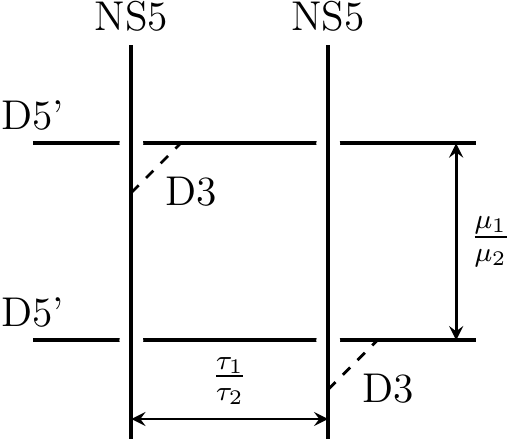}
\hspace{0.5cm} & \includegraphics
    {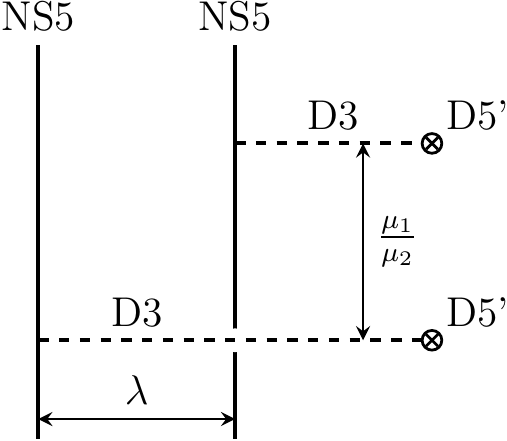} &  \hspace{0.5cm}    \includegraphics
    {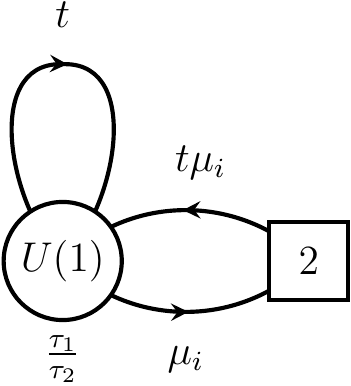}
\\
    c) \hspace{2cm} & d)& e)
  \end{tabular}

 \caption{a) $(p,q)$-brane web $\mathfrak{S}$, corresponding to the
    $5d$ $U(2)$ square gauge theory. The picture can also be viewed as
    the toric diagram of the CY. Compactifying M-theory on this CY one
    arrives at the same gauge theory. b) One NS5 brane is separated in
    the transverse direction, so that the gauge theory goes onto the
    Higgs branch. The dashed lines denote two stacks of $M_1$ and
    $M_2$ D3 branes respectively stretching between the NS5 and D5'
    branes. One can also view this picture as describing the CY
    background after the geometric transition. c) The brane setup
    corresponding to the $3d$ $FT[SU(2)]$ theory where $M_1=1$ and
    $M_2=0$.  The masses $\mu_{1,2}$ and FI parameters $\tau_{1,2}$ of
    the gauge theory are encoded in the positions of the branes as
    shown.  Notice that the second NS5 brane is also detached from the
    web and a single D3 is stretched between it and the second D5'
    brane. This imposes the condition $\bar\mu_i=t \mu_i$.  d) Another
    projection of the brane setup c). The distance between the NS5
    branes in the transversal directions $\lambda$ corresponds to the
    coupling constant of the $3d$ theory which does not enter the
    holomorphic block. e) The $3d$ $FT[SU(2)]$ gauge theory obtained
    form the brane setup c), d). The deformation parameter $t$ giving
    mass to the adjoint multiplet originates from the
    $\Omega$-background in the $\mathbb{R}^4_{q,t}$ part of the $10d$
    geometry.}
  \label{fig:6}
\end{figure}

We can realize $3d$ and $5d$ theories in terms of $(p,q)$-brane webs
in Type IIB string theory~\cite{Aharony:1997bh}.  The $(p,q)$-brane
web $\mathfrak{S}$ for the $U(N)^{N-1}$ theory consists of $N$ NS5
branes (vertical) and $N$ D5' branes (horizontal)\footnote{We can take the NS5 branes extending in directions $012789$ and the D5' in $012478$.} as shown in
Fig.~\ref{fig:6},~a). The NS5 and D5' branes fuse to form
$(1,1)$-branes, which are diagonal in Fig.~\ref{fig:6}. The tensions
of the branes are balanced regardless of their relative positions,
therefore the system has moduli, corresponding to the parameters of
the gauge theory. The $5d$ theory obtained in this way lies on the
Coulomb branch, so some of the brane moduli are Coulomb
parameters. Others correspond to masses and gauge
couplings. Concretely, changing the Coulomb moduli means changing the
positions of the \emph{internal} branes, while fixing the
semi-infinite ones.

\newpage

Where is the Higgs branch in the brane setup? The origin of the Higgs
branch appears when at least one NS5 brane does not fuse with any of
the D5' branes passing vertically through the whole picture. The NS5
brane can then be separated from the rest of the $(p,q)$-web in the
directions perpendicular to the plane of the picture, as shown in
Fig.~\ref{fig:6},~b). The position of the NS5 brane in these
directions corresponds to the Higgs branch moduli. The vortex strings
appearing in the Higgs phase of the $5d$ theory correspond to the D3
branes stretching between the D5' branes and the separated NS5 brane.
$3d$ $FT[SU(N)]$ theory is obtained by further tuning the positions
of the branes as shown in Fig.~\ref{fig:6}.

Notice that we need also to detach the last NS5 brane from the web and
stretch a single D3 between it and one of the D5' brane. This imposes
the condition on the fundamental chiral masses $\bar\mu_i=t
\mu_i$. Notice the resulting diagonal pattern of the D3 branes.

To find the matter content of the $3d$ theory on the D3 branes it is
instructive to look at a different projection of the brane setup shown
in Fig.~\ref{fig:6},~d). The D3 segment between the two NS5 branes
supports $4d$ $U(1)$ theory, which in the IR becomes $3d$, since the
length of the segment is finite, while open strings stretching between D3 branes across NS5 branes 
give rise to bifundamental filelds.
 The flipping fields arise, because the D3 branes
  can move along the five-branes in the directions perpendicular to
  those drawn in Fig.~\ref{fig:6}.
  
  The distance between the D5' branes
$\frac{\mu_1}{\mu_2}$ determines the masses of the fundamental
multiplets while the distance between the NS5 branes in the
``Coulomb'' direction $\frac{\tau_1}{\tau_2}$ gives the FI
parameter of the $3d$ $FT[SU(2)]$ gauge theory.\\

We can now observe that the brane web in Fig.~\ref{fig:6},~c) under
the action of Type IIB $S$-duality which exchanges the NS5 and D5'
branes, thus effectively taking the mirror image along the diagonal,
is sent into an identical web diagram with mass and FI parameters
exchanged. This is due to two properties.
\begin{enumerate}
\item We have $N$ NS5 and $N$ D5' branes. This is the reason why we
  call it \emph{square} theory.
  
\item The number of D3 branes sitting at each intersection is tuned so
  that the whole collection is symmetric along the diagonal.
\end{enumerate}

This construction indicates that the $3d$ spectral self-duality of the
$FT[SU(N)]$ theory follows from Type IIB $S$-duality.

We can see this very explicitly if we transform our Type IIB
$(p,q)$-brane web into a purely geometric background of M-theory
without any five-branes (this technique is known as geometric
engineering of gauge theories).  The background is a toric CY
three-fold $\mathfrak{S}$ with toric diagram copied after the
$(p,q)$-brane web. One can then compute the partition function of
M-theory on $\mathfrak{S} \times \mathbb{R}^4_{q,t} \times S^1$ by
computing the refined (with $(q,t)$-deformation) topological string
partition function $Z_{\mathrm{top}}[X]$
\cite{Aganagic:2003db,Iqbal:2007ii}.

The positions of the five-branes become complexified K\"ahler
parameters of the CY $\mathfrak{S}$. It will be natural for us to
trade the K\"ahler parameters of the compact two-cycles on CY for the
so-called spectral parameters living on the edges of the diagram. They
are defined so that for two parallel lines on the diagram with
spectral parameters $z$ and $w$ the K\"ahler parameter of the
two-cycle between the lines is given by $\frac{z}{w}$:
\begin{equation}
  \label{eq:6}
\includegraphics{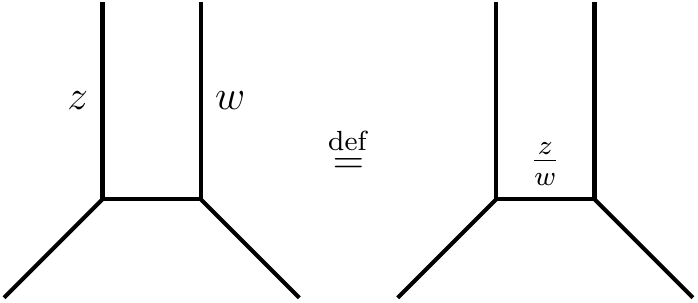}  
\end{equation}
and are conserved at the brane junctions:
\begin{equation}
  \label{eq:24}
  \includegraphics{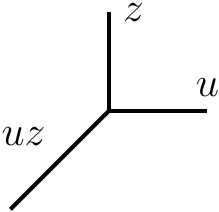}
\end{equation}

\begin{figure}[h]
  \centering
  
  \begin{tabular}{cc}
\hspace{0.25cm}
    \includegraphics
  {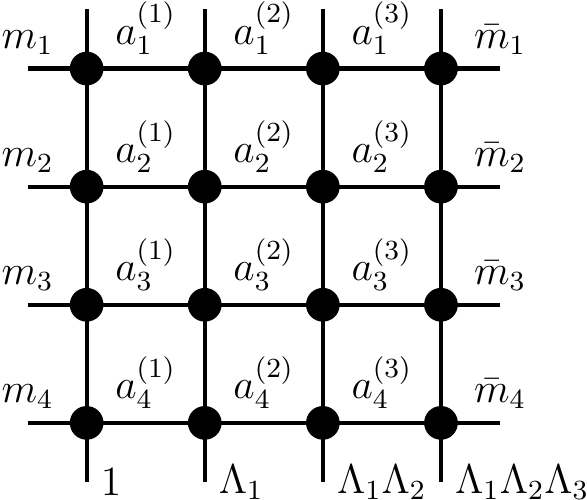} 
\hspace{0.25cm}
& \hspace{0.25cm} \includegraphics
{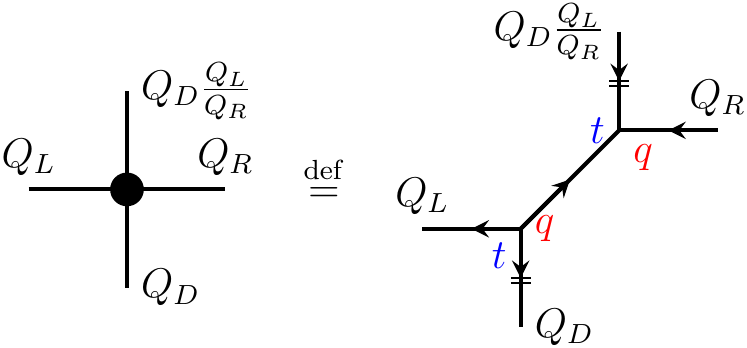}\hspace{0.25cm} \\
  a) & b)
\end{tabular}

\caption{a) The toric diagram $\mathfrak{S}$ producing the square
  gauge theory from Fig.~\ref{fig:1}. The labels of the lines
  correspond to the spectral parameters on the toric diagram and
  encode the parameters of the gauge theory. b) The shorthand notation
  for the crossings: black circles denote the resolved conifold
  geometries with general K\"ahler parameters. Notice that the
  spectral parameter on the upper vertical leg is determined by the
  ``conservation law''.}
  \label{fig:4}
\end{figure}

The toric diagram of the CY background $\mathfrak{S}$ corresponding to
the $5d$ $U(4)^3$ square gauge theory is shown in Fig.~\ref{fig:4} (we
use the shorthand notation for the resolved conifold pieces of the
geometry, as shown in b)). The Higgs branch of the $5d$ gauge theory
appears when all the conifold resolutions along one of the vertical
lines become degenerate, i.e.\ their K\"ahler parameters vanish. In
this case the CY can be \emph{deformed,} so that each crossing looks
like a deformed conifold geometry, locally a $T^{*}S^3$. Resolved and
deformed backgrounds of (refined) topological strings are related by
the geometric transition~\cite{Gopakumar:1998ki, Aganagic:2012hs},
i.e.\ at quantized values of the conifold K\"ahler parameter $Q=
\sqrt{\frac{q}{t}} t^N$ the resolution is equivalent to the deformed
geometry with a stack of $N$ Lagrangian branes wrapped over the
compact three-cycle. The background after the geometric transition can
be illustrated by Fig.~\ref{fig:6} b) and c) with dashed lines now
playing the role of Lagrangian branes. Quantized values of the
K\"ahler parameters correspond to the points~\eqref{eq:18} on the
Coulomb branch of the $5d$ gauge theory, while the deformed geometry
with Lagrangian branes corresponds to $5d$ theory on the Higgs branch
with a collection of vortices, on which the $3d$ $FT[SU(N)]$ theory
leaves. We call the Higgsed version of CY $\mathfrak{S}$ by
$\mathcal{S}$ (so that $\mathcal{S}$ represents a particular point
in the K\"ahler moduli space of $\mathfrak{S}$). In this way, geometric
transition explains the Higgsing procedure described above.

\begin{figure}[h]
  \centering

  \begin{tabular}{cc}
    \hspace{0.25cm}
  \includegraphics
  {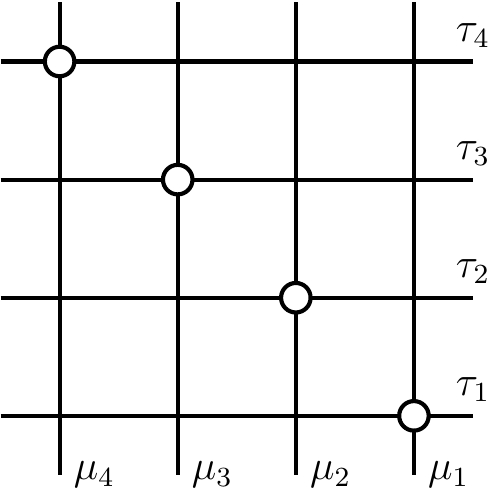}
\hspace{0.25cm} & \hspace{0.25cm}      \includegraphics
    {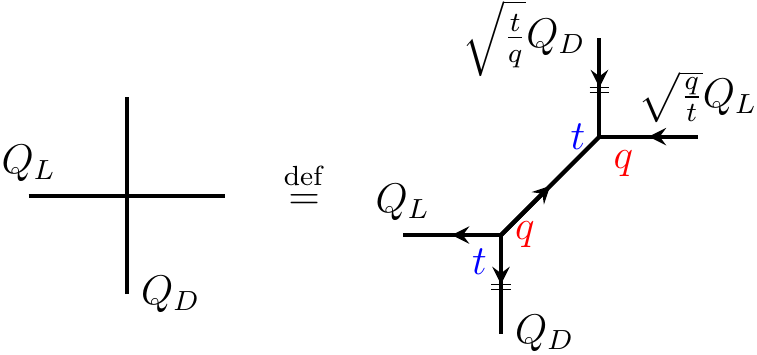} \hspace{0.25cm}
 \\
  a) & b)
\end{tabular}

  \includegraphics
  {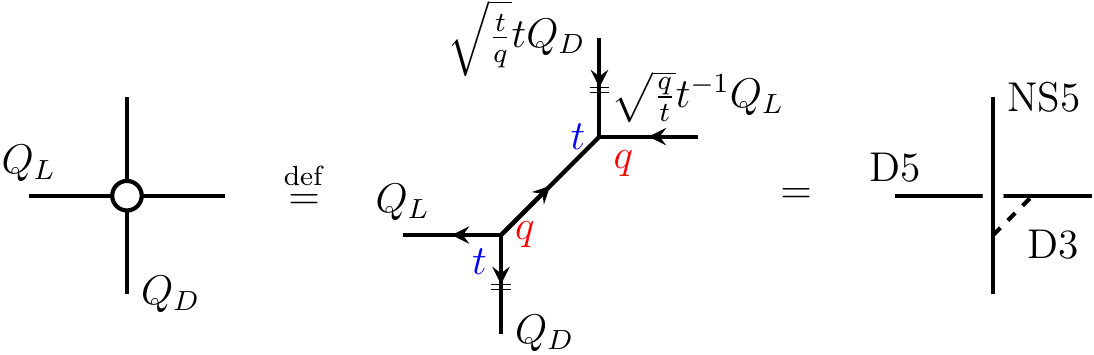}

c)

\caption{a) The toric diagram $\mathcal{S}$ for the higgsed gauge
  theory, i.e.\ for parameters tuned as in Eq.~(\ref{eq:222}). We have
  rotated the picture compared to Fig.~\ref{fig:6} by $\frac{\pi}{2}$
  to aid explicit refined topological string computations. b) The
  ``empty crossing'' denotes the fully degenerate refined conifold
  amplitude. c) The ``higgsed crossing'' denote refined conifold
  amplitude corresponding to the geometric transition of the deformed
  geometry with a single brane wrapping $S^3$.}
  \label{fig:5}
\end{figure}

We can then calculate the closed topological string partition
$Z_{\mathrm{top}}(\vec{\mu}, \vec{\tau}, q,t)$ for the CY background $\mathcal{S}$
with tuned  K\"ahler paramters in Fig.~\ref{fig:5} using
the refined topological vertex \cite{Aganagic:2003db,Iqbal:2007ii} or
using the techniques of~\cite{Awata:2016riz, Mironov:2016yue} and
check that it reproduces the vortex plus one loop factor of the
holomorphic block ${\cal B}^{D_2\times S^1}_{FT[SU(N)]}(\vec{\mu},
\vec{\tau}, q, t)$:
\begin{equation}
  \label{eq:63}
\prod_{i<j} \frac{\left( \frac{q}{t}
      \frac{\mu_i}{\mu_j};q\right)_{\infty}}{\left( t
      \frac{\mu_i}{\mu_j};q\right)_{\infty} }  Z_{1\mathrm{-loop}}^{3d,\,(\alpha_0)}(\vec{\mu}, \vec{\tau}, q, t
  ) Z_{\mathrm{vort}}^{3d,\,(\alpha_0)}(\vec{\mu}, \vec{\tau}, q, t
  ) = Z_{\mathrm{top}}(\vec{\mu},\vec{\tau}, q, t)
\end{equation}
where we have omitted an overall constant independent of $\vec{\mu}$
and $\vec{\tau}$ in $Z_{\mathrm{top}}$. The first factor in the
l.h.s.\ of Eq.~\eqref{eq:63} is the contribution of the flipping
fields (it is essentially $f(\vec{\mu},q,t)^{-1}$ up to a power
factor). 

Notice how the two sides of Eq.~\eqref{eq:63} behave in the unrefined limit $t=q$. The topological
string partition function for $t=q$ simplifies, and in particular
empty crossings become really non-interacting, so that the whole
diagram in Fig.~\ref{fig:5}~a) splits into a product of
non-interacting resolved conifold pieces, so that
$Z_{\mathrm{top}}(\vec{\mu},\vec{\tau}, q, q) = 1$. This agrees with
the behavior of the one-loop and vortex partition functions we have
observed in Eqs.~\eqref{eq:81},~\eqref{eq:61}.

\subsection{Fiber-base and spectral duality}
\label{sec:fiber-base-spectral}
Finally we discuss how the spectral self-duality of the $FT[SU(N)]$
holomorphic block appears from the geometric engineering
perspective. The CY background $\mathcal{S}$ in Fig.~\ref{fig:5} is
invariant under the action of the fiber-base duality (reflection along
the diagonal) which swaps fiber and base K\"ahler parameters or,
equivalently, exchanges $\mu_i$ with $\tau_i$. So is the
corresponding refined topological string partition function which
satisfies\footnote{Notice that in the brane web there is a so-called
  \emph{preferred direction}. When the mirror image is taken the
  direction is modified but the closed string amplitudes are invariant
  under this change.  As we discuss in the next section this can be
  understood in the algebraic approach to the vertex.  For open string
  amplitudes the situation is more subtle,
  see~\cite{Morozov:2015xya}.}:
\begin{equation}
  \label{eq:26}
Z_{\mathrm{top}}(\vec{\mu}, \vec{\tau}, q, t) =
  Z_{\mathrm{top}}(\vec{\tau}, \vec{\mu}, q, t)\,.
\end{equation}
Notice that the parameters $q$ and $t$ of the refined topological
string are left invariant by the action of the fiber-base duality.

Considering the Higgsing relation~(\ref{eq:63}) we see that
Eq.~(\ref{eq:26}) implies
 \begin{multline}
    \label{eq:22}
    \prod_{i<j} \frac{\left( \frac{q}{t}
        \frac{\mu_i}{\mu_j};q\right)_{\infty}}{\left( t
        \frac{\mu_i}{\mu_j};q\right)_{\infty} }
    Z_{1\mathrm{-loop}}^{3d,\,(\alpha_0)}(\vec{\mu}, \vec{\tau}, q, t
    ) Z_{\mathrm{vort}}^{3d,\,(\alpha_0)}(\vec{\mu}, \vec{\tau}, q, t
    ) =\\
    = \prod_{i<j} \frac{\left( \frac{q}{t}
        \frac{\tau_i}{\tau_j};q\right)_{\infty}}{\left( t
        \frac{\tau_i}{\tau_j};q\right)_{\infty} }
    Z_{1\mathrm{-loop}}^{3d,\,(\alpha_0)}(\vec{\tau}, \vec{\mu}, q, t
    ) Z_{\mathrm{vort}}^{3d,\,(\alpha_0)}(\vec{\tau}, \vec{\mu}, q, t
    )\,.
  \end{multline}
We can then easily check that the contact terms in  $f(\vec{\mu},q,t)$ satisfy the following relation
  \begin{equation}
  \label{eq:71}
  e^{-(1-2\beta)\sum_{i=1}^N (i-1) M_i} Z^{3d,\,
    (\alpha_0)}_{\textrm{cl}}(\vec{\mu}, \vec{\tau}, q, t) = e^{-(1-2\beta)\sum_{i=1}^N (i-1) T_i} Z^{3d,\,
    (\alpha_0)}_{\textrm{cl}}(\vec{\tau}, \vec{\mu}, q, t).
\end{equation}
Hence we conclude that 
\begin{multline}
  \label{eq:22} {\cal B}^{D_2\times S^1}_{FT[SU(N)]}\left(\vec{\tau},
    \vec{\mu}, q, t\right) = \prod_{i<j} \frac{\left( \frac{q}{t}
      \frac{\mu_i}{\mu_j};q\right)_{\infty}}{\left( t
      \frac{\mu_i}{\mu_j};q\right)_{\infty} }
  Z_{1\mathrm{-loop}}^{3d,\,(\alpha_0)}(\vec{\mu}, \vec{\tau}, q, t )
  Z_{\mathrm{vort}}^{3d,\,(\alpha_0)}(\vec{\mu}, \vec{\tau}, q, t
  ) =\\
  = \prod_{i<j} \frac{\left( \frac{q}{t}
      \frac{\tau_i}{\tau_j};q\right)_{\infty}}{\left( t
      \frac{\tau_i}{\tau_j};q\right)_{\infty} }
  Z_{1\mathrm{-loop}}^{3d,\,(\alpha_0)}(\vec{\tau}, \vec{\mu}, q, t )
  Z_{\mathrm{vort}}^{3d,\,(\alpha_0)}(\vec{\tau}, \vec{\mu}, q, t
  )={\cal B}^{D_2\times S^1}_{FT[SU(N)]}\left(\vec{\mu}, \vec{\tau},
    q, t\right)
\end{multline}

This is one of our main results: we have an explicit realization of
how the $3d$ spectral duality relation~\eqref{eq:20} follows from the
fiber-base self-duality of the CY background $\mathcal{S}$.

We will provide more examples of this idea in~\cite{APZ}.


\subsubsection{Symmetries of the blocks: the Ding-Iohara-Miki algebra
  approach}
\label{sec:refin-topol-strings}

In this section we briefly discuss the algebraic version of the
topological vertex formalism~\cite{Awata:2011ce} based on the
representation theory of Ding-Iohara-Miki (DIM) algebra
$U_{q,t}(\widehat{\widehat{\mathfrak{gl}}}_1)$~\cite{1996q.alg.....8002D,
  doi:10.1063/1.2823979}. This algebra is a central extension and
quantum deformation of the algebra of double loops in $\mathbb{C}$,
i.e.\ of the polynomials $x^n y^m$, $n,m \in \mathbb{Z}$. The
deformation parameters $q$ and $t$ correspond to the parameters of the
$\Omega$-background in the $5d$ gauge theory, or to the parameters of
the $\mathcal{N}=2$ deformation of the $3d$ $\mathcal{N}=4$ $T[SU(N)]$
theory. The algebra is symmetric under any permutation of the triplet
of parameters $(q,t^{-1}, \frac{t}{q})$. However, the representations
retain only part of this symmetry. The simplest representation is the
representation on the Fock space $\mathcal{F}$ with convenient choice
of basis given by Macdonald polynomials $M_Y^{(q,t)}(a_{-n}) |
\mathrm{vac}\rangle$. It (along with its tensor powers) corresponds to
the action of the algebra on the equivariant cohomology of instanton
moduli space of the $5d$ gauge theory. The representation is invariant
under the exchange of $q$ and $t^{-1}$, provided one maps the creation
operators $a_{-n}$ into $-\frac{1-q^n}{1-t^n} a_{-n}$. In particular,
in the basis of Macdonald polynomials the symmetry corresponds to the
transposition of the Young diagram $Y$:
\begin{equation}
  \label{eq:25}
  M_{Y^{\mathrm{T}}}^{(t^{-1},q^{-1})}(a_{-n}) | \mathrm{vac}\rangle =
  M_Y^{(q,t)}\left(- \frac{1-q^n}{1-t^n} a_{-n}\right) | \mathrm{vac}\rangle
\end{equation}
From physical point of view this symmetry is natural, since $q$ and
$t^{-1}$ are two equivariant parameters acting along two orthogonal
planes in the $\mathbb{R}^4_{q,t}$. 

In the algebraic construction of refined topological strings each leg
of the brane web corresponds to a Fock representation. The direction
of the leg corresponds to vector of two central charges $(k_1, k_2)$
of the DIM algebra. Thus we call Fock representations vertical or
horizontal depending on the value of the central charges.  Brane
junction corresponds to DIM algebra intertwining operator acting from
the tensor product of two Fock spaces (e.g.\ vertical and horizontal)
into a single Fock space (e.g.\ diagonal) or vice versa. Gluing of
vertices corresponds to the composition of intertwiners. Spectral
duality of the brane web corresponds to the \emph{Miki automorphism}
of the DIM algebra, which in particular takes the mirror image of the
central charge vector $(k_1, k_2) \mapsto (k_2, -k_1) $. Mirror image
of charge vectors implies mirror image of all the brane web. Miki
automorphism does not change $q$ and $t$ parameters. Thus, we conclude
that partition function of refined topological string corresponding to
the brane web in Fig.~\ref{fig:5}, a) is invariant under the 
symmetry (\ref{eq:26}).\footnote{There is a subtle part in this argument, because the
  definition of the intertwiner requires the choice of a coproduct in
  the algebra. It turns out that this choice amounts to the choice of
  a direction in the brane web --- the so-called \emph{preferred
    direction}. When the mirror image is taken the direction is
  modified. However, different directions are related by a Drinfeld
  twist and give the same answer for all closed string amplitudes, in
  particular for the partition function.}


When composing two intertwiners (or gluing two vertices in the brane
web) we need to perform the sum over intermediate states belonging to
the Fock representation, i.e.\ over all Young diagrams $Y$. However,
for the specific choice of spectral parameters corresponding to the
higgsed theory, only a subset of diagrams yields nonzero matrix
elements. In the setup shown in Fig.~\ref{fig:5}. Those are diagrams
with at most one column, i.e.\ $Y=[k]$. The sum over these diagrams
corresponds to the sum over $k_i^{(a)}$ in the vortex partition
function~\eqref{eq:15}. The \emph{subspace} of the Fock representation
$\mathcal{F}$ retains \emph{larger} symmetry of the original DIM
algebra. In particular it turns out that, besides the standard $q
\leftrightarrow t^{-1}$ symmetry, the symmetry $t \leftrightarrow
\frac{q}{t}$ is also secretly preserved in the partition function. A
simple example of such situation occurs in the basis of Macdonald
polynomials. The polynomials corresponding to totally antisymmetric
reps do not depend on $q$ and $t$, so they do not feel the exchange of
$t$ and $\frac{q}{t}$. We plan to return to this point in the future.

\section{Duality web III}
\label{secweb4}
In this section we study the Duality web III depicted in
Fig.~\ref{fig3}.

\subsection{$2d$ GLSM,  Hori-Vafa dual and Toda blocks}
 On the gauge theory side (face~{\color{green!60!black}\textsf{3}}) we consider the limit where we shrink the $S^1$
circle and reduce the $FT[SU(N)]$ theory from $D_2\times S^1$ to the
cigar $D_2$. This corresponds to taking $q=e^\hbar \to 1$ since $\hbar
= R \epsilon$ where $R$ is the circle radius and $\epsilon$ is the
equivariant parameter rotating the cigar which we keep fixed (and
indeed can set its numerical value to one).

As we have already mentioned in the Introduction there are various
ways to take the $2d$ limit, here we consider the limit which is the
ordinary dimensional reduction of the $3d$ $FT[SU(N)]$ theory down to
the theory with the same matter content in $2d$. This limit is called
the Higgs limit in~\cite{Aharony:2017adm, Aganagic:2001uw}, since the
$3d$ FI parameters are large and lift the Coulomb branch while the
matter fields remain light.

In our conventions (where the $3d$ real mass parameters are
dimensionless as they have already been rescaled by $R$), this limit
is implemented by taking $T_a$ finite as $\hbar \to 0$ and
\begin{equation}
  \mu_j=e^{M_j}\equiv e^{\hbar f_j}=q^{f_j},\quad t=q^\beta\,.
\label{3d2dpara}
\end{equation}
We identify $f_j$ and $\beta$ as the (dimensionless) twisted mass
parameters for the $SU(N)\times U(1)_A$ symmetries. We will keep all
these deformations finite to ensure that the theory has N isolated
massive vacua.
 
When we take the limit on the partition functions we also have to consider possible rescaling of the integration variable which can single out the contribution for vacua located at infinite distances.
In the Higgs limit case the vacua remain at finite distances which corresponds to taking:
\begin{equation}
x^{(a)}_i=e^{X^{(a)}_i}=e^{\hbar w^{(a)}_i}=q^{w^{(a)}_i}\,.
\label{integr:scaling}
\end{equation}
With this parameterisation, using the following limit discussed in the Appendix \ref{appendix:limits}
\begin{equation}
\lim\limits_{q\to 1}\frac{(q^x;q)_\infty}{(q;q)_\infty}=\left( -\hbar \right)^{1-x}\frac{1}{\Gamma(x)}\,,
\label{gamma:limit}
\end{equation}
we can  take the $q\to 1$ limit of the block ${\cal B}^{D_2\times S^1}_{\mathrm{FT[SU(N)]}}$  and find:
\begin{multline}
\lim_{q\to 1}{\cal B}^{D_2\times S^1}_{\mathrm{FT[SU(N)]}}=\hbar^{\frac{N(N-1)(\beta-1)}{2}}
(-1)^{\frac{1}{2}N(N-1)\beta}
\prod\limits_{a=1}^{N-1}e^{(1-\beta)a^2 \left(T_{a+1}-T_a\right)}
 \prod_{k < l}^N \frac{\Gamma(f_k-f_l+\beta)}{\Gamma(f_k-f_l+1-\beta)}
\times\\
\int \prod\limits_{a=1}^{N-1}\prod\limits_{i=1}^{a}dw_i^{(a)}\prod\limits_{a=1}^{N-1}
\prod\limits_{i=1}^{a}e^{w_i^{(a)} \left(T_{a}-T_{a+1}\right)}\prod\limits_{a=1}^{N-1}
\frac{\prod\limits_{i, j}^{a}\Gamma\left(\beta+w_i^{(a)}-w_j^{(a)}\right)}
{\prod\limits_{i\neq j}^{a}\Gamma\left( w_i^{(a)}-w_j^{(a)}\right)}\,
\times\\
\times\prod\limits_{a=1}^{N-2}\prod\limits_{i=1}^{a}\prod\limits_{j=1}^{a+1}
\frac{\Gamma\left(w_j^{(a+1)}-w_i^{(a)}\right)}{\Gamma\left( \beta+w_j^{(a+1)}-w_i^{(a)}\right)}\,
\prod\limits_{p=1}^{N}\prod\limits_{i=1}^{N-1}
\frac{\Gamma\left(f_p-w_i^{(N-1)}\right)}
{\Gamma\left(\beta+f_p-w_i^{(N-1)}\right)} = \\
\sim \hbar^{\frac{N(N-1)(\beta-1)}{2}}
{\cal B}^{D_2}_{\mathrm{FT[SU(N)]}}\,.
\label{l1TSUN:2d}
\end{multline}

%

%

The divergent $\hbar$ prefactor in the above expressions is the leading contribution to the saddle point and we will have to match it to analogue divergence arising from the limit of the dual block. 
Then we identified  up to a contact term the $D_2$ partition functions  ${\cal B}^{D_2}_{\mathrm{FT[SU(N)]}}$ of the ${\cal N}=(2,2)$  $FT[SU(N)]$ theory which can be written down following \cite{Hori:2013ika,Honda:2013uca}.
The chiral multiplets contributions to the partition function are now given by Gamma functions which sit in the numerator or in the denominator depending whether they  correspond to  Neumann or Dirichlet boundary conditions as in the $3d$ case. Our symmetric choice of the boundary condition for the  chiral multiplets corresponds to a particular boundary condition.

On the spectral dual side, where the FI and mass parameters are swapped the limit we have just described acts very differently and
it corresponds to the so called Coulomb limit. Indeed now the chirals are massive and the Higgs branch is lifted. 
As before however we keep all the deformations parameters non zero so that the $2d$ theory still has isolated vacua.
This time however the vacua are at infinity. In our convention this means that the $3d$ Coulomb brach parameters
$x^{(a)}_i=e^{X^{(a)}_i}$ stay finite as $\hbar \to 0$.

In this case we will use the following property of $q$-Pochhammer symbols 
\begin{equation}
\lim\limits_{q\to 1}\frac{(q^a x;q)_\infty}{(q^b x;q)_\infty}=(1-x)^{b-a}\,,
\label{limit:qPoc}
\end{equation}
which is proven in the Appendix \ref{appendix:limits} and find that:


 \begin{multline}
\lim_{q\to 1} \hat{\cal B}^{D_2\times S^1}_{FT[SU(N)]}
=(-\hbar)^{\frac{N(N-1)}{2} (\beta-1)}  \Gamma(\beta)^{\frac{N(N-1)}{2}}
 e^{(2\beta-1)\sum_{a=1}^N (a-1) T_a}     \prod_{k < l}^N (1-\tfrac{\tau_k}{\tau_l})^{ {2\beta-1} } \times\\
\times\int\prod\limits_{a=1}^{N-1}\prod\limits_{i=1}^{a} \frac{d x^{(a)}_i}{x_i^{(a)}}
\prod\limits_{a=1}^{N-1}\prod\limits_{i=1}^{a}\left( x_i^{(a)} \right)^{f^{(a)}-f^{(a+1)}-\beta}
\prod\limits_{a=1}^{N-1}\prod\limits_{i\neq j}^{a}\left( 1-\frac{x_j^{(a)}}{x_i^{(a)}}\right)^{\beta}
\times\\
\times\prod_{a=1}^{N-2}\prod\limits_{i=1}^{a}\prod\limits_{j=1}^{a+1}\left( 1-\frac{x_j^{(a+1)}}{x_i^{(a)}}\right)^{-\beta}
\prod\limits_{p=1}^N\prod\limits_{i=1}^{N-1}
\left( 1-\frac{\tau_p}{x_i^{(N-1)}}\right)^{-\beta}\sim\\ 
\sim  \hbar^{\frac{N(N-1)(\beta-1)}{2}}  \hat{DF}^{A_{N-1}}_{N+2}\,.
\label{2d:DF}
\end{multline}

We notice first of all that the divergent prefactor in the above expression matches the one we found by taking the limit on the spectral dual
side, which guarantees  that we are comparing the right set of vacua on both sides of the duality.

In the last equality we identified the integral block $\hat{\mathrm{DF}}^{A_{N-1}}_{N+2}$ of $\left( N+2 \right)$ vertex operators in $A_{N-1}$ Toda  CFT with screening charges $N_a=a$ and the following identification of parameters:
\begin{equation}
	\hat{\vec{\alpha}}^{(p)}=\beta\vec{\omega}_{N-1}\,,
	\quad \hat{\alpha}_{a}^{(0)}=f^{(a)}+(1-\beta)a\,,\quad \hat{z}_p=\tau_p \,
,\quad \hat{\beta}=\beta\,.
\label{2d:corr:dict}
\end{equation}
As before we put $\sim$ in (\ref{2d:DF}) because we omitted the
overall $z_p$ dependent factor in the Toda conformal block~(\ref{correlator1:toda})\footnote{The prefactor $\prod_{k < l}^N
  \left(1-\frac{\tau_k}{\tau_l}\right)^{ {2\beta-1} }$ in
  Eq.~(\ref{2d:DF}) has a different power of $\beta$ compared to the
  contribution we would get from the normal ordering of vertices from
  Eq.~(\ref{vertex:operator:toda}). This is due to the fact that, as
  we have mentioned earlier, our deformed vertices naturally
  incorporate the contribution of the central (also called the $U(1)$)
  part, whereas the conventional undeformed Toda vertices do not.}.  

%
%
Thus we have obtained the red diagonal link in the web \ref{fig3} which relates the $2d$ gauge theory to the CFT block.
Notice that the map (\ref{2d:corr:dict}) between the parameters of the gauge theory and Toda block is consistent 
with the $\hbar\to 0$ limit of the  previously derived gauge/CFT correspondence map (see Table \ref{table:map}) after
the spectral duality transformation $\tau_p\leftrightarrow \mu_p$, $\beta\rightarrow \beta$.

To make contact with the Hori-Vafa dual theory of twisted chiral
multiplets which we expect to find on the bottom right corner of
face~{\color{green!60!black}\textsf{3}} in Fig.~\ref{fig3} we simply
need to exponentiate the integrand $\mathcal{I}$ in eq. (\ref{2d:DF})
as $\mathcal{I}=\exp(\log\mathcal{I})$ and identify
$\log(\mathcal{I})$ with $\mathcal{W}$ the twisted superpotential
contribution to the $D_2$ partition function of the Hori-Vafa dual
theory. The dual theory also has $N(N-1)/2$ un-gauged chiral
multiplets which yield the $\Gamma(\beta)$ factors.

Notice that we keep all the FI and the twisted mass deformations
on. This is necessary for the convergence of the integrals (and to
relate them to CFT) so the match of the $D_2$ partition functions is a
check of the duality for the mass deformed theories with isolated
vacua.  As recently discussed in
\cite{Aharony:2016jki,Aharony:2017adm} it is quite subtle to
understand what happens when these deformations are lifted and
generically we are not guaranteed to find a proper IR duality for
massless theories.


\subsection{$2d$ GLSM and the $d$-Virasoro algebra}
\label{sec:2d-glsm-d}

Finally the remaining corner of face~{\color{green!60!black}\textsf{2}}, labelled
$d$DF$^{A_{N-1}}_{N+2}$ is to be interpreted as a conformal block
 of  an \emph{unconventional limit} of the $q$-$W_N$ algebra.
 Here we briefly sketch the  construction of this theory restricting ourselves to the case $N=2$.
We then start from the $q$-Virasoro algebra which is generated by the
current $T(z)$ satisfying the quadratic relation:
\begin{equation}
  \label{zeq:1}
  f \left(\frac{w}{z}\right) T(z) T(w) - f \left( \frac{z}{w} \right)
  T(w) T(z) = - \frac{(1-q)(1-t^{-1})}{1- \frac{q}{t}} \left( \delta_{\times}
    \left( \frac{q}{t} \frac{w}{z} \right) - \delta_{\times}
    \left( \frac{t}{q} \frac{w}{z} \right) \right)
\end{equation}
where
\begin{equation}
  \label{zeq:3}
  f(x) = \exp \left[ \sum_{n \geq 1} \frac{(1-q^n)(1-t^{-n})}{1 +
      \left( \frac{q}{t} \right)^n} \frac{x^n}{n} \right]=
  \frac{ \left( q x; \frac{q^2}{t^2}
    \right)_{\infty} \left( t^{-1} x;\frac{q^2}{t^2}
    \right)_{\infty}}{(1 - x) \left( \frac{q^2}{t} x;\frac{q^2}{t^2}
    \right)_{\infty} \left( \frac{q}{t^2} x;\frac{q^2}{t^2} \right)_{\infty}}
\end{equation}
and  $\delta_{\times} (x) = \sum_{n \in \mathbb{Z}} x^n$ is the
multiplicative delta-function. One can understand $\delta_{\times} (x)$ as the delta function on the unit circle, where $x = e^{i \phi}$, since
\[
\sum_{n \in \mathbb{Z}} e^{i n \phi} = \sum_{m \in \mathbb{Z}} \delta_+ (\phi - 2\pi m),
\]
and $\delta_+(u)$ is the standard (additive) Dirac delta delta-function.

The $q$-Virasoro algebra in the {\it familiar} limit
 $q = e^{\hbar} \to 1$ and $t=q^{\beta}$ with fixed $\beta$ reduces to the Virasoro algebra.
 This can be seen by taking the above limit in  eq. (\ref{zeq:1}) 
keeping fixed also the positions of the currents $z$ and $w$. In this case one recovers the 
quadratic relation for the  ordinary Virasoro algebra. The current $T(z)$ also
 reduces to the Virasoro current $L(z)$:
  \begin{equation}
    \label{zeq:2}
    T(z) = 2 + \beta \hbar^2 \left( z^2 L(z) + \frac{1}{4} \left(
        \sqrt{\beta} - \frac{1}{\sqrt{\beta}} \right)^2 \right) + \ldots
  \end{equation}

We can also take an {\it unconventional} limit of the quadratic relation (\ref{zeq:1}) 
where positions $z$ and $w$ scale as powers of $q$:
  \begin{equation}
    \label{zeq:4}
    z = q^u,\qquad w = q^v,
  \end{equation}
  and the current $t(u) = \lim_{q \to 1} T(q^u)$ remains finite 
  then the relations of the algebra become
  \begin{equation}
    \label{zeq:5}
    g(v-u) t(u) t(v) - g(u-v) t(v) t(u) = - \frac{\beta}{1-\beta}
    \left( \delta_{+} (v-u+1 -\beta) - \delta_{+} (v-u-1 +\beta) \right)
  \end{equation}
  where  the structure
  function becomes
  \begin{equation}
    \label{zeq:6}
    g(u) = \frac{2(1-\beta)}{u} \frac{\Gamma \left(
        \frac{u+2-\beta}{2(1-\beta)} \right) \Gamma \left(
        \frac{u+1-2\beta}{2(1-\beta)} \right)}{\Gamma \left(
        \frac{u+1}{2(1-\beta)} \right) \Gamma \left( \frac{u - \beta}{2(1-\beta)} \right)}.
  \end{equation}
  The main effect of the limit is that the $q$-Pochhammer symbols in
  the definition of the $q$-Virasoro structure function (\ref{zeq:3}) turn into Euler gamma
  functions.  Essentially this algebra, which we will call
  $d$-Virasoro algebra, is the \emph{additive analogue} of the
  $q$-Virasoro algebra.

We claim that conformal blocks of the $d$-Virasoro algebra have the DF
representation which coincides with the GLSM localization
integrals. Moreover, these blocks are spectral dual to the ordinary
CFT conformal blocks, so that the positions of the vertex operators in
$d$-Virasoro become momenta in the dual CFT and vice versa.

The algebra~(\ref{zeq:5}) can be bosonized as follows. We express the current as:
\begin{equation}
  \label{zeq:7}
  t(u) = \Lambda_1(u) + \Lambda_2(u),
\end{equation}
where
\begin{multline}
  \label{eq:8}
  \Lambda_1(u) = e^{\tilde{Q}} \left(g(u) \right)^{\tilde{P}}
  \normord{\exp \Bigl[ \sum_{n \geq 1} \frac{1}{n} c_{-n} \left( u^n -
      (u-\beta)^n
    \right) -\\
    - \sum_{k \geq 0} \sum_{n \geq 1} \frac{1}{n} c_n (-1)^k \left(
      (u-(1-\beta)k)^{-n} - (u - 1 - (1-\beta)k)^{-n} \right)\Bigr]},
\end{multline}
\begin{multline}
  \label{eq:9}
  \Lambda_2(u) = e^{-\tilde{Q}} \left(g(u) \right)^{-\tilde{P}}
  \normord{\exp \Bigl[ - \sum_{n \geq 1} \frac{1}{n} c_{-n} \left( (u
      + \beta - 1)^n - (u - 1)^n
    \right) +\\
    + \sum_{k \geq 0} \sum_{n \geq 1} \frac{1}{n} c_n (-1)^k \left\{
      (u + \beta - 1 - (1-\beta)k)^{-n} - (u + \beta - 2 -
      (1-\beta)k)^{-n} \right\}\Bigr]}
\end{multline}
and the generators $c_n$, $\tilde{Q}$ and $\tilde{P}$ obey the
Heisenberg algebra. Notice that the sums over $k$ in the exponentials
converge for generic $u$.

The screening current commuting with the generator $t(u)$ up to total
difference is given by
\begin{multline}
  \label{zeq:10}
  s(x) = e^{\tilde{Q}+Q} \left( \frac{\Gamma(\beta-u)
      \Gamma(1-u)}{\Gamma(-u)\Gamma(1-\beta-u)} \right)^{\tilde{P}}
  e^{\beta x P} \normord{\exp \Bigl[ - \sum_{n\geq 1} \frac{u^n}{n}
    c_{-n}+\\ + \sum_{n\geq 1} \sum_{k \geq 0} \frac{1}{n} c_n \left\{
      (u-k)^{-n} - (u-\beta-k)^{-n} + (u+\beta-1-k)^{-n} -
      (u-1-k)^{-n} \right\} \Bigr]}
\end{multline}
where we have introduced an additional pair of zero modes $P$ and $Q$,
which commute with the Heisenberg algebra formed by $\tilde{P}$,
$\tilde{Q}$ and $c_n$. 

We can immediately check that the normal ordering of the screening currents
correctly reproduces the gamma function measure of the GLSM integral
measure:\footnote{The ratio of sines in the second line of Eq.~(\ref{eq:11}) is a
periodic function with period $1$ and will factor out of the
integral block. This happens for the same reason as in the $q$-deformed
case: the residues of the integrand which is a product of gamma
functions appear in strings with period $1$.}
\begin{multline}
  \label{eq:11}
  \prod_{i=1}^n s(w_i) = \normord{\prod_{i=1}^n s(w_i)} \prod_{i < j}
  \frac{\Gamma(w_j - w_i + \beta) \Gamma(w_j - w_i + 1)}{\Gamma(w_j -
    w_i) \Gamma(w_j - w_i + 1 - \beta)} =\\
  = \normord{\prod_{i=1}^n s(w_i)} \prod_{i<j} \frac{\sin (\pi (w_i -
    w_j + \beta))}{\sin (\pi (w_i - w_j))} \prod_{i \neq j}
  \frac{\Gamma(w_j - w_i + \beta)}{\Gamma(w_j - w_i)}\,.
\end{multline}
We then introduce  vertex operators:
\begin{multline}
  \label{eq:12}
  v_{\alpha}(x) = e^{\alpha \tilde{Q} + \alpha Q} \left(
    \frac{\Gamma(-\alpha-x)}{\Gamma(-x)}
  \right)^{\tilde{P}} e^{ \alpha x P} \times\\
  \times \normord{\exp \left[ \sum_{n\geq 1} \frac{x^n}{n} c_{-n} -
      \sum_{n\geq 1} \sum_{k \geq 0} \frac{1}{n} c_n \left\{
        (x-k)^{-n} - (x + \alpha -k)^{-n}\right\} \right]}\,,
\end{multline}
and assume that the initial state $|\alpha_{(0)}\rangle$ of the
conformal block is annihilated by $\tilde{P}$ and is the eigenfunction
of $P$:
\begin{equation}
  \label{eq:14}
  \tilde{P} |\alpha_{(0)}\rangle = 0,\qquad   P |\alpha_{(0)}\rangle = \alpha_{(0)}|\alpha_{(0)}\rangle.
\end{equation}
We  can then combine all the pieces and calculate our $d$-DF integral for $(n+2)$-point conformal block which as expected reproduces the 
$FT[SU(N)]$ partition function Eq.~\eqref{l1TSUN:2d}:
\begin{multline}
  \label{eq:13}
  d\mathrm{DF}_{n+2}^{A_1}(u_1,\ldots ,u_n, \alpha_{(0)},
  \alpha_{(1)}, \ldots, \alpha_{(n)},N, \beta)
  \stackrel{\mathrm{def}}{=}\\
  \stackrel{\mathrm{def}}{=} \langle \alpha_{(\infty)} |
  v_{\alpha_{(1)}} (u_1) \cdots v_{\alpha_{(n)}} (u_n)
  \left( \oint dw\, s(w) \right)^N |\alpha_{(0)} \rangle \sim\\
  \sim \int d^N w\, e^{(\alpha_{(0)} + N) \sum_{i=1}^{N} w_i}
  \prod_{i\neq j} \frac{\Gamma(w_i - w_j + \beta)}{\Gamma(w_i - w_j)}
  \prod_{i=1}^N \prod_{p=1}^n \frac{\Gamma(w_i - u_p)}{\Gamma(w_i -
    u_p - \alpha_{(p)})}.
\end{multline}
The Duality web in (face~{\color{green!60!black}\textsf{4}}) Fig.~\ref{fig3} indicates
that the $d$-DF blocks are dual to the  DF block of the ordinary $W_n$ algebra.
This is a consequence of the spectral duality for deformed Toda correlators. In particular 
in the $N=2$ case we have a duality between the  four-point $d$-Virasoro block and the 4-points  ordinary Virasoro block.  
Notice that while the evaluation of the  DF blocks is quite intricate (even in the {\it simple} cases involving vertices with degenerate momenta) the evaluation  $d$-DF blocks can be performed quite easily on contours encircling  the poles of the $\Gamma$ functions.
One can than regard the map of ordinary DF blocks to  $d$-DF blocks or to GLSM partition functions as an efficient computational strategy.
We will continue this discussion in ~\cite{vertex:future}.

\section{Conclusions and Outlook}

In this work we have studied several webs of dualities for the
$FT[SU(N)]$ quiver theory: the spectral duality, the $q$-deformed
Dotsenko-Fateev representation and its realisation via Higgsing.  We
have proven these dualities and correspondences focusing on the
$D_2\times S^1$ partition function $\mathcal{B}^{D_2\times
  S^1}_{FT[SU(N)]}$.

The main results of our paper are:
\begin{enumerate}
\item derivation of the $3d$ spectral duality for the $FT[SU(N)]$
  theory from fiber-base duality of $5d$ gauge theories,

\item identification of the gauge/Toda
  correspondence~\cite{Gomis:2014eya,Gomis:2016ljm,Doroud:2012xw}
  between the ${\cal N}=(2,2)$ $FT[SU(N)]$ theory and the
  Dotsenko-Fateev block with $(N+2)$ vertex operators in $A_{N-1}$
  Toda CFT as a limit of our $3d$ spectral duality.
\end{enumerate}

Our results open up a vista full of possible directions for future
research. Below we propose a number of projects which can provide
better understanding and further expand our conjectures.

First of all in our paper we have focused on the $FT[SU(N)]$ theory,
however via Higgsing, we can generate infinitely many $3d$ spectral
dual pairs (some examples will be given in~\cite{APZ}). For each of
them we could construct duality webs similar to those considered in
above. In particular, by taking the $q\to 1$ limits we should obtain
pairs of $2d$ GLSMs and Dotsenko-Fateev blocks related by the standard
GLSM/CFT correspondence.

%
%

The duality web III shown in Fig. \ref{fig3} has two more corners
which we have not discussed much in our paper.  One corner contains
partition function ${\cal B}_{LG}^{D^2}$ of the Landau-Ginzburg theory
on $D_2$. According to the logic of the duality web III it should be
connected to the DF integral in Toda CFT by simple identification of
the parameters. However at the moment this interesting connection
between two seemingly distinct objects seems not completely obvious.

Another corner of the web contains what we called
$d\mathrm{DF}^{A_{N-1}}_{N+2}$ integrals. These integrals correspond
to conformal blocks with $(N+2)$ primary vertex operators of the
$d$-$W_N$ algebra. In Sec.\ref{secweb4} we have described this algebra
for the case of $N=2$, wrote down its bosonization and conformal
blocks. It would be interesting to generalize this construction to the
case of $d$-$W_N$ algebra with general $N$ and study its properties
and possible relations to integrable models. We plan to do it
in~\cite{vertex:future}.

\section*{Acknowledgements}
We are very grateful to Francesco Aprile, Sergio Benvenuti, Johanna
Knapp, Noppadol Mekareeya, Andrey Mironov, Alexey Morozov, Fabrizio Nieri, Matteo
Sacchi, Alessandro Torrielli and Alberto Zaffaroni, for enlightening
discussions.  S.P. and Y.Z. are partially supported by the ERC-STG
grant 637844-HBQFTNCER and by the INFN. Y.Z. is partially supported by the 
grants  RFBR  16-02-01021, RFBR-India 16-51-45029-Ind, RFBR-Taiwan 15-51-52031-NSC,
RFBR-China 16-51-53034-GFEN, RFBR-Japan 17-51-50051-YaF.
\appendix
\begin{appendix}

\section{Partition function on $D_2\times S^1$}
\label{appendix:block}

In this appendix we quickly record the steps to obtain the holomorphic block integral for the $T[SU(N)]$ theory from the factorisation of the  $S_b^3$ partition function. For  details and notation we refer the reader to \cite{Beem:2012mb,Pasquetti:2016dyl}.
The key point is the following chain of relations relating the partition function on a compact three-manifold which can be obtained gluing to solid tori $D_2\times S^1$ with some $SL(2,Z)$ element,  which in the squashed three sphere case   $S_b^3$  is the element $S$ and the $3d$ holomorphic blocks:
\begin{equation}
Z_{S^3_b}=\int \left|\left|\Upsilon\right|\right|^2_S=\sum_\alpha \left|\left|\int_{\Gamma_\alpha}\Upsilon\right|\right|^2_S=\sum_\alpha \left|\left|\mathcal{B}^{D_2\times S^1}_\alpha\right|\right|^2_S\,.
\end{equation}
This very non-trivial chain of identities provides us with a practical way to obtain the block integrand $\Upsilon$ by factorising the integrand of the $S_b^3$  partition function which consists of the classical contribution of the mixed Chern-Simons couplings and the one-loop contribution of the vector and chiral multiplets:
\begin{equation}
Z_{S^3_b}=\int Z_{CS} Z_{vec} Z_{matter}\,.
\end{equation}
The factorisation of the $S_b^3$ integrand follows from the fact that the vector and matter contributions are expressed in terms of the double sine function $S_2(X)$ which can be factorised as:
\begin{equation}
S_2(X)=e^{\frac{i\pi}{2}B_{22}(X)}\left( e^{2\pi i b X};\,e^{2\pi i b^2}\right)_\infty\,
\left( e^{2\pi i b^{-1} X};\, e^{2\pi i b^{-2}}\right)_\infty\\\equiv
e^{\frac{i\pi}{2}B_{22}(X)}\left|\left|\left( e^{2\pi i b X};\,e^{2\pi i b^2} \right)_\infty \right|\right|_S^2\,,
\label{double:sine:def}
\end{equation}
where 
$B_{22}(X)$ stands for the quadratic Bernoulli polynomial 
\begin{equation}
B_{22}(X)=\left( X-\frac{\omega}{2}\right)^2-\frac{1}{12}\left( b^2+b^{-2} \right)\,,
\label{bernoulli}
\end{equation}
where $\omega=b+b^{-1}$. Using this property we can factorise the one-loop contributions to the $T[SU(N)]$ partition function on $S_b^3$:

\begin{enumerate}
\item \textbf{Bifundamental hypermultiplet} of mass $\tilde{m}$
  conneting nodes $a$ and $b$:
\begin{multline}
  Z_{\mathrm{bifund}}^{(a,b)}\left[ S_b^3
  \right]=\prod_{\pm}\prod_{i=1}^{N_a}\prod_{j=1}^{N_b} S_2^{-1}\left(
    \frac{\omega}{4}-i\frac{\tilde{m}}{2}
    \pm i\left(\tilde{X}_i^{(a)}-\tilde{X}_j^{(b)} \right)\right)=\\
  =\prod_{i=1}^{N_a}\prod_{j=1}^{N_b}\frac{S_2\left(
      \frac{3\omega}{4}+i\frac{\tm}{2}+i\left(\tX_i^{(a)}-\tX_j^{(b)}\right)\right)}
  {S_2\left( \frac{\omega}{4}-\frac{i\tm}{2}+i\left(
        \tX_i^{(a)}-\tX_j^{(b)} \right)\right)}\,,
\end{multline}
and using the factorization formula~(\ref{double:sine:def}) can be expressed as:
\begin{multline}
Z_{\mathrm{bifund}}^{(a,b)}\left[ S_b^3 \right]=\prod_{i=1}^{N_a}\prod_{j=1}^{N_b}e^{\frac{i \pi}{2}\left[B_{22}\left( \frac{3\omega}{4}+\frac{i\tm}{2}+i\left( \tX_i^{(a)}-\tX_j^{(b)} \right)\right) 
-B_{22}\left( \frac{\omega}{4}-\frac{i\tm}{2}+i\left( \tX_i^{(a)}-\tX_j^{(b)} \right)\right) \right]}\times \\
\times \left|\left|\frac{\left( e^{2\pi ib\left[ \frac{3\omega}{4}+\frac{i\tm}{2}+i\left(\tX_i^{(a)}-\tX_j^{(b)}  \right)\right]};\,q \right)_\infty}
{\left( e^{2\pi ib\left[ \frac{\omega}{4}-\frac{i\tm}{2}+i\left(\tX_i^{(a)}-\tX_j^{(b)}
\right)\right]};\,q \right)_\infty} \right|\right|_S^2
\end{multline}

This general expression can be significantly simplified in for the
matter content of $T[SU(N)]$ theory.  In this case only two adjacent
nodes are connected with the bifundamental hypermultiplet so that we
should take $b=a+1$ in the expression above. Also we fix the ranks of
the gauge groups in the following form $N_a=a$.  Then we can write the
contribution of all bifundamental hypers in $T[SU(N)]$ quiver in the
following form:
\begin{multline}
Z_{\mathrm{bifund}}\left[ S_b^3 \right]=\prod_{a=1}^{N-2}Z_{bifund}^{(a,a+1)}\left[ S_b^3 \right]=e^{-\frac{\hbar}{12}N(N-1)(N-2)\beta(1-\beta)}
\prod\limits_{j=1}^{N-1}t^{-(N-2)X_j^{(N-1)}/2\hbar}\times\\
\times \prod\limits_{a=1}^{N-2}\prod\limits_{i=1}^a t^{X_i^{(a)}/\hbar}
\prod\limits_{a=1}^{N-1}\prod\limits_{i,j=1}^{a}\left|\left|\frac{\left( tx_j^{(a+1)}/x_i^{(a)};\,q \right)_\infty}
{\left(x_j^{(a+1)}/x_i^{(a)};\,q  \right)_\infty}\right|\right|^2_S\,,
\label{bifundamental}
\end{multline}
where we made the following identification with the holomorphic block variables:
\begin{equation}
e^{2\pi i b^2}\equiv q=e^\hbar\,, \quad e^{2\pi b \tX_i^{(a)}}e^{2\pi iba\left( \frac{\omega}{4}-i\frac{\tm}{2}\right)}\equiv e^{X_i^{(a)}}=x_i^{(a)}\,,\quad
e^{2\pi i b\left( \frac{\omega}{2}+i\tm \right)}\equiv-q^{1/2}e^{-m'}=t\,.\nonumber\\
\label{parametrizations}
\end{equation}
So we identify $2\pi b \tilde m\equiv m'$ where $\tilde m$ are the dimensionless real masses parameters (use $S^3_b$ radius) entering in the $S^3_b$ partition function while $m'=R m^{3d}$ is the dimensionless axial mass appearing in the holomorphic block. And the $S^3_b$ and $D_2\times S^1$ Coulomb branch variables $2\pi b \tX_i^{(a)}\equiv {X'}_i^{(a)}$ which is then further shifted to $X_i^{(a)}$.

\item \textbf{$N_f$ hypers of masses $\tM_p$ ($p=1,\dots,N_f^{(a)}$) connected to the $U(N-1)$ node.}
Factorising the double sine  as in the previous case and expressing the result in terms of the shifted exponentiated variables  we find:
\begin{equation}
  Z_{\mathrm{fund}}\left[ S_b^3 \right]=e^{-\frac{\hbar}{4}N(N^2-1)\beta(1-\beta)}
  \prod\limits_{i=1}^{N-1}t^{N X_i^{(N-1)}/2\hbar}
  \prod\limits_{i=1}^{N-1}\prod\limits_{p=1}^{N} \left|\left|
  \frac{\left( t\mu_p/x_i^{(N-1)};\,q \right)_\infty}{\left(\mu_p/x_i^{(N-1)};\,q  \right)_\infty}\right|\right|^2_S\,,
\label{fundamental}
\end{equation}
where we also introduced 
\begin{equation}
\mu_p=e^{2\pi b \tM_p}e^{2\pi ibN\left( \frac{\omega}{4}-i\frac{\tm}{2}\right)}=e^{M_p^{(a)}}\,.
\label{mass:parametrization}
\end{equation}
again we have the identification $2\pi b \tM_p=M'_p$ between the dimensionless mass parameters $\tilde M_p$ on $S^3_b$ and the dimensionless mass parameters $M'_p=RM^{3d}_p$ on $D_2\times S^1$.

\item \textbf{Vector+adjoint multiplet of mass $\tm$ at node $a$.} 
Finally the contribution of vector and adjoint 
hypers is given by:
\begin{equation}
Z_{\mathrm{vec}+ \mathrm{adj}}\left[ S_b^3 \right]=e^{\frac{\hbar}{4}N(N-1)\beta(1-\beta)}e^{-\frac{{m'}^2}{4\hbar}N}
\left|\left|\prod_{a=1}^{N-1}
\frac{\prod\limits_{i\neq j}^{N_a}\left( x_j^{(a)}/x_i^{(a)};\,q \right)_\infty}
{\prod\limits_{i,j}^{N_a}\left( tx_j^{(a)}/x_i^{(a)};\,q \right)_\infty}\right|\right|_S^2\,.
\label{vector:contribution}
\end{equation}

\item \textbf{Mixed Chern-Simon terms.} Finally we need to discuss the contribution of the mixed Chern-Simon terms.
In the $T[SU(N)]$ theory we have turned on real masses $T_{a+1}-T_a$ for the topological symmetry of the $a$-th gauge node. These mixed Chern-Simons terms contribute to partition function as: 
\begin{equation}
  Z_{\mathrm{FI}}^{(a)}\left[ S_b^3 \right]=e^{\frac{a N_a}{2}(1-\beta)\left( T_{a+1}-T_a \right)}
  \prod\limits_{i=1}^{N_a}e^{X_i^{(a)}(T_a-T_{a+1})/\hbar}\,,
\end{equation}
where the first factor comes from the change of variables from $\tX_i^{(a)}$ to $X_i^{(a)}$.

Other mixed gauge-flavor Chern-Simons coupling are induced by the factorisation of the chiral multiplets (linear in $X_i^{(a)}$) in~(\ref{bifundamental}),~(\ref{fundamental}). Finally  all the remaining exponential terms  in~(\ref{bifundamental}),~(\ref{fundamental}) and~(\ref{vector:contribution})  are mixed background Chern-Simons contributions. 

\end{enumerate}

At this point we should express these Chern-Simons contributions as squares. To do so one can use the following rewriting of the modular transformation of the Jacobi-theta function:
\begin{equation}
e^{-\tfrac{(X-(i\pi+\hbar/2) )^2}{2\hbar}}= \theta_q(x)\theta_{\tilde q}(\tilde x) =\left|\left|\theta_q(x)\right|\right|_S^2
\end{equation}
where $\theta_q(x)=(q x^{-1};q)_\infty(x;q)_\infty$. Using this identity we can convert quadratic exponential into squares of theta functions and deduce the combination of theta functions which represent the contribution of the Chern-Simons coupling to the block integral. For more details we refer the reader to \cite{Beem:2012mb}.
In \cite{Yoshida:2014ssa} the theta functions appearing in the block integrals have been shown to arise as one-loop contributions of $2d$ multiplets on the boundary torus.

\section{Calculation of free field correlators}
\label{sec:calc-free-field}
In this appendix we show how to get the DF representation of the
conformal blocks in Toda theory and its $q$-deformed version. 

\subsection{Toda conformal block}
\label{sec:toda-theory}
To calculate free field correlators of the form~(\ref{correlator:toda}) we normal
order all our expressions using the standard normal ordering identity
valid for the operators $v_i$ commuting on a $c$-number:
\begin{equation}
\prod\limits_i \normord{e^{v_i}}= \normord{\prod\limits_i e^{v_i}}\,\prod\limits_{i<j}e^{[v_i^{+},\,v_j^{-}]}.
\label{ordering:rule}
\end{equation}
Then using the Heisenberg algebra~(\ref{heisenbeg:algebra}) it is
straightforward to obtain the following relations for the normal
ordering of the screening currents~(\ref{screening:current:toda}) and
vertex operators~(\ref{vertex:operator:toda}).
\begin{enumerate}
\item \textbf{Normal ordering the screening currents from the same sector}
\begin{gather}
\prod\limits_{i=1}^{N_a} S_{(a)}\left( x_i^{(a)} \right)=C^{(a,a)}\left( x^{(a)} \right)
\normord{\prod\limits_{i=1}^{N_a} S_{(a)}\left( x_i^{(a)} \right)},\\
  C^{(a,a)}\left( x^{(a)} \right)=\prod_i \left( x_i^{(a)} \right)^{\beta(N-1)}\prod\limits_{i\neq j}
  \left( 1-\frac{x_j^{(a)}}{x_i^{(a)}}\right)^\beta.
\label{screening:order:1:toda}
\end{gather}

\item \textbf{Normal ordering the screening currents from different
    sectors}
\begin{multline}
\normord{\prod\limits_{i=1}^{N_a} S_{(a)}\left( x_i^{(a)} \right):\; :\prod\limits_{j=1}^{N_b} S_{(b)}\left( x_j^{(b)} \right)}=\\
=C^{(a,b)}\left( x^{(a)},x^{(b)} \right)
\normord{\prod\limits_{i=1}^{N_a} S_{(a)}\left( x_i^{(a)} \right)\prod\limits_{j=1}^{N_b} S_{(b)}\left( x_j^{(b)} \right)},\notag
\end{multline}
where
\begin{equation}
  C^{(a,b)}\left( x^{(a)},\,x^{(b)} \right) = (\delta_{b,a+1}+\delta_{b,a-1})
\prod\limits_{i=1}^{N_a}\left( x_i^{(a)} \right)^{-\beta\,N_{b}}
\prod\limits_{i=1}^{N_a}\prod\limits_{j=1}^{N_{b}}\left( 1-\frac{x_j^{(b)}}{x_i^{(a)}} \right)^{-\beta}.
\label{screening:order:2:toda}
\end{equation}

\item \textbf{Normal ordering the screening currents and vertex
    operators}
\begin{gather}
  V_{\vec{\alpha}}\left( z \right)\prod\limits_{i=1}^{N_a}
  S_{(a)}\left( x_{i}^{(a)} \right) =
  C^{(a)}\left(x^{(a)},z,\vec{\alpha} \right)
  \normord{V_{\vec{\alpha}}\left( z \right) \prod\limits_{i=1}^{N_a} S_{(a)}\left( x_{i}^{(a)} \right)},\notag\\
  C^{(a)}\left( x^{(a)}, z, \vec{\alpha} \right)\sim
  \prod\limits_{i=1}^{N_a}\left(
    1-\frac{x_i^{(a)}}{z}\right)^{\alpha_a-\alpha_{a+1}},
\label{vertex:ordering:toda}
\end{gather}
where we have omitted an overall constant factor.

\item \textbf{Normal ordering of different vertex operators} 
 \be
        \prod\limits_{p=1}^l V^T_{\vec{\alpha}^{(p)}}(z_p)=C_{\mathrm{vert.}}(\vec{\alpha},\,z)\, \normord{\prod\limits_{p=1}^l V^T_{\vec{\alpha}^{(p)}}(z_p)}\,,
	\ee
	where
	\be
	C_{\mathrm{vert.}}\left( \vec{\alpha},\,z \right)=\prod\limits_{p<k}^l\left( z_p-z_k\right)^{\frac{1}{\beta}\left(\vec{\alpha}^{(p)},\,\vec{\alpha}^{(k)}\right)}\,.
        \label{toda:order:prim:vertex}
   \ee

\item \textbf{Vertex operator at the origin.} To evaluate it we notice
  that after the normal ordering under the integral we get the term of
  the form
\begin{equation}
  \prod\limits_{a=1}^n\prod\limits_{i=1}^{N_a}\left( x_i^{(a)} \right)^{\sqrt{\beta}\left(P^{(a)}-P^{(a+1)}  \right)}
  |\vec{\alpha}^{(0)}\rangle = \prod\limits_{a=1}^n\prod\limits_{i=1}^{N_a}\left( x_i^{(a)} \right)^{\left(\alpha^{(0)}_{a}-
      \alpha^{(0)}_{a+1}\right)}
  |\vec{\alpha}^{(0)}\rangle.
\label{zero:modes:contrib}
\end{equation}
Another way to obtain the factor~\eqref{zero:modes:contrib} is to take
$z\to 0$ limit of Eq.~(\ref{vertex:ordering:toda}).
\end{enumerate}

Finally collecting all the factors we have derived above we find that the free field correlator
(\ref{correlator:toda}) becomes the following matrix integral:
\begin{multline}
  {\rm DF}^{A_n}_{l+2}\sim \prod\limits_{p<k}^l\left( z_p-z_k\right)^{\frac{1}{\beta}\left(\vec{\alpha}^{(p)},\,\vec{\alpha}^{(k)}\right)}\oint
  \prod\limits_{a=1}^n\prod\limits_{i=1}^{N_a} d x^{(a)}_i
  \prod\limits_{a=1}^n\prod\limits_{i=1}^{N_a}\left( x_i^{(a)}
  \right)^{\beta(N_a-N_{a+1}-1)+
    (\alpha_a^{(0)}-\alpha_{a+1}^{(0)})}\times\\
  \times \prod\limits_{a=1}^n\prod\limits_{i\neq j}^{N_a}\left(
    1-\frac{x_j^{(a)}}{x_i^{(a)}}\right)^\beta
  \prod_{a=1}^{n-1}\prod\limits_{i=1}^{N_a}\prod\limits_{j=1}^{N_{a+1}}\left(
    1-\frac{x_j^{(a+1)}}{x_i^{(a)}}\right)^{-\beta}
  \prod\limits_{p=1}^l\prod\limits_{a=1}^n\prod\limits_{i=1}^{N_a}
  \left(
    1-\frac{x_i^{(a)}}{z_p}\right)^{\alpha_a^{(p)}-\alpha_{a+1}^{(p)}}
\end{multline}
where we have omitted some of the coefficients of the conformal block, that 
stands in front of the integral. In general this coefficient depends on the
coordinates of the vertex operators insertions.

\subsection{$q$-Toda conformal block}
\label{sec:q-toad-theory}

Repeating the normal ordering calculation of the previous section for
the screening currents~(\ref{screening:current}) and vertex
operators~(\ref{vertex:operator}) we obtain the following relations:
\begin{enumerate}
\item \textbf{Normal ordering the screening currents from the same
    sector}
\begin{equation} 
  \prod\limits_{i=1}^{N_a} S^q_{(a)}\left( x_i^{(a)} \right)=\left[\prod\limits_{i<j}
    \left(\frac{x_i^{(a)}}{x_j^{(a)} } \right)^\beta \left( 1-\frac{x_j^{(a)}}{x^{(a)}_i}\right)
    \frac{\left(\frac{q}{t} \frac{x^{(a)}_j}{x_i^{(a)}};q
      \right)_\infty}{\left(t \frac{x_j^{(a)}}{x_i^{(a)}} ;q \right)_\infty}\right]
  \normord{\prod\limits_{i=1}^{N_a} S^q_{(a)}\left( x_i^{(a)} \right)}
\end{equation}
We notice that the function
\begin{equation}
\prod\limits_{i<j}\left( \frac{x_i}{x_j} \right)^{\beta}
\frac{\left( t x_i\,x_j^{-1};q \right)_\infty\,\left( qt^{-1}x_j\,x_i^{-1};q \right)_\infty}
{\left(x_i\,x_j^{-1};q \right)_\infty\,\left( q x_j\,x_i^{-1};q \right)_\infty}\,,
\end{equation}
is $q$-periodic and thus yields an overall constant in front of the
integral. We can then rewrite previous expression in a more convenient
form
\begin{gather}
\prod\limits_{i=1}^{N_a} S^q_{(a)}\left( x_i^{(a)} \right)=C_q^{(a,a)}\left( x^{(a)} \right)
\normord{\prod\limits_{i=1}^{N_a} S^q_{(a)}\left( x_i^{(a)} \right)},\\
C_q^{(a,a)}\left( x^{(a)} \right)=\prod\limits_{i=1}^{N_a} \left(x_{i}^{(a)}\right)^{\beta(N_a-1)}
\prod\limits_{i\neq j}^{N_a}\frac{\left(\frac{x_j^{(a)}}{x_i^{(a)}};q  \right)_\infty}
{\left(t \frac{x_j^{(a)}}{x_i^{(a)}};q  \right)_\infty}\,.
\label{screening:order:1}
\end{gather}

\item \textbf{Normal ordering the screening currents from different
    sectors}
\begin{equation}
\normord{\prod\limits_{i=1}^{N_a} S^q_{(a)}\left( x_i^{(a)} \right):\; :\prod\limits_{j=1}^{N_b} S^q_{(b)}\left( x_j^{(b)} \right)}=
C_q^{(a,b)}\left( x^{(a)}, x^{(b)} \right)
:\prod\limits_{i=1}^{N_a} S^q_{(a)}\left( x_i^{(a)}
\right)\prod\limits_{j=1}^{N_b} S^q_{(b)}\left( x_j^{(b)} \right):\,,
\end{equation}
where
\begin{equation}
  C_q^{(a,b)}\left( x^{(a)},\,x^{(b)} \right)=\left(\delta^{(b,a+1)}+\delta^{(b,a-1)}\right)\,
  \prod\limits_{i=1}^{N_a}\left( x_i^{(a)} \right)^{-\beta\,N_{b}}
  \prod\limits_{i=1}^{N_a}\prod\limits_{j=1}^{N_{b}}\frac{\left( u
      \frac{x_j^{(b)}}{x_i^{(a)}}; q \right)_\infty}
  {\left( v \frac{x_j^{(b)}}{x_i^{(a)}};q\right)_\infty}\,,
\label{screening:order:2}
\end{equation}
and $v=\sqrt{qt^{-1}}, \,u=\sqrt{qt}$.

\item \textbf{Normal ordering the screening currents and vertex
    operators}
\begin{equation}
  V^q_{\vec{\alpha}}( z ) \prod\limits_{i=1}^{N_a} S^q_{(a)}\left(
    x_{i}^{(a)} \right) = C_q^{(a)}\left(x^{(a)}, z, \vec{\alpha}\right)
  \,:V^q_{\vec{\alpha}}( z )\,\prod\limits_{i=1}^{N_a}
  S^q_{(a)}\left( x_{i}^{(a)} \right):\,,
\end{equation}
where
\begin{multline}
  C_q^{(a)}\left( x^{(a)}, z, \vec{\alpha} \right)\sim
  \prod\limits_{i=1}^{N_a} \frac{\left( q^{\alpha_{a+1}}
      v^{-a}\frac{x_i^{(a)}}{z}; q \right)_\infty}
  {\left( q^{\alpha_{a}}v^{-a} \frac{x_i^{(a)}}{z}; q \right)_\infty}=\\
  =\prod\limits_{i=1}^{N_a}c_q\left( x_i^{(a)}, \vec{\alpha}, z
  \right) \left( x_i^{(a)} \right)^{\alpha_a-\alpha_{a+1}}
  \frac{\left( q^{1-\alpha_a} v^a \frac{z}{x_i^{(a)}};q
    \right)_\infty}{\left( q^{1-\alpha_{a+1}} v^{a}
      \frac{z}{x_i^{(a)}}; q \right)_\infty}\,,
\label{vertex:ordering}
\end{multline}
where we have omitted an overall $z$-dependent factor  and we have also introduced
the following $q$-periodic function:
\begin{equation}
c_q\left( x_i^{(a)}, \vec{\alpha}, z \right)=\left( x_i^{(a)} \right)^{\alpha_{a+1}-\alpha_a}
\frac{\theta_q\left( q^{\alpha_{a+1}}v^{-a} \frac{x_i^{(a)}}{z} \right)}
{\theta_q\left( q^{\alpha_a}v^{-a}\frac{x_i^{(a)}}{z} \right)}
\end{equation}

\item \textbf{Normal ordering of different vertex operators}
\be
\prod_{p=1}^l V_{\vec{\alpha}^{(p)} }=C_{\rm{vert}}^q\left(z,\vec{\alpha} \right)\normord{\prod_{p=1}^l V_{\vec{\alpha}^{(p)} }}\,,
\ee
where
\be
C_{\rm{vert}}^q\left(z,\vec{\alpha} \right)=\prod\limits_{p<r}^l\exp\left[ \sum\limits_{a=1}^n\sum\limits_{k>0}
\frac{q^k\left( q^{k\alpha_a^{(p)}}-1 \right)\left( q^{-k\alpha_a^{(r)}}-v^{2k(n-a-1) } \right)}{\left( 1-q^k \right)\left( 1-t^k \right)}
\times\right.\nonumber\\\left.
\frac{1}{k}\left( \frac{z_r}{z_p} \right)^k\right]\prod\limits_{p<r}^lz_p^{\frac{1}{\beta}\left( \vec{\alpha}^{(p)},\vec{\alpha}^{(r)} \right)}
=       \prod\limits_{a=1}^n\prod\limits_{p<r}^l \frac{\qPoc{q^{1-\alpha_a^{(r)}}\frac{z_r}{z_p};q,t}}
{\qPoc{q^{1+\alpha_a^{(p)}-\alpha_a^{(r)}}\frac{z_r}{z_p};q,t}}\times\nonumber\\
\frac{\qPoc{q^{1+\alpha_a^{(p)}}v^{2(n-a-1)}\frac{z_r}{z_p};q,t}}{\qPoc{qv^{2(n-a-1) } \frac{z_r}{z_p};q,t}}
\prod_{p<r}^lz_p^{\frac{1}{\beta}\left( \vec{\alpha}^{(p)},\vec{\alpha}^{(r)} \right) }\,,
\label{q:toda:order:vert}
\ee
where we have used the definition of $q$-Pochhammer symbol with multiple  parameters defined as follows
\be
\left( x;\,q_1,\dots,\,q_n \right)\stackrel{\mathrm{def}}{=} \prod\limits_{k_1,\dots,\,k_n\geq 1}
\left(1-x\,q_1^{k_1}\cdots q_n^{k_n}   \right)\,.
\label{qPoc:mult:def}
\ee

\item \textbf{Initial and final states} yield the same
  factor~(\ref{zero:modes:contrib}) as in the non-deformed
  case. Notice however, that in the $q$-deformed case one cannot use
  the operator-state correspondence to argue that the initial and
  final states are limits of the vertex
  operators~(\ref{vertex:ordering:toda}) for $z \to 0$ and $z \to
  \infty$ respectively.  We need simply to define the initial
  state~(\ref{state}) separately as the momentum operator eigenstate.
\end{enumerate}

Collecting all the factors we have obtained above, the free field correlator of
$(l+2)$ vertex operators is given by the following matrix integral:
\begin{multline}
  q{\rm DF}^{A_n}_{l+2} \sim C_{{\rm vert}}^{q}\left( \vec{\alpha},z \right)\prod\limits_p^lz_p^{\frac{1}{\beta}\left( \vec{\alpha}^{(p)},\vec{\alpha}^{(0)} \right)+
\sum\limits_{a=1}^nN_a\left( \alpha_a^{(p)}-\alpha_{a+1}^{(p)} \right)}\times\\
\oint
  \prod\limits_{a=1}^n\prod\limits_{i=1}^{N_a} d x^{(a)}_i
  \prod\limits_{a=1}^n\prod\limits_{i=1}^{N_a}\left( x_i^{(a)}
  \right)^{\beta(N_a-N_{a+1}-1)+(\alpha_a^{(0)}-\alpha_{a+1}^{(0)})
    +\sum\limits_{p=1}^l\left( \alpha_a^{(p)}-\alpha_{a+1}^{(p)}\right)}\times\\
  \times\prod\limits_{a=1}^n\prod\limits_{i\neq j}^{N_a}\frac{\left(
      \frac{x_j^{(a)}}{x_i^{(a)}};q \right)_\infty}
  {\left(t\frac{x_j^{(a)}}{x_i^{(a)}};q \right)_\infty}
  \prod_{a=1}^{n-1}\prod\limits_{i=1}^{N_a}\prod\limits_{j=1}^{N_{a+1}}\frac{\left(
      u \frac{x_j^{(a+1)}}{x_i^{(a)}}; q \right)_\infty}
  {\left( v \frac{x_j^{(a+1)}}{x_i^{(a)}};q\right)_\infty} \prod_{p=1}^l \prod_{a=1}^n\prod_{i=1}^{N_a}
  \frac{\left( q^{1-\alpha^{(p)}_{a}}v^a\frac{z_p}{x_i^{(a)}}; q
    \right)_\infty} {\left(
      q^{1-\alpha^{(p)}_{a+1}}v^{a}\frac{z_p}{x_i^{(a)}};q
    \right)_\infty}\,,
\end{multline}
where we have omitted a $q$-periodic function of $z_p$ in front of the
integral. Notice that in the $q\to 1$ limit the expression above
reduces to Eq.~(\ref{correlator1:toda}). To see this we should employ
the following identity for the $q\to 1$ limit of the ratio of two
$q$-Pochhammer symbols:
\begin{equation}
\lim_{q\to 1} \frac{\left(q^c x; q\right)_\infty}{\left( x; q
  \right)_\infty} = ( 1-x )^{-c},
\label{q:poc:limit}
\end{equation}
which we derive in the Appendix~\ref{appendix:limits}. Using this
formula get
\begin{multline}
  \lim_{q\to 1} q{\rm DF}^{A_n}_{l+2} \sim \prod\limits_{p<r} z_p^{\frac{1}{\beta}\left( \vec{\alpha}^{(p)},\vec{\alpha}^{(r)} \right)}
  \left(1-\frac{z_r}{z_p}\right)^{\frac{1}{\beta}\left( \vec{\alpha}^{(p)},\vec{\alpha}^{(r)} \right)+\sum_{a=1}^n\left(\beta^{-1}-1 \right)(n-a-1) \alpha_a^{(p)}  }\\
  \oint
  \prod\limits_{a=1}^n\prod\limits_{i=1}^{N_a} d x^{(a)}_i
  \prod\limits_{a=1}^n\prod\limits_{i=1}^{N_a}\left( x_i^{(a)}
  \right)^{\beta(N_a-N_{a+1}-1)+
    (\alpha_a^{(0)}-\alpha_{a+1}^{(0)})+\sum\limits_{p=1}^l\left( \alpha_a^{(p)}-\alpha_{a+1}^{(p)} \right)}\times\\
  \times\prod\limits_{a=1}^n\prod\limits_{i\neq j}^{N_a}\left(
    1-\frac{x_j^{(a)}}{x_i^{(a)}}\right)^\beta
  \prod_{a=1}^{n-1}\prod\limits_{i=1}^{N_a}\prod\limits_{j=1}^{N_{a+1}}\left(
    1-\frac{x_j^{(a+1)}}{x_i^{(a)}}\right)^{-\beta}
  \prod\limits_{p=1}^l\prod\limits_{a=1}^n\prod\limits_{i=1}^{N_a}
  \left( 1-\frac{z_p}{x_i^{(a)}}\right)^{\alpha_a^{(p)}-\alpha_{a+1}^{(p)}}\sim\\
\sim {\rm DF}^{A_n}_{l+2}\,,
\end{multline}
which coincides with the ordinary Toda conformal block~(\ref{correlator1:toda}) up to $z$-dependent prefactor coming from the normal ordering of 
vertices. This discrepancy happens since $q$-deformed vertex we defined in (\ref{screening:current}) includes so called $U(1)$ factor, which is required 
to match Toda conformal blocks with Nekrasov partition functions. For details see discussion after Eq.(\ref{2d:DF}) and \cite{toda:paper}.

\section{$q\to 1$ limits}
\label{appendix:limits}

In our work we use various formulas for the $q\to 1$ limit of $q$-Pochhammer symbols. 
In this appendix we give proofs for these formulas. 

We start with the derivation of the  standard formula for the following limit:
\begin{equation}
\lim_{q\to 1}\frac{\left( q^a\, x;\,q \right)_\infty}{\left(q^b\, x;\,q \right)_\infty}=\left( 1-x \right)^{b-a}\,,
\label{qto1limit}
\end{equation}
with $x$ variable held fixed during the limit. To prove this formula we need to take logarithm of the right hand 
side, use $q$-Pochhammer definition and perform expansion of the logarithms:
\begin{equation}
\lim_{q\to 1}\sum\limits_{n=0}^{\infty}\log\left(\frac{1-q^{a+n}x}{1-q^{b+n}x}\right)=\lim_{q\to 1} \sum\limits_{n=0}^\infty\sum\limits_{k=1}^\infty
\frac{q^{bk}-q^{ak}}{k}x^k\frac{1}{1-q^k}=\nonumber\\
\sum\limits_{k=1}^{\infty}(a-b)\frac{x^k}{k}=\log(1-x)^{b-a}\,,	
\end{equation}
which completes the proof of~(\ref{qto1limit}).

Second formula we would like to discuss is given in~(\ref{gamma:limit}):
\begin{equation}
\lim\limits_{q\to 1^{-}}\frac{(q;q)_\infty}{(q^x;q)_\infty}=\left( -\hbar \right)^{x-1}\Gamma(x)\,.
\end{equation}

To prove this relation we need to use the definition of $q$-Gamma function:
\begin{equation}
\Gamma_q(x)\equiv \frac{(q;q)_\infty}{(q^x;q)_\infty}(1-q)^{-x}\,,
\label{qGamma:def}
\end{equation}
Then it is known that 
\begin{equation}
\lim\limits_{q\to1^{-}}\Gamma_q(x)=\Gamma(x)\,.
\end{equation}

Here we provide the short proof of this limit due to Gasper \cite{Gasper,Andrews}. First of all 
we notice
\begin{equation}
\prod\limits_{n=1}^\infty\left( \frac{1-q^{n+1}}{1-q^n}\right)^x=
\frac{1}{(1-q)^x}\,.
\end{equation}
Using this identity we can rewrite the limit of $\Gamma_q(x+1)$ in the following form:
\begin{equation}
\lim\limits_{q\to 1^{-}}\Gamma_q(x+1)=\lim\limits_{q\to 1^{-}}
\prod\limits_{n=1}^\infty\frac{(1-q^n)(1-q^{n+1})^x}{(1-q^{n+x})(1-q^n)^x}=\nonumber\\
\prod\limits_{n=1}^\infty\left( 1+\frac{x}{n}\right)^{-1}\left( 1+\frac{1}{n}\right)^x=x\Gamma(x)=\Gamma(x+1)\,,
\end{equation}
which completes the proof of~(\ref{qGamma:def}).

\section{RS Hamiltonians and $T[SU(N)]$ holomorphic blocks}
\label{sec:ruijs-hamilt-tun}
In this Appendix we prove that $T[SU(N)]$ holomorphic block $ {\cal
  B}^{D_2\times S^1,\,(\alpha)}_{T[SU(N)]}(\vec{\mu},
\vec{\tau}, q, t)$ is an eigenfunction of the first RS Hamiltonian and
its $p$-$q$ dual. In other words, we prove Eq.~\eqref{eq:30} and
Eq.~\eqref{eq:55} for $r=1$.

The integral representation of the holomorphic
block~\eqref{TSUN:partition} has the form\footnote{Here we write the
  prefactor $F(q,t,\vec{\tau})$ in a form, which is
  equivalent if one uses the conditions~\eqref{eq:57} on the sum of
  masses and FI parameters. We also omit the overall constant
  independent of $T_i$ and $M_i$.}
\begin{multline}
  \label{eq:35}
  {\cal B}^{D_2\times S^1,\,(\alpha)}_{T[SU(N)]}(\vec{\mu},
  \vec{\tau}, q, t) =  e^{\frac{T_N \sum_{i=1}^N
      M_i}{\hbar} + (1-\beta)\sum_{i=1}^N \left( \frac{1}{2} - i
    \right) T_i} \int_{\Gamma_{\alpha}} \prod_{a=1}^{N-1}
  \prod_{i=1}^a \left( \frac{dx_i^{(a)}}{x_i^{(a)}}
    (x_i^{(a)})^{\frac{T_a -
        T_{a+1}}{\hbar} - \beta } \right)\times\\
  \times \frac{\Delta^{(q,t)}(\vec{x}^{(2)})\cdots
    \Delta^{(q,t)}(\vec{x}^{(N-1)})}{ \bar{\Delta}^{(q,t)}
    (\vec{x}^{(1)},\vec{x}^{(2)})\cdots \bar{\Delta}^{(q,t)}
    (\vec{x}^{(N-2)},\vec{x}^{(N-1)}) \bar{\Delta}^{(q,t)}
    (\vec{x}^{(N-1)},\vec{\mu})}
\end{multline}
where
\begin{equation}
  \label{eq:36}
  \Delta^{(q,t)}(\vec{x}^{(a)}) = \frac{1}{(t;q)_{\infty}^a} \prod_{i\neq j}^a \frac{\left(
      \frac{x_i^{(a)}}{x_j^{(a)}} ;q \right)_{\infty}}{\left( t
      \frac{x_i^{(a)}}{x_j^{(a)}} ;q \right)_{\infty}},\qquad
  \bar{\Delta}^{(q,t)}(\vec{x}^{(a)}, \vec{x}^{(a+1)}) = \prod_{i=1}^{a+1}
  \prod_{j=1}^a \frac{\left(
      \frac{x_i^{(a+1)}}{x_j^{(a)}} ;q \right)_{\infty}}{\left( 
      t \frac{x_i^{(a+1)}}{x_j^{(a)}} ;q \right)_{\infty}}
\end{equation}

\subsection{Hamiltonian in $\vec{\mu}$ variables}
\label{sec:hamilt-vecmu-vari}
The Hamiltonian is given by
\begin{equation}
  \label{eq:34}
  H_1 (\mu_i, q^{\mu_i \partial_{\mu_i}}, q, t) = \sum_{i=1}^N
  \prod_{j \neq i}^N \frac{t \mu_i - \mu_j}{\mu_i - \mu_j} q^{\mu_i \partial_{\mu_i}}.
\end{equation}

When the Hamiltonian acts on $ {\cal B}^{D_2\times
  S^1,\,(\alpha)}_{T[SU(N)]}(\vec{\mu}, \vec{\tau}, q, t)$
it acts on the last factor $\bar{\Delta}^{(q,t)}
(\vec{x}^{(N-1)},\vec{\mu})$. The result is
\begin{equation}
  \label{eq:37}
  H_1(\mu_i, q^{\mu_i \partial_{\mu_i}}, q, t) \frac{1}{\bar{\Delta}^{(q,t)}
(\vec{x}^{(N-1)},\vec{\mu})} = \frac{1}{\bar{\Delta}^{(q,t)}
(\vec{x}^{(N-1)},\vec{\mu})} \sum_{i=1}^N
  \prod_{j \neq i}^N \frac{t \mu_i - \mu_j}{\mu_i - \mu_j}
  \prod_{l=1}^{N-1} \frac{x_l^{(N-1)} - \mu_i}{x_l^{(N-1)} - t \mu_i}
\end{equation}
Next we use we write the r.h.s.\ of Eq.~\eqref{eq:37} as a contour integral
in auxiliary variable $z$:
\begin{equation}
  \label{eq:38}
  \sum_{i=1}^N
  \prod_{j \neq i}^N \frac{t \mu_i - \mu_j}{\mu_i - \mu_j}
  \prod_{l=1}^{N-1} \frac{x_l^{(N-1)} - \mu_i}{x_l^{(N-1)} - t \mu_i}
  = \frac{1}{t-1}\oint_{\mathcal{C}_{\mu}} \frac{dz}{z} \prod_{j=1}^N
  \frac{t z - \mu_j}{z- \mu_j} \prod_{l=1}^{N-1} \frac{x_l^{(N-1)} -
    z}{x_l^{(N-1)} - t z}
\end{equation}
where $\mathcal{C}_{\mu}$ encircles all the points $\mu_i$. We can now
deform the contour $\mathcal{C}_{\mu}$ so that it encircles all the
other poles of the integral. Those are located at
\begin{enumerate}
\item $z=0$:
  \begin{equation}
    \label{eq:39}
    - \frac{1}{t-1} \oint_{\mathcal{C}_0}\frac{dz}{z} \prod_{j=1}^N
  \frac{t z - \mu_j}{z- \mu_j} \prod_{l=1}^{N-1} \frac{x_l^{(N-1)} -
    z}{x_l^{(N-1)} - t z} = - \frac{1}{t-1}
  \end{equation}

\item $z =\infty$:
  \begin{equation}
    \label{eq:40}
     - \frac{1}{t-1}\oint_{\mathcal{C}_{\infty}}\frac{dz}{z} \prod_{j=1}^N
  \frac{t z - \mu_j}{z- \mu_j} \prod_{l=1}^{N-1} \frac{x_l^{(N-1)} -
    z}{x_l^{(N-1)} - t z} = - \frac{t}{t-1}
  \end{equation}

\item $z=\frac{x_l^{(N-1)}}{t}$ for $l=1,\ldots N-1$:
  \begin{multline}
    \label{eq:41}
     - \frac{1}{t-1} \oint_{\mathcal{C}_{\frac{x}{t}}}\frac{dz}{z} \prod_{j=1}^N
  \frac{t z - \mu_j}{z- \mu_j} \prod_{l=1}^{N-1} \frac{x_l^{(N-1)} -
    z}{x_l^{(N-1)} - t z} =\\
  =t \sum_{i=1}^N \prod_{j\neq i}^{N-1}
  \frac{x_i^{(N-1)} - t x_j^{(N-1)}}{x_i^{(N-1)} - x_j^{(N-1)}}
  \prod_{p=1}^N \frac{x_i^{(N-1)} - \mu_p}{x_i^{(N-1)} - t \mu_p}
  \end{multline}
\end{enumerate}
Summing the residues we get an identity:
\begin{multline}
  \label{eq:42}
  \sum_{i=1}^N
  \prod_{j \neq i}^N \frac{t \mu_i - \mu_j}{\mu_i - \mu_j}
  \prod_{j=1}^{N-1} \frac{x_j^{(N-1)} - \mu_i}{x_j^{(N-1)} - t
    \mu_i}=\\
  = 1 + t \sum_{i=1}^{N-1} \prod_{j\neq i}^{N-1}
  \frac{x_i^{(N-1)} - t x_j^{(N-1)}}{x_i^{(N-1)} - x_j^{(N-1)}}
  \prod_{p=1}^N \frac{x_i^{(N-1)} - \mu_p}{x_i^{(N-1)} - t \mu_p}.
\end{multline}
On the next step we express the last product in the r.h.s.\ of
Eq.~\eqref{eq:42} as the action of difference operator in
$x_i^{(N-1)}$ on $\bar{\Delta}^{(q,t)}
(\vec{x}^{(N-1)},\vec{\mu})$:
\begin{equation}
  \label{eq:43}
   \prod_{p=1}^N \frac{x_i^{(N-1)} - \mu_p}{x_i^{(N-1)} - t \mu_p} \frac{1}{\bar{\Delta}^{(q,t)}
(\vec{x}^{(N-1)},\vec{\mu})} = q^{-x_i^{(N-1)} \partial_{x_i^{(N-1)}}} \frac{1}{\bar{\Delta}^{(q,t)}
(\vec{x}^{(N-1)},\vec{\mu})}
\end{equation}
We can now shift the integration variable $x_i^{(N-1)}$ by $q$ under
the integral (provided that the new contour $q \Gamma_{\alpha}$
contains the same poles as $\Gamma_{\alpha}$). The shift can be
understood as integration by parts:
\begin{equation}
  \label{eq:44}
  \oint_{\Gamma} \frac{dx}{x} f(x) q^{-x \partial_x} g(x) =
  \oint_{q \Gamma} \frac{dx}{x} \left( q^{x \partial_x} f(x) \right) g(x).
\end{equation}
The shift operator $q^{x_i^{(N-1)} \partial_{x_i^{(N-1)}}}$ thus acts
on all the $x_i^{(N-1)}$-dependent parts of the integrand except
$\bar{\Delta}^{(q,t)} (\vec{x}^{(N-1)},\vec{\mu})$:
\begin{enumerate}
\item The FI parameters:
  \begin{equation}
    \label{eq:46}
     q^{x_i^{(N-1)} \partial_{x_i^{(N-1)}}} (x_i^{(N-1)})^{\frac{T_{N-1} -
        T_{N}}{\hbar} - \beta } = \frac{\tau_{N-1}}{t \tau_N} (x_i^{(N-1)})^{\frac{T_{N-1} -
        T_{N}}{\hbar} - \beta }
  \end{equation}

\item The remaining terms in the Hamiltonian:
  \begin{equation}
    \label{eq:47}
    q^{x_i^{(N-1)} \partial_{x_i^{(N-1)}}} \prod_{j\neq i}^{N-1}
    \frac{x_i^{(N-1)} - t x_j^{(N-1)}}{x_i^{(N-1)} - x_j^{(N-1)}} = \prod_{j\neq i}^{N-1}
    \frac{q x_i^{(N-1)} - t x_j^{(N-1)}}{q x_i^{(N-1)} - x_j^{(N-1)}}
  \end{equation}
  
\item The $q$-Vandermond determinant $\Delta^{(q,t)}(\vec{x}^{N-1})$:
\begin{equation}
  \label{eq:45}
  q^{x_i^{(N-1)} \partial_{x_i^{(N-1)}}}
  \Delta^{(q,t)}(\vec{x}^{N-1}) = \prod_{j\neq i}^{N-1}
  \frac{(q x^{(N-1)}_i - x^{(N-1)}_j)(t x^{(N-1)}_i - x^{(N-1)}_j)}{(x^{(N-1)}_i - x^{(N-1)}_j)(q x^{(N-1)}_i - t x^{(N-1)}_j)}   \Delta^{(q,t)}(\vec{x}^{N-1})
\end{equation}

\item The interaction term between $x_i^{(N-1)}$ and $x_i^{(N-2)}$
  $\bar{\Delta}^{(q,t)} (\vec{x}^{(N-2)},\vec{x}^{(N-1)})$. We will
  leave this unevaluated for a moment.
\end{enumerate}
Notice that there is a cancellation between the
contributions~\eqref{eq:47} and~\eqref{eq:45}. Finally we get:
\begin{multline}
  \label{eq:32}
  H_1(\mu_i, q^{\mu_i \partial_{\mu_i}}, q, t) {\cal B}^{D_2\times
    S^1,\,(\alpha)}_{T[SU(N)]}(\vec{\mu}, \vec{\tau}, q, t)
  = \\
  = F(q,t,\vec{\tau}) \int_{\Gamma_{\alpha}} \prod_{a=1}^{N-1} \prod_{i=1}^a
  \left( \frac{dx_i^{(a)}}{x_i^{(a)}} (x_i^{(a)})^{\frac{T_a -
        T_{a+1}}{\hbar} - \beta } \right)   \frac{\Delta^{(q,t)}(\vec{x}^{(N-1)})}{ \bar{\Delta}^{(q,t)}(\vec{x}^{(N-1)},\vec{\mu})}\times\\
  \times \left( \tau_N + \tau_{N-1} \sum_{i=1}^{N-1} \prod_{j\neq
      i}^{N-1} \frac{t x^{(N-1)}_i - x^{(N-1)}_j}{x^{(N-1)}_i -
      x^{(N-1)}_j} q^{x_i^{(N-1)} \partial_{x_i^{(N-1)}}} \right)
  \frac{\Delta^{(q,t)}(\vec{x}^{(2)})\cdots
    \Delta^{(q,t)}(\vec{x}^{(N-2)})}{ \bar{\Delta}^{(q,t)}
    (\vec{x}^{(1)},\vec{x}^{(2)})\cdots
    \bar{\Delta}^{(q,t)}(\vec{x}^{(N-2)},\vec{x}^{(N-1)})}=\\
  = F(q,t,\vec{\tau}) \int_{\Gamma_{\alpha}}
  \prod_{a=1}^{N-1} \prod_{i=1}^a \left( \frac{dx_i^{(a)}}{x_i^{(a)}}
    (x_i^{(a)})^{\frac{T_a -
        T_{a+1}}{\hbar} - \beta } \right)   \frac{\Delta^{(q,t)}(\vec{x}^{(N-1)})}{ \bar{\Delta}^{(q,t)}(\vec{x}^{(N-1)},\vec{\mu})}\times\\
  \times \left( \tau_N + \tau_{N-1} H_1 \left(x_i^{(N-1)},
      q^{x_i^{(N-1)} \partial_{x_i^{(N-1)}}}, q, t\right) \right)
  \frac{\Delta^{(q,t)}(\vec{x}^{(2)})\cdots
    \Delta^{(q,t)}(\vec{x}^{(N-2)})}{ \bar{\Delta}^{(q,t)}
    (\vec{x}^{(1)},\vec{x}^{(2)})\cdots
    \bar{\Delta}^{(q,t)}(\vec{x}^{(N-2)},\vec{x}^{(N-1)})}.
\end{multline}
We have obtained the same operator $H_1$, but now acting on the
variables $\vec{x}^{(N-1)}$! Thus the recursion begins. We can move
the operator $H_1$ on the variables $\vec{x}^{(N-2)}$ and so on until
we reach $x_1^{(1)}$ and get a trivial result. What remains is the
eigenvalue and the initial holomorphic block:
\begin{multline}
  \label{eq:48}
  H_1(\mu_i, q^{\mu_i \partial_{\mu_i}}, q, t) {\cal B}^{D_2\times
    S^1,\,(\alpha)}_{T[SU(N)]}(\vec{\mu}, \vec{\tau}, q,
  t)=\\
  = \left( \tau_N + \tau_{N-1} + \tau_{N-2}
    + \cdots + \tau_1 \right) {\cal B}^{D_2\times
    S^1,\,(\alpha)}_{T[SU(N)]}(\vec{\mu}, \vec{\tau}, q, t)
  = e_1 (\vec{\tau}) {\cal B}^{D_2\times
    S^1,\,(\alpha)}_{T[SU(N)]}(\vec{\mu}, \vec{\tau}, q, t)
\end{multline}

\subsection{$p$-$q$ dual Hamiltonian in $\vec{\tau}$ variables for
  $T[U(2)]$ theory}
\label{sec:hamilt-vect-vari}
The Hamiltonian which is $p$-$q$ dual to Eq.~\eqref{eq:37} reads
\begin{equation}
  \label{eq:49}
    H_1 \left(\tau_i, q^{\tau_i \partial_{\tau_i}}, q, \frac{q}{t} \right) = \sum_{i=1}^N
  \prod_{j \neq i}^N \frac{\frac{q}{t} \tau_i - \tau_j}{\tau_i - \tau_j} q^{\tau_i \partial_{\tau_i}}.
\end{equation}
We limit ourselves to the $T[U(2)]$ holomorphic block:
\begin{equation}
  \label{eq:50}
  {\cal B}^{D_2\times
    S^1,\,(\alpha)}_{T\left[U(2)\right]}(\vec{\mu}, \vec{\tau}, q, t)
  = e^{\frac{T_2 (M_1 + M_2 + (\beta-1)\hbar)}{\hbar}} \int_{\Gamma_{\alpha}} 
  \frac{dx}{x}  x^{\frac{T_1 -
      T_2}{\hbar} - \beta }  \frac{\left( t \frac{\mu_1}{x};q
    \right)_{\infty} \left( t \frac{\mu_2}{x};q
    \right)_{\infty} }{\left(  \frac{\mu_1}{x};q
    \right)_{\infty} \left(  \frac{\mu_2}{x};q
    \right)_{\infty}}
\end{equation}
The Hamiltonian~\eqref{eq:49} acts on the power of $x$ producing the
following in integral
\begin{multline}
  \label{eq:51}
  H_1 \left(\tau_i, q^{\tau_i \partial_{\tau_i}}, q, \frac{q}{t}
  \right) {\cal B}^{D_2\times
    S^1,\,(\alpha)}_{T\left[U(2)\right]}(\vec{\mu}, \vec{\tau}, q, t)
  = \\
  =e^{\frac{T_2 (M_1 + M_2 + (\beta-1)\hbar)}{\hbar}} \int_{\Gamma_{\alpha}}
  \frac{dx}{x} \left[ \frac{\frac{q}{t} \tau_1 - \tau_2}{\tau_1 -
      \tau_2} x + \frac{\frac{q}{t} \tau_1 - \tau_2}{\tau_1 - \tau_2}
    \frac{t\mu_1 \mu_2}{q x} \right] x^{\frac{T_1 - T_2}{\hbar} - \beta }
  \frac{\left( t \frac{\mu_1}{x};q \right)_{\infty} \left( t
      \frac{\mu_2}{x};q \right)_{\infty} }{\left( \frac{\mu_1}{x};q
    \right)_{\infty} \left( \frac{\mu_2}{x};q \right)_{\infty}}
\end{multline}
We need to evaluate the integral with the insertion of the term in the
square brackets. To this end consider the the following integral of a
particular total difference:
\begin{equation}
  \label{eq:52}
  0 = \int_{\Gamma_{\alpha}}
  \frac{dx}{x} \left( 1 - q^{x \partial_x} \right) \left\{ x \left( 1 -
    \frac{\mu_1}{x} \right) \left( 1 - \frac{\mu_2}{x} \right) x^{\frac{T_1 - T_2}{\hbar} - \beta }
  \frac{\left( t \frac{\mu_1}{x};q \right)_{\infty} \left( t
      \frac{\mu_2}{x};q \right)_{\infty} }{\left( \frac{\mu_1}{x};q
    \right)_{\infty} \left( \frac{\mu_2}{x};q \right)_{\infty}} \right\}.
\end{equation}
Acting with the shift operator on every term in the curly brackets one
gets the following:
\begin{multline}
  \label{eq:53}
  0 = \int_{\Gamma_{\alpha}}
  \frac{dx}{x} \left[ x \left( 1 -
    \frac{\mu_1}{x} \right) \left( 1 - \frac{\mu_2}{x} \right) -
  \frac{q\tau_1}{t\tau_2} x \left( 1 -
    \frac{t\mu_1}{qx} \right) \left( 1 - \frac{t \mu_2}{qx} \right)\right] x^{\frac{T_1 - T_2}{\hbar} - \beta }
  \frac{\left( t \frac{\mu_1}{x};q \right)_{\infty} \left( t
      \frac{\mu_2}{x};q \right)_{\infty} }{\left( \frac{\mu_1}{x};q
    \right)_{\infty} \left( \frac{\mu_2}{x};q \right)_{\infty}} =\\
  \int_{\Gamma_{\alpha}}
  \frac{dx}{x} \left[ x \left( 1 - \frac{q \tau_1}{t \tau_2} \right) -
  (\mu_1 + \mu_2) \left( 1- \frac{\tau_1}{\tau_2} \right) + \frac{\mu_1 \mu_2}{x} \left( 1 - \frac{t \tau_1}{q
    \tau_2} \right)  \right] x^{\frac{T_1 - T_2}{\hbar} - \beta }
  \frac{\left( t \frac{\mu_1}{x};q \right)_{\infty} \left( t
      \frac{\mu_2}{x};q \right)_{\infty} }{\left( \frac{\mu_1}{x};q
    \right)_{\infty} \left( \frac{\mu_2}{x};q \right)_{\infty}}.
\end{multline}
We observe that miraculously the expression in square brackets
contains exactly the combination of powers of $x$ appearing in
Eq.~\eqref{eq:51} with the right coefficients. What remains is the
scalar factor, which gives the eigenvalue. We thus get
\begin{equation}
  \label{eq:54}
   H_1 \left(\tau_i, q^{\tau_i \partial_{\tau_i}}, q, \frac{q}{t}
  \right) {\cal B}^{D_2\times
    S^1,\,(\alpha)}_{T\left[U(2)\right]}(\vec{\mu}, \vec{\tau}, q, t)
  = (\mu_1 + \mu_2) {\cal B}^{D_2\times
    S^1,\,(\alpha)}_{T\left[U(2)\right]}(\vec{\mu}, \vec{\tau}, q, t).
\end{equation}
Finally one can notice that
\begin{multline}
  \label{eq:64}
  H_2 \left(\tau_i, q^{\tau_i \partial_{\tau_i}}, q, \frac{q}{t}
  \right) {\cal B}^{D_2\times
    S^1,\,(\alpha)}_{T\left[U(2)\right]}(\vec{\mu}, \vec{\tau}, q, t)
  = \frac{q}{t} q^{\tau_1 \partial_{\tau_1} +
    \tau_2 \partial_{\tau_2}} {\cal B}^{D_2\times
    S^1,\,(\alpha)}_{T\left[U(2)\right]}(\vec{\mu}, \vec{\tau}, q,
  t)=\\
  = \mu_1 \mu_2 {\cal B}^{D_2\times
    S^1,\,(\alpha)}_{T\left[U(2)\right]}(\vec{\mu}, \vec{\tau}, q, t).
\end{multline}
\end{appendix}

\bibliographystyle{JHEP}
\bibliography{vortex}

\end{document}